\documentclass[11pt,3p,review,authoryear]{elsarticle}

\biboptions{longnamesfirst}
\usepackage[dvipsnames]{xcolor}
\usepackage{color} 
\usepackage{soul,framed} 
\usepackage{mathtools}
\usepackage{lscape}
\usepackage{multirow}
\usepackage{booktabs}
\colorlet{shadecolor}{yellow}
\usepackage{natbib}
\usepackage[ruled,vlined]{algorithm2e}
\usepackage{amsmath}
\usepackage{amsfonts}
\usepackage{amssymb}
\usepackage{subcaption}
\usepackage{hyperref}
\usepackage{lscape}
\usepackage{array}

\usepackage[utf8]{inputenc}
\usepackage{placeins}

\usepackage{blindtext}
\usepackage{dirtytalk}
\usepackage{adjustbox, xcolor}
\usepackage{pdfpages} 
\usepackage{pgffor} 
\usepackage{xr}
\makeatletter
\AtBeginDocument{\let\LS@rot\@undefined}
\cleardoublepage\makeatother

\usepackage[shortlabels]{enumitem}
\usepackage{mdwtab}
\usepackage{eqparbox}
\usepackage{url}

\usepackage{mathtools}
\usepackage{dsfont}
\usepackage{bbold}
\usepackage[english]{babel} 
\usepackage{blindtext}
\usepackage{epstopdf}
\DeclareGraphicsExtensions{.png,.pdf,.jpg}
\usepackage{float}
\usepackage{hyperref}
\hypersetup{
    colorlinks=true,
    linkcolor=blue,
    filecolor=magenta,      
    urlcolor=cyan,
    pdftitle={Overleaf Example},
    pdfpagemode=FullScreen,
    }
\usepackage[utf8]{inputenc}
\usepackage{blindtext}
\usepackage{appendix}
\usepackage{chngcntr}

\makeatletter
\def\ps@pprintTitle{%
  \let\@oddhead\@empty
  \let\@evenhead\@empty
  \let\@oddfoot\@empty
  \let\@evenfoot\@empty
}
\makeatother

\begin{document}
\begin{frontmatter}

\title{Macroeconomic Forecasting for the G7 countries under Uncertainty Shocks}

\author{Shovon Sengupta$^{1,2,3}$, 
Sunny Kumar Singh$^{2}$, 
Tanujit Chakraborty$^{1,4}$\\
{\scriptsize \textsuperscript{1} SAFIR, Sorbonne University Abu Dhabi, United Arab Emirates.}\\
{\scriptsize \textsuperscript{2} BITS Pilani, Hyderabad Campus, India.} \\
{\scriptsize \textsuperscript{3} Fidelity Investments, Boston, USA.}
\\
{\scriptsize \textsuperscript{4}Sorbonne Center for Artificial Intelligence, Sorbonne University, Paris, France}}


\begin{abstract}
Accurate macroeconomic forecasting has become harder amid geopolitical disruptions, policy reversals, and volatile financial markets. Conventional vector autoregressions (VARs) overfit in high-dimensional settings, while threshold VARs struggle with time-varying interdependencies and complex parameter structures. We address these limitations by extending the Sims–Zha Bayesian VAR with exogenous variables (SZBVARx) to incorporate domain-informed shrinkage and four newspaper-based uncertainty shocks—economic policy uncertainty, geopolitical risk, US equity market volatility, and US monetary policy uncertainty. The framework improves structural interpretability, mitigates dimensionality, and imposes empirically guided regularization. Using G7 data, we study spillovers from uncertainty shocks to five core variables—unemployment, real broad effective exchange rates, short-term rates, oil prices, and CPI inflation—combining wavelet coherence (time–frequency dynamics) with nonlinear local projections (state-dependent impulse responses). Out-of-sample results at 12- and 24-month horizons show that SZBVARx outperforms 14 benchmarks, including classical VARs and leading machine-learning models, as confirmed by Murphy difference diagrams, multivariate Diebold–Mariano tests, and Giacomini–White predictability tests. Credible Bayesian prediction intervals deliver robust uncertainty quantification for scenario analysis and risk management. The proposed SZBVARx offers G7 policymakers a transparent, well-calibrated tool for modern macroeconomic forecasting under pervasive uncertainty.
\end{abstract}

\end{frontmatter}

\section{Introduction}
Uncertainty is deeply ingrained in the growth and development process of the G7 economies: Canada, France, Germany, Italy, Japan, the United Kingdom, and the United States. The economic importance of accurate macroeconomic forecasting for these countries is self-evident. Collectively, the G7 economies account for about 28–30\% of global GDP, more than 40\% of global household wealth, and are home to the deepest and most liquid financial markets in the world \citep{imf2024weo, creditsuisse2023globalwealth}. The G7 also host four of the world's five largest currencies, with the US dollar, euro, yen, and pound sterling together accounting for about 89\% of disclosed official foreign-exchange reserves. For these reasons, the G7 economies are by far the largest anchors of the international monetary system \citep{imf2024cofer,gopinath2020dollar}. The G7 is also home to the majority of systemically important financial institutions and is the world's largest source and recipient of cross-border capital flows \citep{rey2015dilemma, mirandaagrippino2021transmission}. Given this significant footprint in world trade, finance, and monetary conditions, shocks that originate in the G7 or are transmitted through it have sizable spillover effects on emerging markets and the global economy more broadly \citep{bloom2020really}. The G7's growth outlook continues to be buffeted by a perfect storm of adverse global shocks (including financial crises, political instability, technological change, wars, pandemics, and more), of which uncertainty, shocks, and their transmission are key \citep{baker2016measuring, caldara2022measuring, bloom2009impact}. At the heart of this storm lie the network interactions among five key macroeconomic variables (unemployment rate, real broad effective exchange rate (REER), short-term interest rates (SIR), oil prices, and consumer price index (CPI) inflation), which are not only highly interrelated but also extremely sensitive to exogenous uncertainty shocks of all types (e.g., economic and monetary policy uncertainties, geopolitical risk, financial market volatility etc.). The shocks have been becoming more frequent and larger in magnitude, which has severe and measurable adverse impacts on expectations, financial stability, inflation, and real activity in the G7 economies. Forecasting frameworks can no longer be based only on fundamentals but must explicitly account for high-frequency information on uncertainty, which continues to be generated at an unprecedented speed and scale. Empirical studies in recent literature indicate that uncertainty shocks feed into each other (i.e., so-called ``uncertainty multipliers"), which change the well-documented causal relationships among key macroeconomic variables, destabilize labor markets and investment, and result in volatile swings in exchange rates, flows, and prices \citep{caggiano2014uncertainty, smales2022influence, rossi2013exchange, yang2024economic}. We treat them as core drivers of G7 forecasts.

In this paper, we examine the response of five macroeconomic variables—unemployment rate, REER, SIR, the oil price, and CPI inflation—to four sources of uncertainty shocks: economic policy uncertainty (EPU), geopolitical risk (GPR), US equity market volatility (USEMV), and US monetary policy uncertainty (USMPU). Such shocks, which encompass a range of uncertainty from policy-related disagreements to geopolitical tensions, are recognized as major driving forces of the outlook in advanced economies, and particularly G7 countries \citep{jones2013time, baker2016measuring, caldara2022measuring}. Our work, therefore, is timely, as their increasing prominence in the economy suggests that they should be included in the forecasting system to understand and monitor the macroeconomic implications of global uncertainty and, potentially, to inform policy. The ongoing process of deepening interdependence across countries is a major motivation for multivariate forecasting as a key tool for G7 policymakers to monitor the interdependent co-movements of unemployment, inflation, exchange rates, interest rates, and commodity prices. 

This article presents a structured platform to link the different uncertainty shocks to outcomes by emphasizing the G7 economy. The labor market, foreign exchange market, monetary policy, oil market, and CPI inflation outcomes are observed to be highly responsive to the jointly moving EPU, GPR, USEMV, and USMPU variables in the G7 \citep{baker2016measuring, caldara2022measuring, bloom2009impact, caggiano2014uncertainty, smales2022influence, jurado2015measuring}. At the same time, pronounced spikes in EPU and GPR are linked to deteriorations in employment, volatile inflation and interest rates, shifting risk premia, and volatile exchange rates, particularly in the context of systemic cross-country shocks \citep{baker2016measuring, caldara2022measuring, bloom2009impact, caggiano2014uncertainty, smales2022influence, jurado2015measuring}. These can propagate quickly through financial and trade linkages, inducing capital flows and currency realignments \citep{forbes2012capital, rossi2013exchange, ahir2022world, bekaert2013risk}. Labor market responses are also particularly important, with hiring and investment often retrenching in the face of rising uncertainty \citep{bloom2009impact, caggiano2014uncertainty}. By contrast, exchange rates tend to track sudden, uncertainty-induced capital flows, \citep{forbes2012capital, smales2022influence}. The recent USMPU can also transmit the interest rate, and particularly so when mixed with EPU and GPR \citep{husted2020monetary, tillmann2020monetary}. Oil price shocks are another potentially strong source of inflation and growth uncertainty, which have been found to respond strongly to policy and geopolitical news and interact with financial speculation \citep{kilian2009not, baumeister2016forty, kilian2014role, kang2017oil}. In addition, the latest inflation evidence in the G7 shows it is now more sensitive to external uncertainty shocks and policy changes than to domestic slack \citep{coibion2019inflation, fajgelbaum2020return}. Recent developments in the cross-country modeling frontier are therefore especially relevant for modeling the global nature of the interactions and support the use of multiple sources of uncertainty when forecasting the policy-relevant variables \citep{bai2022macroeconomic, baker2016measuring, caldara2022measuring}.

In a data-rich environment like the G7, Bayesian VAR models, proposed by \cite{litterman1986forecasting, sims1983forecasting, sims1998bayesian} have long been recognized as a well-performing benchmark. Recent developments, including hierarchical shrinkage priors \citep{banbura2010large}, stochastic volatility priors \citep{koop2013forecasting}, and extensions to multi-country or high-dimensional macro systems \citep{miranda2019bayesian, bai2022macroeconomic,besher2025forecasting} continue to improve the accuracy and flexibility of these models \citep{carriero2019large, feldkircher2022approximate, gefang2023forecasting}. On the empirical front, there is the need to include external drivers into the forecasting models, such as EPU \citep{baker2016measuring}, GPR \citep{caldara2022measuring}, and financial or monetary shocks \citep{husted2020monetary, tang2015uncertainty}, as they have a significant impact on inflation, exchange rates, and other macro dynamics \citep{carriero2019comprehensive, sengupta2025forecasting, bojer2022understanding,chakraborty2025neural}. At the same time, a rich set of methodological advances, such as Bayesian Global VARs \citep{canova2004forecasting, pesaran2009forecasting, cuaresma2016forecasting}, time-varying parameter VARs \citep{koop2013large, bekiros2014forecasting}, and deep learning models for multi-horizon prediction \citep{laborda2023multi}, are being used to capture a range of complex features in the data, such as nonlinearities, regime changes, and spillovers. 

Despite these advancements, classical and Bayesian VARs remain prominent forecasting \\ workhorses but face limitations in capturing nonlinear dynamics, structural breaks, and uncertainty shock propagation \citep{sims1998bayesian, koop2013large, canova2011methods, kilian2017structural}. Threshold and Markov-switching VARs introduce regime shifts, yet their complexity constrains practical use in high-dimensional G7 environments \citep{balke1997threshold, lo2001threshold}. Even recent time-varying parameter Bayesian VARs with structural identification can overlook key external policy and financial shocks \citep{koop2013large, del2008forming, carriero2019large}. The Sims-Zha Bayesian VAR with exogenous variables (SZBVARx) framework directly incorporates major uncertainty sources \citep{baker2016measuring, caldara2022measuring, husted2020monetary}. Unlike previous VAR approaches to uncertainty, SZBVARx allows these shocks to have unrestricted effects on all of the core macroeconomic variables. 
In contrast to some large-scale multi-country models, SZBVARx allows for a good degree of country-level granularity but also explicitly introduces cross-country spillovers using a common set of exogenous variables \citep{pesaran2009forecasting, cuaresma2016forecasting}. Bayesian shrinkage in the SZBVARx model helps to ensure reliable estimation, especially in high-dimensional environments \citep{belmonte2014hierarchical}. 

Against this background, the present paper contributes to the existing literature on three main fronts. \textit{First}, on methods and theory, we propose a framework in which uncertainty is explicitly modeled as an autonomous and structural driver of the G7 macroeconomy. We directly incorporate four externally measured uncertainty indices in a Sims–Zha Bayesian VAR with exogenous variables \citep{jurado2015measuring,ludvigson2021uncertainty,baker2016measuring,caldara2022measuring}, and treat uncertainty as a predictive state variable that augments more standard shocks. We couple nonlinear local projections \citep{carriero2019comprehensive} to recover regime-dependent impulse responses with shrinkage-based Bayesian estimation for robust inference in moderately large systems. We use Bayesian posterior sampling to construct predictive credible intervals that account for forecast uncertainty, providing empirically calibrated credible intervals \citep{banbura2010large, carriero2019comprehensive, koop2013forecasting}. 
\textit{Second}, on evidence, we consider five major macroeconomic variables for G7 countries. Long-run wavelet coherence confirms that uncertainty indices systematically lead macroeconomic variables along the business cycle and during major events like the global financial crisis and COVID-19, highlighting their value as truly exogenous drivers. State-dependent impulse responses reveal a number of regime asymmetries, with uncertainty shocks being both more impactful on prices and real quantities under tight monetary conditions and more persistent in easy regimes, especially through external channels for open economies. A thorough forecast evaluation against fourteen classical, machine-learning, and deep-learning benchmarks at 12- and 24-month horizons further highlights consistent gains in accuracy across a number of standard loss functions, validated by Multiple Comparisons with the Best, multivariate Diebold–Mariano and Giacomini–White tests, as well as Murphy-difference diagrams. \textit{Third}, on application, our approach delivers output in a form that can be directly utilized by central banks and other policy institutions, including regime-conditioned impulse responses, coherent medium-term projections, and thus ready to be used in real time to assess, monitor, and report measures of forecast uncertainty, aiding in risk management, forward guidance, and scenario planning, as needed in environments of geopolitical or financial uncertainty 
\citep{bernanke2015courage, adrian2019vulnerable, fernandez2011risk, huber2023nowcasting}. This paper bridges the gap between state-of-the-art econometric methods and the real-world need for uncertainty-aware forecasting of advanced economies.


The article is organized as follows. Section~\ref{Section_Data_and_Methodology} presents the data and empirical setup. Section~\ref{sec:Causal_anal_wca_irfs} provides wavelet-coherence evidence on G7 uncertainty transmissions and an impulse-response analysis under different monetary regimes. Section~\ref{Section_SZBVARx_Model} details the SZBVARx framework, assumptions, and treatment of exogenous shocks. Section~\ref{Section_Experimental_Evaluation} reports the forecasting study such as baselines, performance metrics, results and experimental comparisons, significance and robustness, and uncertainty quantification via credible intervals. Section~\ref{Sec_Policy_Implications} discusses policy implications. Section~\ref{Section_Conclusion} concludes with contributions, limitations, and future work.

\section{Data and Preliminary Analysis}\label{Section_Data_and_Methodology}
This study is concerned with the macroeconomic dynamics of the G7 countries. In particular, we jointly analyze five macro variables—Unemployment Rate, Real Broad Effective Exchange Rate (REER), Short-term Interest Rate (SIR), Oil Prices proxied by the benchmark West Texas Intermediate (WTI), and consumer price index (CPI) Inflation—together with four newspaper-based uncertainty measures: log-transformed country-wise Economic Policy Uncertainty (EPU), country-wise Geopolitical Risk (GPR), US Equity Market Volatility (USEMV), and US Monetary Policy Uncertainty (USMPU). The sample spans January 1995 to March 2024, a window that includes multiple business-cycle phases, a Global Financial Crisis, a zero lower bound era, and the COVID-19 pandemic, a global pandemic shock. We assess forecast performance using a rolling window design at two horizons: 12-month-ahead (``semi-long-term'') and 24-month-ahead (``long-term''). This allows us to contrast medium-run accuracy with longer-run stability in timeframes relevant for policy and investment decisions.
The Section~\ref{Sec_Data_Characteristics} documents the key data features through summary statistics and global characteristics, followed by tests for endogeneity among key macroeconomic variables and structural break detection. 

\subsection{Global Characteristics}\label{Sec_Data_Characteristics}
This section provides a broad description of the data organization involved in our multivariate analysis for the G7 countries. We concentrate on five major macroeconomic variables, which are the unemployment rate, REER, SIR, oil price (WTI), and CPI inflation. These five macroeconomic indicators were chosen as they are key factors for major advanced economies. The unemployment rate is a measure of slack in the labor market and cyclical variation in employment. It is obtained from the Infra-Annual Labor Statistics. The REER is a weighted average of bilateral exchange rates and is normalized by relative consumer prices. It is used as a measure of each economy's external competitiveness. SIR is used as a proxy for the stance of monetary policy and liquidity conditions. We use the three-month interbank rate for this purpose. Oil price (WTI) is included as a summary of global commodity market conditions that feed into both production costs and inflationary pressures. CPI inflation is the year-on-year change in consumer prices, and thus it measures the purchasing power of households and is the most direct indicator of price stability. All the macroeconomic series are sourced from the Federal Reserve Economic Data (FRED) database to ensure methodological consistency and cross-country comparability. A first pass through the trends and variability - plotted for each country-variable pair and presented in Figure~\ref{fig:trend_G7} of Appendix \ref{app:Appendix_trend_plots_G7} of the supplementary material, reveals substantial heterogeneity. Pronounced cyclical swings and crisis-related spikes characterize unemployment and CPI inflation, while exchange rates and oil prices exhibit more structural volatility. SIRs appear to be impacted by distinct monetary policy regimes across the entire sample period, including the extended era of the zero lower bound (a prolonged period, particularly since the global financial crisis, when central banks kept short-term interest rates near zero, limiting conventional policy and leading to widespread adoption of tools such as quantitative easing and forward guidance). The autocorrelation diagnostics (see Appendix~\ref{app:Appendix_acf_pacf}, Figure~\ref{fig:acf_G7}) indicate that most series display strong persistence through a gradually decaying autocorrelation structure. 
This widespread persistence motivates the use of a Bayesian VAR framework, which naturally accommodates persistent dynamics through its autoregressive structure while simultaneously capturing joint dynamics and cross-country spillover effects through the multivariate specification~\citep{hyndman2018forecasting,diebold2009measuring,pesaran2004modeling}. Additionally, four newspaper-based high-frequency measures of uncertainty that are designed to capture different dimensions of global risk and policy uncertainty are utilized in this study. We start with the EPU index of \cite{baker2016measuring}, which is constructed as the frequency of terms related to policy and uncertainty in leading newspapers. We follow \cite{bloom2009impact} and take the natural logarithm of the series (EPU) to reduce heteroscedasticity and the effect of spikes in uncertainty. The country-wise GPR index of \cite{caldara2022measuring} is designed to measure geopolitical risks to the global economy that stem from geopolitical events such as terrorism, armed conflict, and political and economic instability, using a news dataset hand-coded at the international level. It includes both sudden spikes in uncertainty and background levels of geopolitical tensions. USEMV index of \cite{baker2019policy} is the share of news articles that connect economic and financial volatility to U.S. stock market returns and is thus a forward-looking measure of financial market stress. USMPU index of \cite{husted2020monetary} is based on a set of terms relating to monetary policy and uncertainty and constructed as the frequency with which articles in leading financial newspapers use those terms to identify uncertainty in Federal Open Market Committee decisions. These four indices are designed to provide complementary coverage of policy, geopolitical, financial-market, and monetary uncertainty, respectively. 

Standard summary statistics, describing the distribution of the series, namely (minimum, quartiles, median, mean, standard deviation, and maximum) and two complexity measures: the coefficient of variation (CoV) and entropy, to describe the distributions of both macro series and uncertainty indices, are reported in Table~\ref{tab:summary_stats_g7}. The CoV, standard deviation divided by the mean, renders volatility comparable across units of observation. A higher CoV implies that observations are more dispersed, making point forecasting more difficult. Entropy is a measure of randomness in distributions. It can be viewed as a proxy for informational complexity: higher entropy implies distributions are more dispersed and harder to predict. The picture is not uniform across countries. Some labor-market series are characterized by high CoV and entropy consistent with their erratic dynamics. In a number of countries, inflation also displays relatively high complexity. USEMV and USMPU, along with oil prices, appear to be among the more idiosyncratic exogenous drivers, as they display high CoV and entropy. This is also a manifestation of their episodic, shock-driven behavior that also transmits more forecast uncertainty. These findings demonstrate the need for a flexible, robust forecasting framework for the G7 economies.

\begin{table*}
\centering
\caption{Summary statistics of the datasets utilized in this analysis for G7 countries. For each country, statistics are reported for unemployment rate, REER, SIR, CPI Inflation, EPU, and GPR. Global variables include Oil Price (WTI), USEMV, and USMPU.}
\label{tab:summary_stats_g7}
\begin{adjustbox}{width=0.90\textwidth}
\begin{tabular}{llrrrrrrrrr}
\toprule
\textbf{Country} & \textbf{Series} & \textbf{Min Value} & \textbf{Q1} & \textbf{Median} & \textbf{Mean} & \textbf{SD} & \textbf{Q3} & \textbf{Max Value} & \textbf{CoV} & \textbf{Entropy} \\
\midrule
Canada  & Unemployment Rate & 4.50 & 6.30 & 7.20 & 7.30 & 1.39 & 8.10 & 14.30 & 19.17 & 5.91 \\
        & REER    & 88.92 & 99.63 & 102.85 & 106.36 & 10.74 & 115.29 & 132.19 & 10.10 & 5.86 \\
        & SIR     & 0.18 & 1.16 & 2.40 & 2.73 & 1.88 & 4.38 & 8.31 & 68.83 & 5.80 \\
        & CPI Inflation     & -0.95 & 1.27 & 1.99 & 2.13 & 1.32 & 2.57 & 8.13 & 62.30 & 5.94 \\
        & EPU            & 1.48 & 1.92 & 2.16 & 2.15 & 0.28 & 2.38 & 2.83 & 13.15 & 6.08 \\
        & GPR              & 0.06 & 0.13 & 0.17 & 0.21 & 0.16 & 0.24 & 1.72 & 75.64 & 8.03 \\
\midrule
US     & Unemployment Rate & 3.10 & 4.30 & 5.10 & 5.61 & 1.85 & 6.20 & 14.40 & 32.96 & 5.90 \\
        & REER    & 78.43 & 86.79 & 95.02 & 93.99 & 8.25 & 99.25 & 112.99 & 8.79 & 5.86 \\
        & SIR     & 0.09 & 0.32 & 1.87 & 2.62 & 2.24 & 5.24 & 6.73 & 85.55 & 5.64 \\
        & CPI Inflation     & -2.10 & 1.63 & 2.29 & 2.54 & 1.66 & 3.18 & 9.06 & 65.25 & 5.91 \\
        & EPU            & 1.65 & 1.94 & 2.05 & 2.07 & 0.18 & 2.21 & 2.70 & 8.92 & 6.10 \\
        & GPR              & 0.75 & 1.63 & 1.99 & 2.28 & 1.29 & 2.58 & 13.23 & 56.87 & 5.97 \\
\midrule
France  & Unemployment Rate & 6.50 & 8.30 & 9.10 & 9.41 & 1.52 & 10.30 & 13.20 & 16.20 & 5.90 \\
        & REER    & 92.98 & 100.05 & 104.10 & 104.85 & 5.79 & 109.64 & 118.15 & 5.52 & 5.86 \\
        & SIR     & -0.58 & 0.03 & 2.12 & 1.98 & 1.98 & 3.56 & 8.06 & 99.91 & 5.73 \\
        & CPI Inflation     & -0.73 & 0.87 & 1.54 & 1.63 & 1.24 & 2.07 & 6.28 & 76.24 & 5.92 \\
        & EPU            & 1.05 & 1.92 & 2.21 & 2.15 & 0.32 & 2.41 & 2.76 & 14.79 & 6.08 \\
        & GPR              & 0.17 & 0.34 & 0.45 & 0.52 & 0.32 & 0.60 & 2.80 & 60.58 & 6.68 \\
\midrule
Germany & Unemployment Rate & 2.80 & 4.00 & 7.20 & 6.62 & 2.68 & 8.80 & 12.20 & 40.45 & 5.85 \\
        & REER    & 94.46 & 99.89 & 102.85 & 105.03 & 6.68 & 109.39 & 127.17 & 6.37 & 5.86 \\
        & SIR     & -0.58 & 0.03 & 2.12 & 1.89 & 1.82 & 3.45 & 5.16 & 96.43 & 5.77 \\
        & CPI Inflation     & -1.04 & 1.12 & 1.51 & 1.80 & 1.55 & 2.01 & 8.82 & 86.49 & 5.89 \\
        & EPU            & 1.45 & 1.94 & 2.10 & 2.14 & 0.29 & 2.30 & 2.93 & 13.46 & 6.08 \\
        & GPR              & 0.08 & 0.23 & 0.32 & 0.40 & 0.29 & 0.48 & 2.66 & 74.32 & 6.92 \\
\midrule
Japan   & Unemployment Rate & 2.10 & 3.00 & 3.90 & 3.84 & 0.95 & 4.70 & 5.80 & 24.73 & 5.96 \\
        & REER    & 70.72 & 97.85 & 123.75 & 119.24 & 25.43 & 136.97 & 193.97 & 21.33 & 5.84 \\
        & SIR     & -0.08 & -0.02 & 0.04 & 0.14 & 0.29 & 0.11 & 2.25 & 206.35 & 7.15 \\
        & CPI Inflation     & -2.56 & -0.42 & 0.10 & 0.36 & 1.21 & 0.72 & 4.39 & 341.67 & 8.04 \\
        & EPU            & 1.68 & 1.94 & 2.01 & 2.02 & 0.12 & 2.09 & 2.38 & 6.30 & 6.11 \\
        & GPR              & 0.05 & 0.13 & 0.18 & 0.23 & 0.16 & 0.27 & 1.24 & 69.39 & 7.86 \\
\midrule
UK      & Unemployment Rate & 3.50 & 4.70 & 5.30 & 5.76 & 1.47 & 7.10 & 9.00 & 25.53 & 5.92 \\
        & REER    & 94.28 & 101.92 & 109.08 & 113.26 & 12.30 & 126.50 & 135.22 & 10.86 & 5.86 \\
        & SIR     & 0.03 & 0.59 & 3.75 & 3.21 & 2.55 & 5.55 & 7.81 & 79.41 & 5.65 \\
        & CPI Inflation     & 0.20 & 1.50 & 2.09 & 2.41 & 1.65 & 2.69 & 9.61 & 68.36 & 5.88 \\
        & EPU            & 1.68 & 1.98 & 2.16 & 2.16 & 0.21 & 2.33 & 2.64 & 9.84 & 6.09 \\
        & GPR              & 0.40 & 0.71 & 0.93 & 1.06 & 0.66 & 1.16 & 5.99 & 61.56 & 6.20 \\
\midrule
Italy   & Unemployment Rate & 5.30 & 8.10 & 9.60 & 9.59 & 1.93 & 11.20 & 14.40 & 20.21 & 5.89 \\
        & REER    & 86.42 & 100.05 & 102.85 & 103.00 & 4.04 & 106.10 & 110.63 & 3.92 & 5.86 \\
        & SIR     & -0.58 & 0.03 & 2.12 & 2.45 & 2.82 & 3.93 & 11.01 & 114.99 & 5.57 \\
        & CPI Inflation     & -0.58 & 1.07 & 2.00 & 2.23 & 1.95 & 2.73 & 11.84 & 87.66 & 5.77 \\
        & EPU            & 1.68 & 1.98 & 2.16 & 2.16 & 0.21 & 2.33 & 2.64 & 9.84 & 6.09 \\
        & GPR              & 0.03 & 0.08 & 0.12 & 0.14 & 0.09 & 0.17 & 0.65 & 67.34 & 9.25 \\
\midrule
Global  & Oil Price (WTI)   & 11.27 & 28.83 & 54.09 & 55.71 & 28.73 & 77.03 & 133.96 & 51.57 & 5.73 \\
        & USEMV             & 9.57 & 16.24 & 18.93 & 21.18 & 8.02 & 23.97 & 69.83 & 37.90 & 5.82 \\
        & USMPU             & 16.57 & 48.20 & 75.04 & 90.62 & 58.69 & 120.16 & 407.94 & 64.76 & 5.68 \\
\bottomrule
\end{tabular}
\end{adjustbox}
\end{table*}
\FloatBarrier

To provide a comprehensive overview of the properties of the series, considered in this study, we report seven key statistical properties: skewness, kurtosis, nonlinearity, seasonality, stationarity, long-range dependence, and presence of outliers - in Table~\ref{tab:global_char_g7}.
The nonlinearity is captured by Tsay’s and Keenan’s test \citep{tsay1986nonlinearity,keenan1985tukey}, which tests for departure from linear autoregressive structure and can accommodate threshold or regime-switching dynamics. The seasonality is captured by the Ollech–Webel combined procedure \citep{ollech2023random}, which is a robust diagnostic that bundles up several different tests to reduce misclassification. The stationarity is checked by the KPSS test \citep{kwiatkowski1992testing}, which takes stationarity as the null and is used in conjunction with standard unit-root tests. The long-range dependence is captured by the Hurst exponent \citep{shang2020comparison}, which differentiates between mean-reverting, memoryless, and trend-reinforcing behavior. The outliers are detected by Bonferroni-adjusted tests based on studentized residuals \citep{weisberg1982residuals}, which filter the rare but influential shocks. The diagnostics suggest that most of the G7 macroeconomic series are non-stationary, nonlinear, and exhibit long memory. Table~\ref{tab:global_char_g7} also indicates the presence of seasonality, mainly evident in unemployment rates, EPU, and USEMV.
Statistically significant outliers cluster around the global financial crisis and the pandemic. In sum, the evidence suggests complex, persistent, and sometimes volatile dynamics, which call for uncertainty-resilient multivariate models that can accommodate nonlinearity, regime shifts, and persistent shocks.

To investigate potential endogeneity within a block of macroeconomic variables, we implement the Wu–Hausman (WH) test for the five principal variables in each of the G7 economies \citep{wu1973alternative, hausman1978specification, heinisch2025assumption}. The Wu-Hausman (WH) test for endogeneity compares the Ordinary Least Squares (OLS) and Instrumental Variable (IV) estimators: OLS is efficient under exogeneity, but becomes inconsistent if regressors are endogenous, while IV remains consistent in both cases. The WH statistic measures the difference between these estimators, which should be asymptotically zero if all regressors are exogenous, and the null hypothesis follows a chi-squared distribution. Using a 10\% significance level, p-values below 0.10 lead to rejection of exogeneity—indicating possible feedback or simultaneity and confirming the need for IV estimation \citep{hausman1978specification, wu1973alternative}. Country-specific statistics and p-values (presented in Table~\ref{tab:wh_endogeneity_g7}) provide evidence of endogeneity at this threshold, supporting treatment of the block of these macroeconomic variables as a jointly determined system and the use of the four uncertainty indices as external drivers in the forecasting framework.
\begin{table*}[!ht]
\centering
\caption{Global characteristics of the economic time series under study for G7 countries. For each country, statistics are reported for Unemployment Rate, REER, SIR, CPI Inflation, EPU, and GPR. Global variables include Oil Price (WTI), USEMV, and USMPU.}
\label{tab:global_char_g7}
\begin{adjustbox}{width=0.98\textwidth}
\begin{tabular}{llrrrrrrrrr}
\toprule
\textbf{Country} & \textbf{Series} & \textbf{Skewness} & \textbf{Kurtosis} & \textbf{Nonlinearity} & \textbf{Seasonality} & \textbf{Stationarity} & \textbf{Long-Range Dependence} & \textbf{Outlier(s) Detected} \\
\midrule
Canada  & Unemployment Rate & 1.03 & 2.67 & Non-linear & Seasonal & Non-stationary & 0.81 & 4 \\
        & REER    & 0.57 & -0.83 & Linear & Non-seasonal & Non-stationary & 1.02 & 0 \\
        & SIR     & 0.53 & -0.67 & Non-linear & Non-seasonal & Non-stationary & 0.76 & 0 \\
        & CPI Inflation     & 1.63 & 4.43 & Non-linear & Non-seasonal & Non-stationary & 0.84 & 3 \\
        & EPU            & -0.07 & -0.88 & Non-linear & Seasonal & Non-stationary & 0.87 & 0 \\
        & GPR              & 4.92 & 36.34 & Non-linear & Non-seasonal & Trend Stationary & 0.91 & 4 \\
\midrule
US     & Unemployment Rate & 1.35 & 1.84 & Non-linear & Seasonal & Non-stationary & 0.76 & 2 \\
        & REER    & 0.01 & -0.94 & Linear & Non-seasonal & Non-stationary & 1.00 & 0 \\
        & SIR     & 0.34 & -1.51 & Non-linear & Non-seasonal & Non-stationary & 0.86 & 0 \\
        & CPI Inflation     & 1.19 & 3.17 & Non-linear & Non-seasonal & Trend Stationary & 0.67 & 0 \\
        & EPU            & 0.38 & 0.01 & Linear & Seasonal & Non-stationary & 0.78 & 0 \\
        & GPR              & 4.45 & 30.07 & Non-linear & Non-seasonal & Trend Stationary & 0.87 & 4 \\
\midrule
France  & Unemployment Rate & 0.59 & -0.33 & Non-linear & Seasonal & Non-stationary & 1.17 & 0 \\
        & REER    & 0.21 & -0.93 & Linear & Seasonal & Non-stationary & 0.93 & 0 \\
        & SIR     & 0.37 & -0.77 & Non-linear & Non-seasonal & Non-stationary & 0.72 & 0 \\
        & CPI Inflation     & 1.51 & 3.34 & Linear & Non-seasonal & Trend Stationary & 0.71 & 0 \\
        & EPU            & -0.61 & -0.24 & Non-linear & Seasonal & Non-stationary & 0.88 & 1 \\
        & GPR              & 3.09 & 13.99 & Linear & Non-seasonal & Trend Stationary & 0.79 & 6 \\
\midrule
Germany & Unemployment Rate & 0.06 & -1.40 & Non-linear & Seasonal & Non-stationary & 0.88 & 0 \\
        & REER    & 0.95 & 0.58 & Linear & Seasonal & Non-stationary & 0.87 & 0 \\
        & SIR     & 0.10 & -1.43 & Non-linear & Seasonal & Non-stationary & 0.75 & 0 \\
        & CPI Inflation     & 2.40 & 7.12 & Non-linear & Non-seasonal & Non-stationary & 0.66 & 5 \\
        & EPU            & 0.65 & 0.19 & Non-linear & Non-seasonal & Non-stationary & 0.85 & 0 \\
        & GPR              & 3.51 & 19.45 & Linear & Non-seasonal & Non-stationary & 0.87 & 5 \\
\midrule
Japan   & Unemployment Rate & 0.04 & -1.18 & Linear & Seasonal & Non-stationary & 0.99 & 0 \\
        & REER    & 0.21 & -0.51 & Non-linear & Non-seasonal & Non-stationary & 0.82 & 0 \\
        & SIR     & 3.39 & 16.94 & Non-linear & Non-seasonal & Non-stationary & 1.10 & 3 \\
        & CPI Inflation     & 1.04 & 0.94 & Linear & Non-seasonal & Non-stationary & 0.71 & 0 \\
        & EPU            & 0.34 & 0.15 & Non-linear & Non-seasonal & Trend Stationary & 0.62 & 0 \\
        & GPR              & 2.35 & 7.72 & Linear & Non-seasonal & Trend Stationary & 1.03 & 5 \\
\midrule
UK      & Unemployment Rate & 0.54 & -0.93 & Non-linear & Seasonal & Non-stationary & 1.13 & 0 \\
        & REER    & 0.29 & -1.52 & Non-linear & Non-seasonal & Non-stationary & 0.89 & 0 \\
        & SIR     & 0.14 & -1.62 & Non-linear & Non-seasonal & Non-stationary & 0.84 & 0\\
        & CPI Inflation     & 2.35 & 6.45 & Non-linear & Non-seasonal & Non-stationary & 0.87 & 4 \\
        & EPU            & 0.06 & -0.91 & Non-linear & Seasonal & Non-stationary & 0.76 & 0 \\
        & GPR              & 4.19 & 24.36 & Non-linear & Non-seasonal & Trend Stationary & 0.95 & 6 \\
\midrule
Italy   & Unemployment Rate & 0.00 & -0.89 & Non-linear & Seasonal & Non-stationary & 1.22 & 0 \\
        & REER    & -0.24 & 0.22 & Non-linear & Seasonal & Non-stationary & 1.01 & 2 \\
        & SIR     & 1.14 & 0.91 & Non-linear & Non-seasonal & Non-stationary & 0.84 & 0 \\
        & CPI Inflation     & 1.92 & 5.58 & Non-linear & Non-seasonal & Trend Stationary & 0.69 & 4 \\
        & EPU            & 0.06 & -0.91 & Non-linear & Seasonal & Non-stationary & 0.76 & 0 \\
        & GPR              & 2.22 & 6.73 & Non-linear & Non-seasonal & Non-stationary & 0. & 6 \\
\midrule
Global  & Oil Price (WTI)   & 0.32 & -0.92 & Non-linear & Non-seasonal & Non-stationary & 1.06 & 0 \\
        & USEMV             & 2.34 & 8.29 & Non-linear & Seasonal & Trend Stationary & 0.79 & 4 \\
        & USMPU             & 1.66 & 3.93 & Linear & Non-seasonal & Trend Stationary & 0.90 & 2 \\
\bottomrule
\end{tabular}
\end{adjustbox}
\end{table*}
\FloatBarrier

Finally, we assess the structural stability of the VAR model using the cumulative sum (CUSUM) based on the OLS procedure \citep{ploberger1992cusum} over the five macroeconomic variables, with potential implications for both specification and evaluation \citep{stock1996evidence, hansen2001new}. Appendix~\ref{app:Appendix_structural_breakpoints} presents Figure~\ref{fig:OLS_CUSUM_SB_G7}, which reveals heterogeneous patterns. The unemployment rate clearly presents episodes of instability for several economies around some targeted episodes (early 2000s and post-2012 in parts of the sample). REER registers breaks in a few countries in the early 2000s. Oil prices exhibit common breaks across all G7 in 2008–2009 and 2014–2015, consistent with global commodity shocks (although none of the breaks are found to be statistically significant). Most critically, SIRs exhibit systematic structural breaks in Canada, the United States, France, and Japan, consistent with the zero lower bound, quantitative-easing phases, and subsequent normalizations \citep{bernanke2020new,yellen2017inflation}. By contrast, Germany, the United Kingdom, and Italy display CUSUM paths within the confidence bands, consistent with relative parameter stability. CPI inflation is comparatively more stable across the G7, and the few near-significant breakpoints are consistent with credible inflation-targeting frameworks \citep{mishkin2007does,bernanke2000inflation}.

\begin{table*}[!ht]
\centering
\caption{Wu-Hausman test results for endogeneity of the five key macroeconomic variables across G7 countries. The test statistic and p-values are reported for each country. A small p-value (\texorpdfstring{$<$0.10}{<0.10}) indicates rejection of the null hypothesis (\texorpdfstring{$H_0$}{H0}) of exogeneity, providing evidence of endogeneity among regressors.}
\label{tab:wh_endogeneity_g7}
\begin{adjustbox}{width=0.9\textwidth}
\begin{tabular}{lccc} \hline
\toprule
\textbf{Country} & 
\textbf{WH Test Statistic} & 
\textbf{WH Test (p-value)} & 
\textbf{Insights} \\
\midrule
Canada   & 6.655 & 0.00004 & Evidence against exogeneity (endogeneity detected) \\
US      & 3.044 & 0.01735 & Evidence against exogeneity (endogeneity detected) \\
Germany  & 35.893 & 0.00000 & Evidence against exogeneity (endogeneity detected) \\
France   & 12.491 & 0.00000 & Evidence against exogeneity (endogeneity detected) \\
Japan    & 2.064 & 0.08514 & Evidence against exogeneity (endogeneity detected) \\
UK       & 14.620 & 0.00000 & Evidence against exogeneity (endogeneity detected) \\
Italy    & 13.297 & 0.00000 & Evidence against exogeneity (endogeneity detected) \\
\bottomrule
\end{tabular}
\end{adjustbox}
\end{table*}
\FloatBarrier

Overall, the descriptive statistics, structural diagnostics, endogeneity tests, and stability checks jointly motivate our modeling choices. The macro series are persistent, often nonlinear, and subject to structural change; the uncertainty indices are high-frequency and shock-driven; there is evidence of feedback within the block of macroeconomic variables and of regime shifts - particularly in SIRs. These features argue for multivariate, shrinkage-based methods that can handle high-dimensional systems; for state-dependent tools to trace transmission under different monetary environments; and for forecasting designs that evaluate both medium- and long-run horizons. This empirical groundwork justifies the inclusion and treatment of variables in the forecasting system and guides the specification of the multivariate framework applied to the data.

\section{Causality Analysis}\label{sec:Causal_anal_wca_irfs}
The time–frequency causality analysis to map how uncertainty relates to the macro indicators is presented in Section~\ref{Causal_anal}. Then, transmission paths are assessed using impulse response analysis in Section~\ref{impulse_resp}. These diagnostics provide the empirical basis for variable selection, clarify propagation channels, and motivate the multivariate forecasting framework applied in subsequent sections.
\subsection{Wavelet Coherence Evidence for Uncertainty Transmission in the G7}\label{Causal_anal}
In this section, we examine the causal relationships between our focal macroeconomic variables and exogenous uncertainty measures in the G7 economies in the time-frequency space using wavelet coherence analysis (WCA), which was originally proposed by \cite{grinsted2004application} and subsequently extended with scale-wise False Discovery Rate (FDR) correction by \cite{aguiar2014continuous}. WCA is used as a descriptive time-frequency diagnostic to identify co-movement and lead-lag patterns whose strength and direction may differ across horizons and over time; as such, the resulting maps are interpreted as evidence on transmission timing and relevant frequency bands, rather than as structural tests of causality. Wavelet coherence plots visualize time-frequency information, with information on the lead-lag (phase) between the two variables being analyzed represented by the direction of the arrows in the plot. Arrows pointing to the right (left) represent an in-phase (anti-phase) relation between the pair of variables, whereas upward (downward) pointing arrows represent a leading (lagging) relation in time. The cone of influence boundaries, by identifying the areas where edge effects could undermine statistical reliability, ensure the reliable interpretation of the coherence patterns throughout the analysis. The scale-wise application of FDR correction, using the Benjamini-Hochberg procedure, systematically removes statistically unreliable coherence regions at each temporal scale independently.

As a result, only the segments of the estimated coherence function that are unlikely to be affected by spurious inferences of significant deviations from zero will be apparent in the resulting maps. In these maps, shades of red (coherence values near 1 on a 0-1 scale) indicate the time points and frequencies at which the lead-lag relationship between uncertainty shocks and fundamentals is most likely to be statistically significant. The scale-wise FDR screen removes visually attractive but fragile patches, thereby increasing confidence that remaining shaded regions reflect genuine time–frequency associations rather than multiple-testing artifacts.

We compute bivariate coherence between each uncertainty index and macro series for each country using the complex Morlet wavelet ($\omega_{o}=6$), which is implemented through the \texttt{biwavelet} \textbf{R} package. To prevent look-ahead bias, the WCA is calculated only in the in-sample period (Jan, 1995 – Mar, 2022), which is distinct from the 24 months reserved for out-of-sample forecast validation. Statistical significance is determined using 1,000 Monte Carlo replications and controlled at each scale with a Benjamini–Hochberg FDR adjustment set at $\alpha=10\%$; only regions that remain significant after FDR correction are shaded (see Appendix~\ref{app:Appendix_causal_analysis} for a more technical discussion of the FDR-corrected WCA). 

\begin{figure}[!htbp]
\centering
\includegraphics[width=\textwidth]
{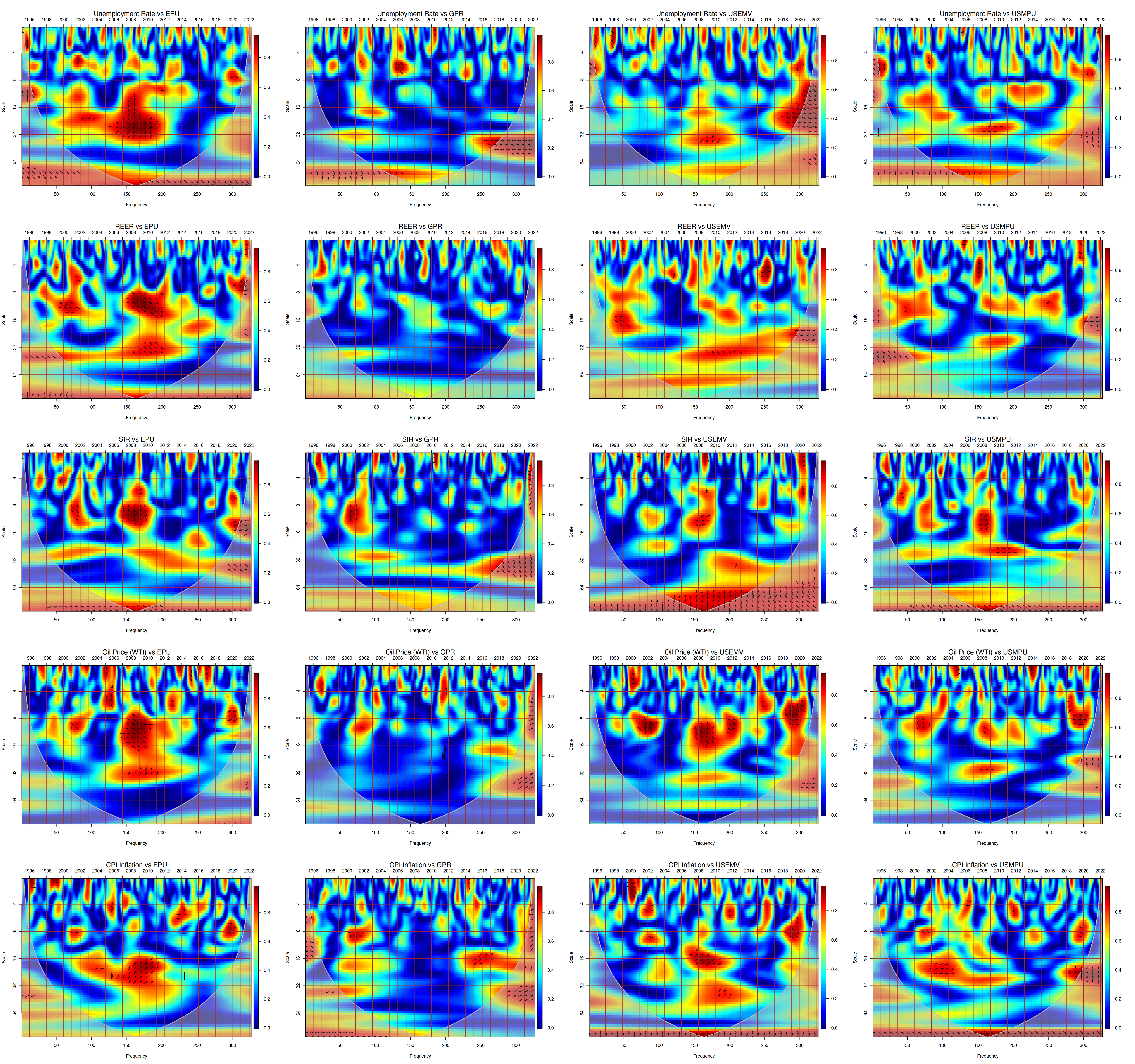}
\caption{Scale-wise FDR-corrected wavelet coherence spectra for Canada (Jan, 1995 – Mar, 2022), displaying Unemployment Rate, REER, SIR, Oil Price (WTI), and CPI Inflation against EPU, GPR, USEMV, and USMPU. Each subplot visualizes the time-frequency coherence between a macroeconomic variable (y-axis: Scale; x-axis: Frequency) and an uncertainty index, with warm colors (red/yellow) indicating intervals of high, statistically significant co-movement after FDR correction. The direction of phase arrows indicates lead-lag relationships: rightward arrows mean the series move in phase; leftward, in anti-phase; upward arrows signify the uncertainty index leads; downward, that it lags. Shaded regions inside the cone of influence reflect reliable coherence, while grid consistency enables robust cross-variable and cross-period comparison. The level of significance ($\alpha$) for the FDR adjustment is set at 10\%.}
\label{fig:WCA_canada}
\end{figure}

Figure~\ref{fig:WCA_canada} shows a canonical example (Canada), and similar grids of WCA for the remaining G7 appear in the Appendix~\ref{app:Appendix_causal_analysis} (Figures~\ref{fig:WCA_usa} – \ref{fig:WCA_italy}). Three patterns stand out in the case of Canada. First, coherence is strongest at intermediate periodicities (multi-quarter to multi-year bands) and exhibits pronounced clustering around 2008–2009 and 2020–2021 (reflecting the global financial crisis and the COVID-19 pandemic, respectively). This dominance of the medium term is consistent with the shock-driven nature of the uncertainty indices as well as the stylized volatility documented for oil prices and exchange rates. Second, coherence is stronger in times of stress relative to tranquil periods, suggesting that uncertainty and macroeconomic aggregates co-move more closely when policy and financial conditions are tight. Phase information frequently reveals the uncertainty series leading the macroeconomic variables in these frequency bands. Third, frequency localization varies across variables: SIR and REER co-move with USEMV and USMPU at short-to-medium periodicities (consistent with policy and financial-expectations channels), whereas unemployment rate and CPI inflation exhibit more medium-to-long periodicities vis-à-vis EPU and GPR, suggesting a slower speed of real-side adjustment. Cross-country heterogeneity is nontrivial. For the United States, coherence between oil prices and uncertainty is especially strong; for the European economies, inflation–uncertainty coherence is pronounced; and for Japan, low-frequency REER synchronization is observed with US-centered uncertainty measures. 

The existence of FDR-robust regions in which uncertainty drives macroeconomic variables justifies treating EPU, GPR, USEMV, and USMPU as exogenous drivers in the multivariate forecasting framework and thus complements the endogeneity evidence in the block of macroeconomic variables documented in Section~\ref{Sec_Data_Characteristics} (Table~\ref{tab:wh_endogeneity_g7}). Overall, FDR-corrected WCA offers evidence on when and at what frequencies uncertainty shocks are associated with G7 macroeconomic aggregates, thus motivating an exogenous uncertainty block, stance conditioning, and horizon choice in the forecasting framework.

\subsection{Impulse response analysis}\label{impulse_resp}
This section focuses on the responses of the five key macroeconomic variables to the four newspaper-based uncertainty measures. The choice of the uncertainty drivers is based on the evidence presented in Section \ref{Causal_anal}. This identification of exogenous uncertainty drivers informs the design of our empirical impulse-response analysis. We estimate impulse response functions using nonlinear local projections, as in \cite{jorda2005estimation} (See Appendix~\ref{app:Appendix_IRFS_NLP} for details). This methodology runs horizon-specific regressions that directly map innovations in the uncertainty measures to future values of each macro variable, without the need to invert a full VAR system or to impose strong parametric restrictions. We use the nonlinear extension described in \cite{RJ-2019-052}, which employs a logistic transition function to generate state-contingent responses; observations are smoothly weighted by proximity to each regime, allowing both the size and timing of effects to depend on the monetary environment. As discussed in Section~\ref{Causal_anal}, the predominance of statistically significant coherence at medium-to-long periodicities suggests that dynamic interactions between uncertainty indices and financial or policy variables are most salient at these horizons. Guided by this frequency-localized coherence, we set both our key forecast horizons (12 and 24 months, corresponding to semi-long-term and long-term windows) and the maximal horizon for our impulse response analysis to align with the bands in which the strongest and most persistent co-movements and lead-lag relationships are identified. This ensures that our forecasting and IRF exercises are empirically grounded and tied directly to cycles supported by the underlying time-frequency evidence. 

Lag orders of the endogenous macroeconomic variables (6 months) and exogenous uncertainty indicators (4 months) are chosen by using the Bayesian Information Criterion to account for persistence and delayed effects, respectively. Newey-West robust standard errors \citep{newey1987hypothesis} are used to compute the 95\% confidence intervals, which are able to cope with time-varying volatility, especially during periods of crisis. Regime switching is lagged (\texttt{lag-switching = TRUE}) to ensure that the switching coincides with the pre-shock monetary environment. The IRF plots distinguish high-rate (blue) and low-rate (orange) regimes, while shaded regions represent 95\% confidence bands. We compute horizon-by-horizon impulse responses up to 24 months. The smooth regime-switching is done by setting the regime probability following a logistic function, with curvature parameter $\gamma=3$ to have a compromise between smooth transition and separation of the two regimes. Deterministic trends are turned off (\texttt{trend = 0}) to focus on cyclical features, and standardization of the short-term interest rate is done for each country to compare the magnitude across countries and over time. The country-specific IRFs are documented for Canada in Figures \ref{fig:irf_figure1_canada} - \ref{fig:irf_figure2_canada} and for the other six G7 economies in the Appendix~\ref{app:Appendix_IRFS_USA_ITA} (Figures \ref{fig:irf_figure1_usa} - \ref{fig:irf_figure2_ita}), the United States (Figures \ref{fig:irf_figure1_usa} and \ref{fig:irf_figure2_usa}), France (Figures \ref{fig:irf_figure1_fra} and \ref{fig:irf_figure2_fra}), Germany (Figures \ref{fig:irf_figure1_ger} and \ref{fig:irf_figure2_ger}), Japan (Figures \ref{fig:irf_figure1_jap} and \ref{fig:irf_figure2_jap}), the United Kingdom (Figures \ref{fig:irf_figure1_uk} and \ref{fig:irf_figure2_uk}) and Italy (Figures \ref{fig:irf_figure1_ita} and \ref{fig:irf_figure2_ita}). 

\begin{figure}[!htbp]
    \centering
\includegraphics[width=\textwidth,height=0.80\textheight,keepaspectratio]
{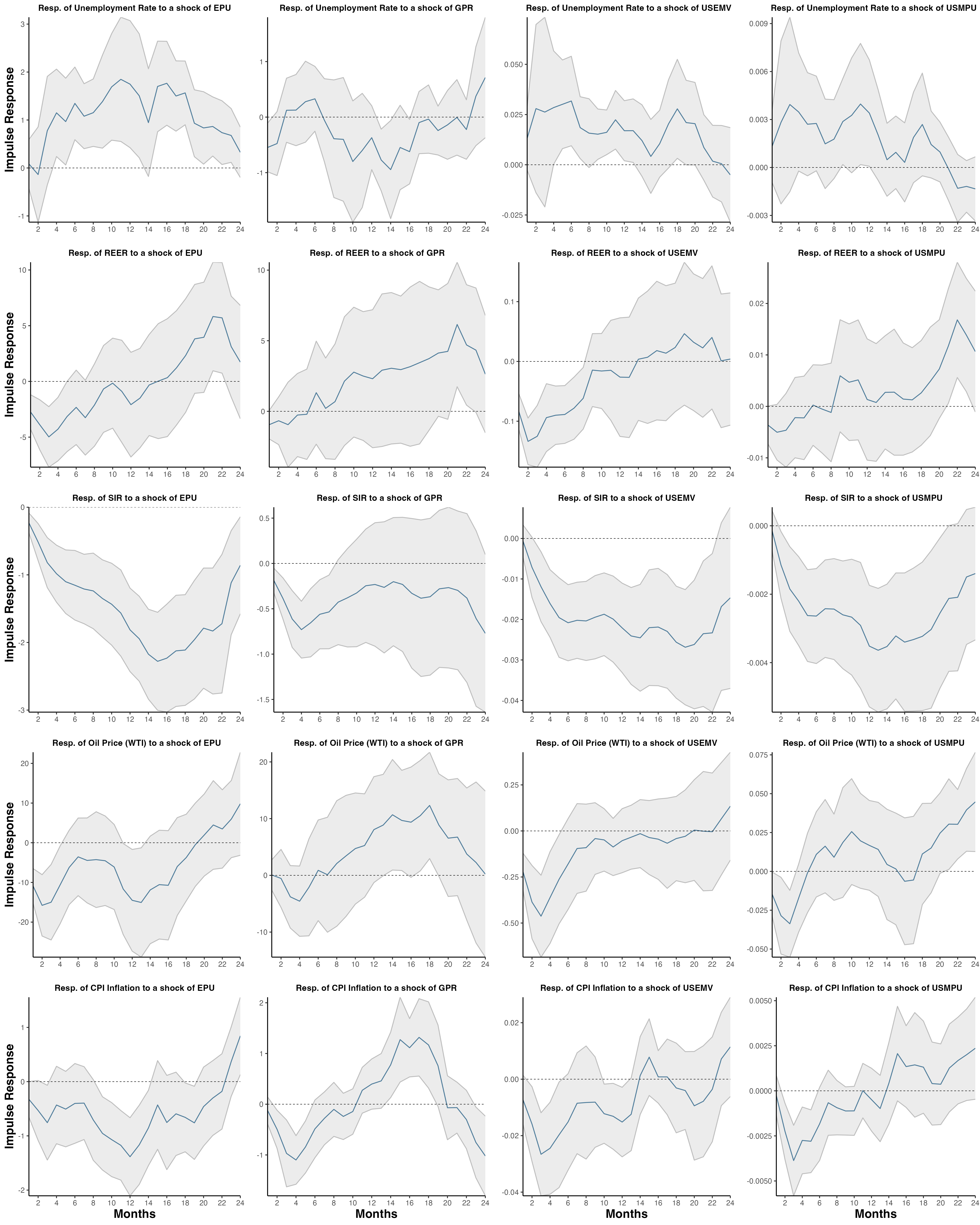}
    \caption{Impulse response functions (nonlinear local projections) of Unemployment Rate, REER, SIR, Oil Price (WTI), and CPI Inflation to EPU, GPR, USEMV, and USMPU shocks for Canada (high-rate regime). Each row shows responses of a macro variable; each column, a distinct uncertainty shock. Solid \textbf{\textcolor{blue}{blue}} lines depict IRFs with 95\% confidence bands (grey). The x-axis is horizon (months), the y-axis the response magnitude. Compare responses by reading across rows (by variable) or down columns (by shock); zero response is marked by the dotted line.}
    \label{fig:irf_figure1_canada}
\end{figure}
\begin{figure}[!htbp]
\centering
\includegraphics[width=\textwidth,height=0.85\textheight,keepaspectratio]
{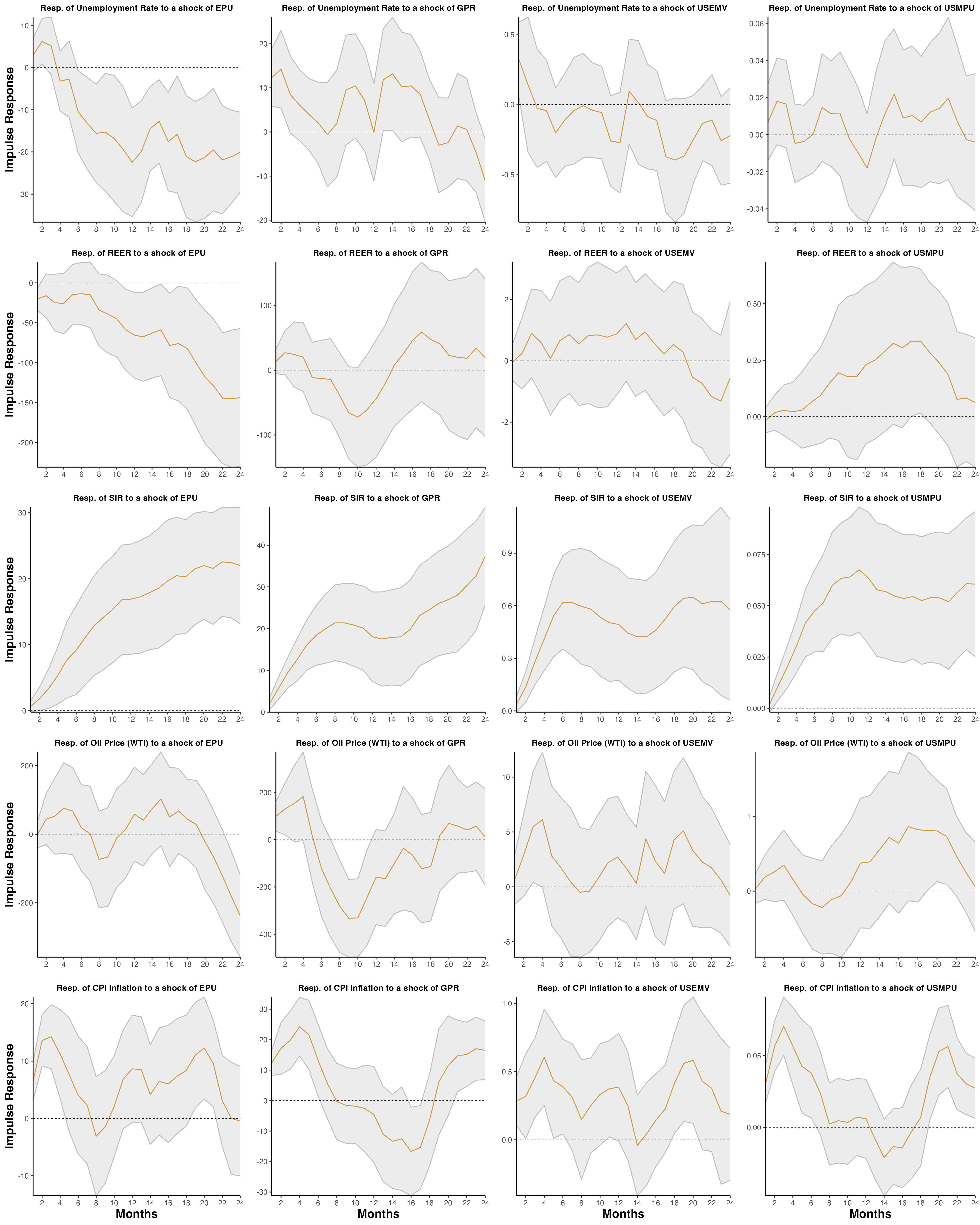}
\caption{IImpulse response functions (nonlinear local projections) of Unemployment Rate, REER, SIR, Oil Price (WTI), and CPI Inflation to EPU, GPR, USEMV, and USMPU shocks for Canada (low-rate regime). Each row shows responses of a macro variable; each column, a distinct uncertainty shock. Solid \textbf{\textcolor{orange}{orange}} lines depict IRFs with 95\% confidence bands (grey). The x-axis is horizon (months), the y-axis the response magnitude. Compare responses by reading across rows (by variable) or down columns (by shock); zero response is marked by the dotted line.}
\label{fig:irf_figure2_canada}
\end{figure}
\FloatBarrier

The country-specific plots show responses (EPU, GPR, USEMV, USMPU) of uncertainty shocks for five macroeconomic variables. Dotted horizontal lines represent zero baselines, and shaded bands are 95\% confidence intervals. The sample period is January, 1995 –Mar, 2024, where gradual changes in policy transmission are also captured through state-contingent responses across different monetary regimes \citep{jorda2005estimation}. The country grids display the impulse responses of the unemployment rate, REER, SIR, oil price (WTI), and CPI inflation to EPU, GPR, USEMV, and USMPU under the high- and low-rate regimes. We use the SIR as the regime-switching variable in the local projections framework for three reasons. First, following the documented evidence on the G7 in aggregate, the main episodes of monetary policy actions in each country before and after 2020 have also been associated with clear structural breaks in each country's short rate dynamics, both in the mean-reversion and in the volatility properties of short rates. The same is true for both the COVID shock and recovery episodes, where reallocation in rate dynamics and regime-dependent volatility are particularly large (think, e.g., of the US taper tantrum episode or the ECB PEPP announcement) \citep{darracq2024monetary}. Second, the available evidence on nonlinear rate adjustments, and in particular on the documented asymmetries in short-term rate responses to inflation, output, and the term spread, are better accounted for by regime-switching specifications rather than linear models \citep{haug2006behavior,ang2002regime}. Third, the OLS-CUSUM diagnostic suggests the presence of clear structural breaks in several G7 countries' short rates. The zero-lower-bound and quantitative easing episodes, as well as the subsequent policy normalizations, broadly line up with those. Overall, these pieces of evidence jointly provide strong support for the use of the short rate as the regime threshold in the nonlinear local projections design.

For Canada (Figures~\ref{fig:irf_figure1_canada} and \ref{fig:irf_figure2_canada}), IRFs are characterized by pronounced regime asymmetry. In high-rate states, EPU leads to transitory unemployment spikes alongside REER depreciation and a tightening bias in SIR, whereas GPR reinforces this pattern with sizable rate hikes and more volatile CPI inflation. In low-rate environments, similar shocks are met with accommodation and currency appreciation, but unemployment increases. USEMV effects are more muted, whereas USMPU highlights Canada's financial integration with the United States through divergent REER and CPI inflation paths. The sign of CPI inflation responses also reverses with stance—positive in high-rate states and negative in low-rate states—pointing to regime-dependent expectations around imported prices and domestic slack; this is in line with asymmetric transmission in Canada \citep{bhuiyan2007real}. Stance dependence is salient for the United States (Figures \ref{fig:irf_figure1_usa} and \ref{fig:irf_figure2_usa}). At high rates, EPU raises unemployment at its peak in month 14, coincides with large negative oil price responses during months 14–16, and drives near-term increases in CPI inflation with peaks at months 5–7 and again near month 10. GPR appreciates the REER, particularly via safe-haven flows. Low rates lead to amplified and more persistent real effects of EPU, keeping unemployment higher through month 16 while also generating persistent negative responses in CPI inflation. Interest rates respond contractionarily to GPR at high rates but accommodate at low rates—consistent with \cite{lhuissier2021regime} and the ``toxic interaction" between uncertainty and tight financial conditions \citep{caldara2016macroeconomic}. The impulse responses for France (Figures~\ref{fig:irf_figure1_fra} and  \ref{fig:irf_figure2_fra}) indicate that in the high-rate states, EPU leads to higher unemployment (peaking around months 18--20), higher REER (appreciation), and lower oil prices (peaking around month 12). In addition, CPI inflation disinflates initially but turns modestly positive after month 16. GPR shocks result in higher policy rates and lower CPI inflation, with the peak (negative) inflation effects materializing in months 2-4. In contrast, in the low-rate regime, unemployment's response to EPU shocks is more gradual, yet it remains elevated through month 20. Policy rates fall more noticeably during months 12-15. Furthermore, CPI inflation responses are consistently negative throughout the forecast horizon. These results, which echo those for the euro area as a whole, are consistent with those reported by \cite{castelnuovo2019domestic} for European countries, where they observe that uncertainty appears to have different transmission channels depending on the current policy rate environment. The impulse responses for the German economy (Figures~\ref{fig:irf_figure1_ger} and \ref{fig:irf_figure2_ger}) are also consistent with safe-haven effects. An increase in GPR raises unemployment, and this effect is persistent (remaining positive until month 20). The REER appreciates, peaking at about month 10. The response of interest rates to GPR turns positive after an initial period of accommodation and persists for the whole sample period. The response of CPI inflation is negative at first but becomes positive after month 10. All these features are consistent with a safe-haven role, also characterized by a credible commitment to price stability. The impulse responses for the low-rate state are instead different: unemployment jumps up following a shock to EPU, with the largest response at its peak, and then declining. The response of the REER to EPU is hump-shaped, peaking at months 10–12. The response of oil prices to EPU turns positive after an initial period of negative values. The response of CPI inflation is mostly countercyclical in the low-rate state, but it turns procyclical in the high-rate state \citep{Berg_2019,grimme2018measuring}. The IRFs for Japan (Figures \ref{fig:irf_figure1_jap} and \ref{fig:irf_figure2_jap}) show state-dependent transmission and strong exchange-rate channels. High-rate episodes are characterized by EPU-induced oscillatory unemployment dynamics, a peak effect around month 18, and sizeable delayed REER appreciation after month 12, all in line with the safe-haven properties of the yen and the Bank of Japan's gradualist policy stance. Low-rate episodes are instead associated with more muted unemployment dynamics, more contained USEMV responses, and a switch from positive to negative CPI inflation responses—aligning with Japan's deflationary bias and the structural labor-market frictions highlighted by \cite{miyamoto2016uncertainty}. The larger responses in size and volatility under tight policy conditions are also consistent with \cite{lhuissier2021regime}'s result that financial frictions amplify the effects of uncertainty when monetary policy has little room to maneuver. In the United Kingdom (Figures~\ref{fig:irf_figure1_uk} and \ref{fig:irf_figure2_uk}), an exchange-rate–inflation nexus is quite evident. In the high-rate regime, EPU increases unemployment (highest response around month 14) and is associated with persistent REER depreciation; CPI inflation displays an initial inflationary pulse followed by persistent disinflation (lasting to month 12). Interest rates act in a counter-cyclical way, with persistent negative responses, which can be interpreted as consistent with a ``stabilization first'' rather than a pure inflation-targeting policy in the face of geopolitical tensions. In the low-rate regime, unemployment responds to EPU more persistently, USEMV elicits a hump-shaped oil price response (with the maximum effect around month 16), and CPI inflation disinflates throughout; GPR displays persistent negative pressure on prices. These asymmetries are also consistent with the United Kingdom's nonlinear projections in \citep{hauzenberger2025gaussian} and with a heightened sensitivity of the United Kingdom's financial conditions, and their impact on the economy, in situations where policy space is more limited \citep{cesa2020monetary}. For Italy (Figures \ref{fig:irf_figure1_ita} and \ref{fig:irf_figure2_ita}), the difference between regimes is more in the nature of real-side adjustment: under high rates, EPU causes unemployment to increase moderately and persistently over months 12–16, the REER to appreciate, and oil prices to fall with large negative responses. There is some negative response for CPI inflation initially before a mild recovery. Low-rate effects of the same shocks have more persistent unemployment (running through to month 16), REER depreciation, more muted negative interest rate responses, and larger oil price declines; CPI inflation is negative throughout with only a tentative recovery at the end. This is in line with recent euro-area evidence on real effects of uncertainty and limited offset from accommodation \citep{bobasu2024tracking}.

To summarize, the IRF grids reveal three broad patterns that are consistent across G7 economies. First, monetary stance matters: conditional on a given uncertainty shock, responses are typically larger and faster in the high-rate regime than in the low-rate regime, a pattern that WCA analysis also suggests by revealing clusters of statistically significant coherence around episodes with sharp movements in the SIR. This finding further justifies the use of the SIR as a transition variable in the state-dependent impulse-response design. Second, the policy-distinct measures that incorporate external risks (GPR, USEMV) have strong transmission through REER and oil price (WTI) with secondary effects on CPI inflation and unemployment at medium horizons; the policy-focused measures (EPU, USMPU) exhibit clearer connections to SIR, CPI inflation, and unemployment rate. Third, cross-country heterogeneity is significant - exchange rate channels are particularly important where financial-market or energy exposures are large, while real-side adjustments predominate elsewhere. These patterns motivate the forecasting design in what follows: incorporating the same set of uncertainty measures as exogenous drivers in the multivariate model and conditioning the evaluations on the monetary environment to capture the state-dependent transmission documented here.
\FloatBarrier

\section{SZBVARx Model: Methodology}\label{Section_SZBVARx_Model}
SZBVARx provides a flexible structure for multivariate forecasting, extending the classical VAR approach by integrating Bayesian inference and allowing for exogenous variables. The model jointly estimates parameters and latent components, ensuring robust forecast performance in the presence of complex economic dynamics and uncertainty.
The methodology builds upon the seminal work of \cite{sims1998bayesian} and extends the traditional Bayesian VAR framework to incorporate exogenous uncertainty shocks, following recent advances in \cite{carriero2015bayesian} and \cite{chan2020large}. We employ the SZBVARx framework for multivariate macroeconomic forecasting across G7 economies. 

We specify the SZBVARx model as a reduced-form vector autoregression that jointly models $m$ endogenous macroeconomic variables while incorporating $k$ exogenous uncertainty measures. The model captures dynamic interdependencies among macroeconomic variables through autoregressive lags of order $p$, while allowing uncertainty shocks to contemporaneously affect all endogenous variables. The general specification takes the form:
\begin{equation}\label{eq:szbvarx_model}
y_t = \mu + \sum_{i=1}^{p} \Phi_i y_{t-i} + \Gamma x_t + u_t, \qquad u_t \sim \mathcal{N}(0, \Sigma).
\end{equation}
In this context, $y_t=[y_{1,t},\ldots,y_{m,t}]^{\top}\in \mathbb{R}^{m\times1}$ denotes the $m$-dimensional vector of endogenous macroeconomic variables, $\mu\in \mathbb{R}^{m}$ represents the $m$-dimensional vector of intercept terms, $x_t=[x_{1,t},\ldots,x_{k,t}]^{\top}\in \mathbb{R}^{k\times1}$ is a $k$-dimensional vector of exogenous uncertainty measures, and $u_t\in \mathbb{R}^{m}$ represents a white noise vector. The autoregressive coefficient matrices $\Phi_i\in \mathbb{R}^{m\times m}$,  $\forall i = \{1,\ldots,p\}$ capture cross-variable spillovers within the macroeconomic system. The matrix $\Gamma\in \mathbb{R}^{m \times k}$ measures the impact of macroeconomic uncertainty shocks on each endogenous variable, such that $\Gamma x_{t}\in \mathbb{R}^{m \times 1}$ represents the contemporaneous effect of the $k$ exogenous uncertainty measures on the $m$ endogenous variables. The covariance matrix $\Sigma \in \mathbb{R}^{m \times m}$ is positive definite and captures the contemporaneous covariance structure of the reduced-form errors.
This specification can be expressed compactly by defining the regressor vector $Z_t = [1, y_{t-1}^{\top}, \ldots, y_{t-p}^{\top}, x_t^{\top}]^{\top} \in \mathbb{R}^{d \times 1}$ and coefficient matrix $B = [\mu^{\top}, \Phi_1^{\top}, \ldots, \Phi_p^{\top}, \Gamma^{\top}]^{\top} \in \mathbb{R}^{d \times m}$, yielding:
\begin{equation}\label{eq:compact_form}
y_t = B^{\top} Z_t + u_t.
\end{equation}
The regressor dimension $d = 1 + mp + k$ reflects the intercept term, $mp$ autoregressive coefficients, and $k$ exogenous variable coefficients. For estimation purposes, we stack observations over $t = 1, \ldots, T$ to obtain the matrix form $Y = ZB + U$, where the sum of squared errors $\text{SSE}(B) = (Y - ZB)^{\top}(Y - ZB)$ forms the basis for likelihood construction. The SZBVARx framework is particularly suited for policy analysis, as it distinguishes between endogenous macroeconomic responses and exogenous uncertainty-driven fluctuations, enabling policymakers to assess the relative importance of domestic versus external uncertainty sources in driving macroeconomic volatility.

\subsection{Prior Specifications}\label{szbvarx_prior}
Bayesian inference in high-dimensional macroeconomic forecasting critically depends on the specification of prior distributions for model parameters. Following \cite{sims1998bayesian}, we place a matrix-normal distribution on the coefficient matrix $B$ conditional on the covariance matrix $\Sigma$, and an inverse-Wishart prior on $\Sigma$. Let $\beta = \text{vec}(B) \in \mathbb{R}^{md \times 1}$ denote the vectorized coefficient matrix, where $d = 1 + mp + k$ is the regressor dimension and $B \in \mathbb{R}^{d \times m}$. The prior specification takes the form:
\begin{align}
    B \mid \Sigma &\sim \mathcal{MN}_{d,m}\big( B_0,\, \lambda_0^2 \Omega_0,\, \Sigma \big) \label{eq:prior_B_matrix}\\
    \Sigma &\sim \mathcal{IW}(\Psi_0,\, \nu_0) \label{eq:prior_sigma}
\end{align}
which is equivalent to the multivariate-normal vectorized form:
\begin{equation}
    \beta \mid \Sigma \sim \mathcal{N}\big( \beta_0,\, \Sigma \otimes \lambda_0^2 \Omega_0 \big) \label{eq:prior_beta}
\end{equation}
where $\beta_0 = \text{vec}(B_0) \in \mathbb{R}^{md \times 1}$ represents the prior mean for the vectorized coefficients, $\Omega_0 \in \mathbb{R}^{d \times d}$ is a positive definite matrix encoding the prior covariance over regressors, $\Psi_0 \in \mathbb{R}^{m \times m}$ is the prior scale matrix for the covariance structure, $\nu_0 > m + 1$ denotes the prior degrees of freedom (ensuring finite prior mean $\mathbb{E}[\Sigma] = \Psi_0/(\nu_0 - m - 1)$), and $\lambda_0 > 0$ is the overall tightness parameter controlling the strength of shrinkage toward the prior mean.

The SZBVARx framework adopts this conjugate \textit{Matrix-Normal--Inverse-Wishart prior} as it provides computational efficiency through closed-form posterior distributions while allowing flexible control over parameter shrinkage. This prior specification is particularly well-suited for high-dimensional macroeconomic forecasting, as it naturally accommodates the structure of the reduced-form VAR system defined in equation \eqref{eq:compact_form}. The Kronecker product structure $\Sigma \otimes \lambda_0^2 \Omega_0$ in equation \eqref{eq:prior_beta} propagates cross-equation dependence via $\Sigma$, while structure in $\Omega_0$ delivers lag-/regressor-specific tuning; equation-specific tuning can be obtained by column-side scaling (e.g., equation-specific $\lambda_j$ or diagonal rescaling of $\Sigma$), all while preserving conjugacy \citep{banbura2010large}. Smaller values of $\lambda_0$ impose tighter shrinkage toward $\beta_0$, effectively regularizing the model in high-dimensional settings. This conjugate prior structure directly shapes the model's ability to balance signal extraction and shrinkage, which is critical for reliable multivariate forecasting in the presence of complex economic interdependencies and uncertainty shocks. Informative, structured priors such as the Matrix-Normal--Inverse-Wishart allow for nuanced modeling of heterogeneity and regularization, enhancing stability and forecast reliability in complex, multi-country analyses. Given its relative strength in balancing computational tractability, forecast accuracy, and robustness to model misspecification, we incorporate this prior specification in the flexible SZBVARx framework for G7 macroeconomic forecasting\footnote{The SZBVARx framework supports three prior specifications: (i) the Matrix-Normal--Inverse-Wishart (MN-IW) prior described above, (ii) a semi-conjugate flat-Gaussian prior, and (iii) a non-informative flat-flat prior. In our empirical implementation for G7 macroeconomic forecasting, the MN-IW prior was unanimously selected via grid-search optimization across all countries and forecast horizons, demonstrating its superior performance for this application. Consequently, the prior type is not treated as a tunable hyperparameter in Section~\ref{szbvarx_hyperparams}, and all results reported herein employ the MN-IW specification.}

\subsection{Hyperparameter Specification}\label{szbvarx_hyperparams}
The SZBVARx framework employs a systematic hyperparameter specification following \cite{sims1998bayesian} to operationalize the prior distributions described above, with parameters selected through grid search optimization detailed in Section~\ref{Sec_exp_results_baseline_comp}. To accommodate the heterogeneous dynamics observed across G7 economies and forecast horizons, all hyperparameters are specified with country and horizon-specific scaling, denoted by superscripts $(c,h)$ where $c$ indexes countries and $h$ denotes forecast horizons. The hyperparameter tuple $(p, \lambda_0, \lambda_1, \lambda_3, \lambda_4, \lambda_5, \mu_5, \mu_6)$ controls key aspects of model specification and prior shrinkage\footnote{Note that $\lambda_2$, which appears in some Minnesota prior specifications to control cross-equation shrinkage, is omitted in the Sims-Zha Normal-Wishart framework as cross-equation dependencies are captured through the Kronecker structure $\Sigma \otimes \lambda_0^2 \Omega_0$ of the prior covariance matrix.}: the lag length $p^{(c,h)} \in \{1, 2, \ldots, p_{\max}\}$ determines the autoregressive order, while $\lambda_0^{(c,h)} \geq 0$ governs overall prior tightness as introduced in the Matrix-Normal--Inverse-Wishart specification above. The relative shrinkage parameters $\lambda_1^{(c,h)} \geq 0$ (own versus cross-variable lag tightness), $\lambda_3^{(c,h)} \geq 0$ (lag decay with harmonic decay when $\lambda_3 = 1$), $\lambda_4^{(c,h)} \geq 0$ (intercept shrinkage), and $\lambda_5^{(c,h)} \geq 0$ (exogenous coefficient shrinkage) provide granular control over different coefficient groups, reflecting the economic intuition that own lags typically matter more than cross-variable effects and that coefficient importance generally declines with lag length. The dummy observation weights $\mu_5^{(c,h)} \geq 0$ and $\mu_6^{(c,h)} \geq 0$ implement additional shrinkage through sum-of-coefficients and initial-conditions constraints, with the former imposing the economically motivated restriction $\sum_{i=1}^{p} \Phi_i \approx I_m$ that shrinks the system toward unit root behavior rather than enforcing row-sum constraints \citep{sims1998bayesian,waggoner2003likelihood}. The Matrix-Normal--Inverse-Wishart prior was employed uniformly across all G7 countries and forecast horizons (semi-long-term and long-term), as it emerged as the superior choice in preliminary grid-search experiments. This prior's dominance reflects its ability to balance computational tractability through conjugacy, flexible shrinkage control via the Kronecker structure $\Sigma \otimes \lambda_0^2 \Omega_0$, and proper uncertainty quantification through the inverse-Wishart specification on $\Sigma$—features particularly valuable for capturing the complex interdependencies and heterogeneous dynamics characteristic of G7 macroeconomic systems.

\subsection{Posterior Distributions and MCMC Inference}\label{szbvarx_posterior}
In Bayesian macroeconomic forecasting, the posterior distribution reflects updated prior beliefs about model parameters after incorporating observed data, providing a natural framework for uncertainty quantification that is particularly valuable in policy applications where parameter uncertainty can significantly affect forecast intervals and decision-making. Unlike classical approaches that treat parameters as fixed unknowns, the Bayesian framework acknowledges that macroeconomic relationships evolve over time and that our knowledge of these relationships is inherently clouded by uncertainty, making posterior distributions essential for robust inference in multivariate forecasting, where the curse of dimensionality amplifies parameter uncertainty. Given data $(Y, Z)$ where $Z \in \mathbb{R}^{T \times d}$ is the regressor matrix and a chosen prior, Bayesian inference for the SZBVARx model proceeds by updating prior beliefs about model parameters using the reduced-form Gaussian likelihood:
\begin{equation}\label{eq:likelihood}
\mathcal{L}(B, \Sigma \mid Y, Z) \propto |\Sigma|^{-T/2} \exp\left[-\frac{1}{2} \text{tr}\left(\Sigma^{-1} (Y - ZB)^{\top}(Y - ZB)\right)\right],
\end{equation}
where $B \in \mathbb{R}^{d \times m}$ and $\Sigma \in \mathbb{R}^{m \times m}$. Under the conjugate Matrix-Normal--Inverse-Wishart prior specified in Equations \eqref{eq:prior_B_matrix}--\eqref{eq:prior_sigma}, the posterior distributions retain the same functional form. Because of full conjugacy, the joint posterior factorizes as
\begin{equation}
p(\Sigma, B \mid Y, Z) = p(\Sigma \mid Y, Z) p(B \mid \Sigma, Y, Z),
\end{equation}
yielding closed-form conditional posteriors:
\begin{align}
B \mid \Sigma, Y, Z &\sim \mathcal{MN}_{d,m}(\bar{B}, \bar{\Omega}, \Sigma), \label{eq:mniw_post_B_matrix} \\
\Sigma \mid Y, Z &\sim \mathcal{IW}(\bar{\Psi}, \bar{\nu}), \label{eq:mniw_post_Sigma}
\end{align}
with posterior hyperparameters:
\begin{equation}
\bar{\Omega}^{-1} = (\lambda_0^2 \Omega_0)^{-1} + Z^{\top}Z, \label{eq:post_Omega}
\end{equation}
representing the posterior precision matrix for the coefficient covariance that combines prior information with the observed regressor cross-product;
\begin{equation}
\bar{B} = \bar{\Omega}\left[(\lambda_0^2 \Omega_0)^{-1}B_0 + Z^{\top}Y\right], \label{eq:post_B}
\end{equation}
the posterior mean coefficient matrix that optimally weights the prior mean $B_0$ and the data-driven OLS estimate;
\begin{equation}
\bar{\Psi} = \Psi_0 + (Y - Z\bar{B})^{\top}(Y - Z\bar{B}) + (\bar{B} - B_0)^{\top}(\lambda_0^2 \Omega_0)^{-1}(\bar{B} - B_0), \label{eq:post_Psi}
\end{equation}
the posterior scale matrix for $\Sigma$ that accumulates prior scale, observed sum-of-squares, and adjustment terms reflecting the updating from prior to posterior mean; and
\begin{equation}
\bar{\nu} = \nu_0 + T, \label{eq:post_nu}
\end{equation}
the posterior degrees of freedom that increase with sample size $T$. Equivalently, in vectorized form:
\begin{equation}
\beta \mid \Sigma, Y, Z \sim \mathcal{N}(\bar{\beta}, \Sigma \otimes \bar{\Omega}), \label{eq:mniw_post_beta_vec}
\end{equation}
where $\beta = \text{vec}(B)$ and $\bar{\beta} = \text{vec}(\bar{B})$.

Markov Chain Monte Carlo (MCMC) inference can proceed via two equivalent approaches. For direct sampling, we draw independently and identically distributed (i.i.d.) posterior samples: for $s = 1, \ldots, S$, draw $\Sigma^{(s)} \sim \mathcal{IW}(\bar{\Psi}, \bar{\nu})$ and then $B^{(s)} \sim \mathcal{MN}_{d,m}(\bar{B}, \bar{\Omega}, \Sigma^{(s)})$, yielding i.i.d. posterior draws. Alternatively, for a Gibbs sampler with conditional dependence on the current draw, we alternate $B^{(s)} \sim \mathcal{MN}_{d,m}(\bar{B}, \bar{\Omega}, \Sigma^{(s-1)})$ and $\Sigma^{(s)} \sim \mathcal{IW}(\Psi_0 + (Y - ZB^{(s)})^{\top}(Y - ZB^{(s)}), \nu_0 + T)$. Both approaches are correct and yield a sequence of parameter draws $\{(B^{(s)}, \Sigma^{(s)})\}_{s=1}^S$ whose empirical distribution approximates the joint posterior, enabling Bayesian credible prediction intervals and posterior predictive inference essential for uncertainty quantification in high-dimensional G7 macroeconomic applications with country and horizon-specific hyperparameters configurations.

\subsection{Forecast Generation and Stability}\label{szbvarx_forecast}
The SZBVARx framework generates forecasts through posterior predictive simulations that naturally incorporate the future shock uncertainty, providing comprehensive forecast distributions essential for policy analysis under uncertainty. This approach leverages the full posterior distribution $\{B^{(s)}, \Sigma^{(s)}\}_{s=1}^S$ obtained from the Gibbs sampling algorithm to produce forecasts that reflect all sources of uncertainty inherent in multivariate macroeconomic systems. For each MCMC draw $s$ and forecast horizon $h$, the recursive forecast equation is:
\begin{equation}\label{eq:forecast_recursion}
y_{T+h \mid T}^{(s)} = \mu^{(s)} + \sum_{i=1}^{p} \Phi_i^{(s)} y_{T+h-i \mid T}^{(s)} + \Gamma^{(s)} x_{T+h} + u_{T+h}^{(s)},
\end{equation}
where $u_{T+h}^{(s)} \sim \mathcal{N}(0, \Sigma^{(s)})$ represents future innovations and $y_{T+j \mid T}^{(s)} = y_{T+j}$ for $j \leq 0$. When future uncertainty measures $x_{T+h}$ are unobserved, we condition on policy scenarios or employ auxiliary forecasting models with uncertainty propagation. The point forecast is computed as the posterior mean:
\begin{equation}\label{eq:posterior_mean}
\hat{y}_{T+h \mid {y}_{T}} = \frac{1}{S} \sum_{s=1}^{S} y_{T+h \mid {y}_{T}}^{(s)},
\end{equation}
while credible intervals utilize empirical quantiles of the forecast distribution $\{y_{T+h \mid {y}_{T}}^{(s)}\}_{s=1}^S$.
System stability is ensured by verifying that all eigenvalues of the companion matrix
\begin{equation*}
\mathbf{C} = \begin{bmatrix}
\Phi_1 & \Phi_2 & \cdots & \Phi_{p-1} & \Phi_p \\
I_m & 0 & \cdots & 0 & 0 \\
0 & I_m & \cdots & 0 & 0 \\
\vdots & \vdots & \ddots & \vdots & \vdots \\
0 & 0 & \cdots & I_m & 0
\end{bmatrix} 
\end{equation*}
satisfy $|\rho_i| < 1,$ $\forall {i}=\{1, \ldots, mp\}$, with forecasts retained only for parameter draws satisfying this stability condition to ensure well-defined long-run dynamics \citep{lutkepohl2005new}. This stability check is particularly important in high-dimensional G7 applications where the interaction between domestic variables and uncertainty shocks can potentially generate explosive behavior without proper regularization through the country and horizon-specific prior specifications.

\section{Experimental Evaluation and Analysis of Results}\label{Section_Experimental_Evaluation}
This section empirically benchmarks the proposed SZBVARx against a wide variety of baseline models, including classical econometric models, machine learning models, and deep learning models. The experimental setup aims to evaluate the generalization capability and effectiveness of each model in a more realistic macroeconomic environment. The target prediction task is to forecast the conditional changes in the next $h$ months ($h \in \{12, 24\}$) of the five endogenous variables (Unemployment Rate, REER, SIR, Oil Price (WTI), and CPI Inflation) in the presence of the four exogenous uncertainty shocks (EPU, GPR, USEMV, and USMPU). The goal is to evaluate the forecasting capability in both the semi-long-term (12 months) and long-term (24 months) horizons to provide a comprehensive evaluation of the model's performance in macroeconomic prediction tasks. The rest of this section is structured as follows. Section~\ref{Sec_baseline_models} introduces the baseline algorithms. Section~\ref{Sec_evaluation_Metric} provides details on the evaluation metrics. Section~\ref{Sec_exp_results_baseline_comp} presents and discusses the experimental results. Section~\ref{Sec_stat_signif} explores the statistical significance of the forecasting performance gaps between different models, while Section~\ref{CPPI_coverage_G7} quantifies the uncertainty in predictive distributions through credible intervals (see Appendix~\ref{app:Appendix_PPI_12M} for credible intervals of the 12-month ahead forecast horizon).

\subsection{Baseline Models}\label{Sec_baseline_models}
To properly evaluate SZBVARx, we select an extended set of baseline models, representing traditional and contemporary approaches used in macroeconomic forecasting that can accommodate exogenous variables.We use \textit{VARx}, \textit{Threshold VAR (TVAR)}, and \textit{Vector exponential smoothing (VES)} as our baselines from the classical econometric literature. To benchmark against modern machine learning and deep learning, we consider \textit{Neural Basis Expansion Analysis for Time Series with exogenous variables (NBEATSx), Neural Hierarchical Interpolation for Time Series Forecasting with exogenous covariates
(NHITSx), CatBoost with exogenous covariates (CatBoostx) and LightGBM with exogenous covariates (LGBMx)}, \textit{Xtreme Gradient Boosting with exogenous covariates (XGBx)}, \textit{Temporal Fusion Transformer with exogenous covariates (TFTx)}, \textit{Time-Series Dense Encoder with exogenous covariates (TiDEx)}, \textit{Block recurrent neural networks with exogenous covariates (BRNNx)}, \textit{Time-Series Mixer with exogenous covariates (TSMIXERx)}, and \textit{DishTS ensembles augmented by CatBoostx and XGBx} (${DTS}\_CBx$, ${DTS}\_{XGBx}$). All the machine learning and neural models can take uncertainty shocks as covariates. We provide additional details and exogenous feature processing for each model in Appendix~\ref{app:Appendix_Baseline}.

\subsection{Evaluation Metrics}\label{Sec_evaluation_Metric}
To rigorously assess the forecasting accuracy and robustness of the SZBVARx, we use five standard metrics: Root Mean Square Error (RMSE), Symmetric Mean Absolute Percentage Error (SMAPE), Mean Absolute Scaled Error (MASE), Theil's $U_1$ (denoted 
as $TU_1$), and Median Absolute Percentage Error (MDAPE). RMSE is scale-dependent; SMAPE and MDAPE are percentage-based; MASE rescales errors by the in-sample seasonal naïve benchmark; and $TU_1$ is a scale-normalized index in $[0,1]$. The mathematical formulations are:
\begin{equation*}
\begin{gathered}
\mathrm{RMSE}=\sqrt{\frac{1}{h}\sum_{t=1}^{h}\bigl(y_t-\hat y_t\bigr)^2};\qquad
\mathrm{MASE}=\frac{\displaystyle \sum_{t=D+1}^{D+h} \lvert \hat y_t-y_t\rvert}{\displaystyle \frac{h}{D-S}\sum_{t=S+1}^{D}\lvert y_t-y_{t-S}\rvert};\\[4pt]
\mathrm{SMAPE}=\frac{1}{h}\sum_{t=1}^{h}\frac{\lvert \hat y_t-y_t\rvert}{(\lvert \hat y_t\rvert+\lvert y_t\rvert)/2}\times 100\%;\qquad
TU_1=\frac{\sqrt{\frac{1}{h}\sum_{t=1}^{h}\bigl(y_t-\hat y_t\bigr)^2}}{\sqrt{\frac{1}{h}\sum_{t=1}^{h}y_t^{2}}+\sqrt{\frac{1}{h}\sum_{t=1}^{h}\hat y_t^{2}}};\\[4pt]
\mathrm{MDAPE}=\operatorname{median}_{t=1,\ldots,h}\!\left(\frac{\lvert y_t-\hat y_t\rvert}{\lvert y_t\rvert}\right)\times 100\%.
\end{gathered}
\end{equation*}
Here, $y_t$ denotes the observed value at time $t$, $\hat y_t$ the forecast, $h$ the forecast horizon, and $D,S$ the in-sample length and seasonal period used in MASE. Because no single metric suffices in macro-casting settings with persistence, breaks, and exogenous shocks, this suite jointly assesses magnitude accuracy (RMSE), percentage accuracy (SMAPE, MDAPE), scale-free performance relative to a seasonal naïve benchmark (MASE), and scale-normalized error independent of units ($TU_1$). Conventionally, the model minimizing a given error metric is regarded as the top performer \citep{hyndman2018forecasting, sengupta2025forecasting, hewamalage2023forecast, komunjer2012multivariate}.

\subsection{Experimental results and baseline comparison} \label{Sec_exp_results_baseline_comp}
We evaluate SZBVARx on the joint dynamics of five stated endogenous indicators augmented with four exogenous uncertainty shocks. Hyperparameters are tuned by an explicit grid search in \textbf{R} using \texttt{MSBVAR}'s \texttt{szbvar} routine and custom orchestration, selecting the tuple $(p, \lambda_0, \lambda_1, \lambda_3, \lambda_4, \\ \lambda_5, \mu_5, \mu_6)$
that minimizes a rolling multivariate RMSE on the training sample under an expanding-window evaluation for each forecast horizon $h \in \{12, 24\}$. For a candidate hyperparameter vector $\theta$, denote $e_{t+h\mid t}(\theta) = y_{t+h} - \hat{y}_{t+h\mid t}(\theta)$ and $\mathcal{T}_h$ the set of training-origin dates used in the grid search. We define the Multivariate Root Mean Square Error (MRMSE) as:
\begin{equation}
\operatorname{MRMSE}_h(\theta)
= \sqrt{\frac{1}{m\, |\mathcal{T}_h|}\, \sum_{t \in \mathcal{T}_h} \big\| e_{t+h\mid t}(\theta) \big\|_2^2 } ,
\end{equation} where $m=5$ is the number of endogenous variables and $\big\|.\big\|_2$ denotes the Euclidean norm.
We select $\hat{\theta}_h = \arg\min_{\theta} \operatorname{MRMSE}_h(\theta)$. This procedure respects the time-series ordering and avoids peek-ahead bias \citep{hamilton1994time,lutkepohl2005new}. The lag length $p$ captures the persistence structure of macroeconomic relationships, while shrinkage parameters $(\lambda_0, \lambda_1, \lambda_3, \lambda_4, \lambda_5)$ control the degree of regularization applied to blocks of coefficients in the reduced-form specification (see Equation~\eqref{eq:szbvarx_model}).

The dummy-observation weights $(\mu_5, \mu_6)$ implement, the sum-of-coefficients (long-run) and initial-conditions/drift controls in the Sims--Zha prior, respectively, which help stabilize medium- and long-horizon dynamics \citep{sims1998bayesian,banbura2010large,giannone2015prior}. This optimization framework allows the Matrix-Normal--Inverse-Wishart (MN--IW) structure to adapt to country-specific features and horizon-specific persistence. Table~\ref{tab:szbvarx_hyperparams_g7} reveals three economically intuitive patterns:
\textit{(i) Medium-horizon parsimony (12-month ahead horizon):} Most G7 countries favor $p=1$ with moderate overall tightness $\lambda_0=0.2$ and light intercept shrinkage $\lambda_4=0.1$, consistent with first-order autoregressive sufficiency at medium horizons. The United States is a notable exception, selecting $p=4$ and stronger lag decay $\lambda_3=3$; it also uses $\lambda_5=0.5$ and nonzero dummy weights $(\mu_5,\mu_6)=(0.5,0.5)$ at 12-month ahead horizon. \textit{(ii) Longer-horizon persistence (24-month ahead horizon):} Several economies transition to higher lag orders and looser overall tightness: France adopts $p=3$ with $\lambda_0=0.8$, while Germany, the United Kingdom, and Italy use $p=4$ (Germany with $\lambda_0=0.4$). Intercept control tightens to $\lambda_4=0.5$ for France, Germany, the United Kingdom; Italy, Canada and the United States remain at $\lambda_4=0.1$. Lag decay increases to $\lambda_3=2$ for Germany (others mostly remain at 1). Relative own-vs.-cross tightness $\lambda_1$ becomes less stringent for the United States ($0.1$) and France ($0.2$) at 24--month horizon, compared with $0.05$ elsewhere. \textit{(iii) Long-run dummies and exogenous channels at 24-month ahead horizon:} The initial-conditions/drift weight $\mu_6$ becomes active at 24--month horizon for Canada ($1$), France ($0.5$), Germany ($0.5$), the United Kingdom ($1$), and Italy ($0.5$), while remaining $0$ for the United States and Japan. The sum-of-coefficients weight $\mu_5$ is already active at 12-month ahead horizon for most countries and at 24-month ahead horizon increases for the United States (to $1$), decreases for Japan (to $0.5$), and remains $1$ elsewhere except France ($\mu_5=0$ at 24-month ahead horizon). Shrinkage on exogenous coefficients is relaxed at 24-month ahead horizon for Canada ($\lambda_5=1$), and Italy ($\lambda_5=0.5$), allowing stronger transmission from uncertainty shocks; other countries retain $\lambda_5=0$ at 24-month ahead horizon. The grid search evaluated three reduced-form prior families: Normal--Wishart (conjugate MN--IW), Flat--Gaussian (Gaussian on $\mathrm{vec}(B)$ with Jeffreys prior on $\Sigma$), and Flat--Flat. The MN--IW specification was unanimously selected across all G7 countries and both horizons (Table~\ref{tab:szbvarx_hyperparams_g7}). Conjugacy yields closed-form posteriors and exact Gibbs updates for $(B,\Sigma)$, promotes coherent cross-equation shrinkage, and tends to improve forecast robustness in medium-dimensional systems \citep{sims1998bayesian,banbura2010large,giannone2015prior,carriero2015bayesian}.

Against the baselines in Section~\ref{Sec_baseline_models}
SZBVARx achieves one of the best and most consistent performances across both horizons (Section~\ref{Sec_stat_signif}). The SZBVARx model was implemented in \textbf{R} via the \texttt{MSBVAR} package, while baseline models used \texttt{tsDyn} and \texttt{legion} in \textbf{R} and \texttt{Darts} in Python. The effectiveness of the proposed SZBVARx model lies in its ability to harness the benefits of hierarchical shrinkage to regularize the VARx system. This hierarchical shrinkage penalizes both variance inflation when increasing the dimension of the system (\(p\)), and allows for cross-equation coherence and structural flexibility to capture near-unit-root processes and level shifts with hyperparameters (\(\mu_5, \mu_6\)). The competing VARx baseline has no regularization and is subject to overfitting in higher dimensions. TVAR alternatives require very large datasets for stable estimation of regimes, but underperform at long horizons. Alternative state-of-the-art machine learning and deep learning models can potentially achieve high performance on long stationary panels, but in general, are over-parameterized in short macroeconomic samples, leading to overfitting. 
Overall, principled regularization of SZBVARx results in robust, interpretable, and superior forecasting performance under various macroeconomic environments.

\begin{table*}[htbp!]
\centering
\caption{
Trained hyper-parameters (\textit{shrinkage, and model controls}) for SZBVARx models across G7 countries and both forecast horizons: 12-month-ahead (12) and 24-month-ahead (24). Parameters are estimated for optimal multivariate macroeconomic forecasting with exogenous uncertainty shocks.
}
\label{tab:szbvarx_hyperparams_g7}
\begin{adjustbox}{max width=0.7\linewidth}
\begin{tabular}{lccccccccc}
\toprule
\textbf{Country} & \textbf{Horizon} & \textbf{$p$} & \textbf{$\lambda_0$} & \textbf{$\lambda_1$} & \textbf{$\lambda_3$} & \textbf{$\lambda_4$} & \textbf{$\lambda_5$} & \textbf{$\mu_5$} & \textbf{$\mu_6$} \\
\midrule
Canada  & 12 & 1 & 0.2 & 0.05 & 1 & 0.1 & 0 & 1 & 0  \\
        & 24 & 1 & 0.6 & 0.05 & 1 & 0.1 & 1 & 1 & 1  \\
\midrule
US      & 12 & 4 & 0.2 & 0.05 & 3 & 0.1 & 0.5 & 0.5 & 0.5  \\
        & 24 & 4 & 0.2 & 0.1  & 1 & 0.1 & 0 & 1 & 0  \\
\midrule
France  & 12 & 1 & 0.2 & 0.05 & 1 & 0.1 & 0 & 1 & 0  \\
        & 24 & 3 & 0.8 & 0.2 & 1 & 0.5 & 0 & 0 & 0.5 \\
\midrule
Germany & 12 & 1 & 0.2 & 0.05 & 1 & 0.1 & 0 & 1 & 0 \\
        & 24 & 4 & 0.4 & 0.05 & 2 & 0.5 & 0 & 1 & 0.5 \\
\midrule
Japan   & 12 & 1 & 0.2 & 0.05 & 1 & 0.1 & 0 & 1 & 0 \\
        & 24 & 2 & 0.2 & 0.05 & 1 & 0.1 & 0 & 0.5 & 0 \\
\midrule
UK      & 12 & 1 & 0.2 & 0.05 & 1 & 0.1 & 0 & 1 & 0 \\
        & 24 & 4 & 0.2 & 0.05 & 1 & 0.5 & 0 & 1 & 1 \\
\midrule
Italy   & 12 & 1 & 0.2 & 0.05 & 1 & 0.1 & 0 & 1 & 0 \\
        & 24 & 4 & 0.2 & 0.05 & 1 & 0.1 & 0.5 & 1 & 0.5 \\
\bottomrule
\end{tabular}
\end{adjustbox}
\end{table*}
\FloatBarrier
The country-horizon patterns in Table~\ref{tab:szbvarx_hyperparams_g7} thus reflect an adaptive, economically disciplined prior: tight where data is informative, permissive where persistence and breaks dominate, a configuration that underwrites SZBVARx's robust, low-variance forecasts for the G7. Across Tables~\ref{tab:can_perf_eval_G7}--\ref{tab:ita_perf_eval_G7}, SZBVARx delivers the most consistent top-tier performance across the five targets, both horizons, and scoring metrics. On the Unemployment Rate, SZBVARx is best at a 24-month-ahead horizon in France, the United Kingdom, and Italy, and is highly competitive elsewhere, while VARx leads in Canada and the United States, and TVAR in Germany; Japan's best is VES. For CPI Inflation at the 24-month-ahead horizon, SZBVARx tops the list in Canada and in the United States, whereas VARx or TVAR dominate elsewhere (France: VARx; Germany: TVAR; United Kingdom: VARx; Italy: TVAR; Japan: VARx). For REER at a 24-month-ahead horizon, SZBVARx ranks first in Canada, France, and Italy, TVAR leading in the United States and Japan, XGBx in Germany, and CatBoostx in the UK. In Oil Price (WTI) at a 24-month-ahead horizon, SZBVARx leads in four countries - Canada, the United States, France, and Japan - with NHITSx winning in Germany and Italy and NBEATSx in the United Kingdom. For SIR at a 24-month-ahead horizon, deep nets generally dominate (e.g., BRNNx: Canada, France, Germany, United Kingdom, Italy); Japan's best is DTS\_CBx, while SZBVARx is often strongest at a 12-month-ahead horizon (e.g., Canada; France; Italy). Scale-normalized diagnostics corroborate these patterns: unemployment rate at 24-month-ahead horizon posts low TU\textsubscript{1} for SZBVARx in France/United Kingdom and Italy; CPI Inflation at 24-month-ahead horizon yields efficient TU\textsubscript{1} for Canada, United States; and oil price (WTI) at 24-month-ahead horizon records uniformly small TU\textsubscript{1} across countries, indicating strong \emph{scale-adjusted} accuracy.

\begin{table*}[!ht]
\caption{Evaluation of the SZBVARx model’s performance relative to 14 baseline forecasters across all forecast horizons (12-month-ahead \& 24-month-ahead) for Canada (\textbf{\underline{best}} and \textbf{\textit{second-best}} results are highlighted).}
\label{tab:can_perf_eval_G7}
\centering
\begin{adjustbox}{width=1\textwidth, height=4.5cm}
{\tiny
\renewcommand{\arraystretch}{0.85}
\begin{tabular}{cclccccccccccccccc}
\toprule
Variables & Horizons & Metrics & \textbf{VARx} & \textbf{TVAR} & \textbf{VES} & \textbf{NBEATSx} & \textbf{LGBMx} & \textbf{DTS\_CBx} & \textbf{DTS\_XGBx} & \textbf{NHITSx} & \textbf{TSMIXERx} & \textbf{BRNNx} & \textbf{XGBx} & \textbf{CatBoostx} & \textbf{TiDEx} & \textbf{TFTx} & \textbf{\textcolor{blue}{SZBVARx}} \\
\midrule
\multirow{10}{*}{Unemployment Rate} & 12M & RMSE & \textbf{\underline{0.37}} & 0.43 & 0.58 & 2.04 & 2.43 & 2.91 & 4.54 & 3.38 & 2.50 & 1.79 & 2.82 & 1.93 & 1.70 & 1.70 & \textbf{\textit{0.39}} \\
 &     & SMAPE & 0.06 & \textbf{\textit{0.06}} & 0.09 & 0.26 & 0.32 & 0.37 & 0.47 & 0.44 & 0.35 & 0.27 & 0.26 & 0.24 & 0.23 & 0.22 & \textbf{\underline{0.05}} \\
 &     & MDAPE & 5.81 & \textbf{\underline{4.06}} & 11.40 & 29.84 & 37.29 & 41.85 & 51.80 & 59.19 & 39.28 & 38.16 & 16.06 & 20.76 & 31.04 & 24.60 & \textbf{\textit{4.11}} \\
 &     & ${\rm TU}_1$ & \textbf{\underline{0.03}} & 0.04 & 0.05 & 0.16 & 0.18 & 0.21 & 0.29 & 0.23 & 0.18 & 0.14 & 0.21 & 0.15 & 0.13 & 0.14 & \textbf{\textit{0.03}} \\
 &     & MASE & 1.04 & \textbf{\textit{1.00}} & 1.51 & 5.27 & 6.77 & 8.04 & 11.82 & 9.83 & 7.33 & 5.27 & 6.01 & 4.90 & 4.52 & 4.40 & \textbf{\underline{0.90}} \\
\cline{3-18}
 & 24M & RMSE & \textbf{\underline{1.20}} & 1.37 & 2.61 & 2.41 & 2.12 & 1.83 & 1.85 & 1.82 & 2.46 & 2.02 & 2.52 & 2.03 & 1.74 & 1.58 & \textbf{\textit{1.32}} \\
 &     & SMAPE & \textbf{\underline{0.18}} & 0.21 & 0.38 & 0.34 & 0.32 & 0.28 & 0.28 & 0.27 & 0.37 & 0.31 & 0.35 & 0.30 & 0.24 & 0.22 & \textbf{\textit{0.20}} \\
\bottomrule
\end{tabular}
}
\end{adjustbox}
\end{table*}
\FloatBarrier

\begin{table*}[!ht]
\caption{Evaluation of the SZBVARx model’s performance relative to 14 baseline forecasters across all forecast horizons (12-month-ahead \& 24-month-ahead) for the US (\textbf{\underline{best}} and \textbf{\textit{second-best}} results are highlighted).}
\label{tab:usa_perf_eval_G7}
\centering
\begin{adjustbox}{width=1\textwidth, height=4.5cm}
{\tiny
\renewcommand{\arraystretch}{0.85}
\begin{tabular}{cclccccccccccccccc}
\toprule
Variables & Horizons & Metrics & \textbf{VARx} & \textbf{TVAR} & \textbf{VES} & \textbf{NBEATSx} & \textbf{LGBMx} & \textbf{DTS\_CBx} & \textbf{DTS\_XGBx} & \textbf{NHITSx} & \textbf{TSMIXERx} & \textbf{BRNNx} & \textbf{XGBx} & \textbf{CatBoostx} & \textbf{TiDEx} & \textbf{TFTx} & \textbf{\textcolor{blue}{SZBVARx}} \\
\midrule
\multirow{10}{*}{Unemployment Rate} & 12M & RMSE & \textbf{\textit{0.56}} & \textbf{\underline{0.51}} & 0.94 & 3.74 & 3.44 & 2.78 & 4.10 & 6.17 & 4.53 & 2.13 & 1.39 & 1.38 & 3.64 & 3.42 & 0.65 \\
 &     & SMAPE & \textbf{\textit{0.13}} & \textbf{\underline{0.11}} & 0.22 & 0.56 & 0.60 & 0.52 & 0.60 & 0.88 & 0.72 & 0.44 & 0.28 & 0.28 & 0.56 & 0.51 & 0.13 \\
 &     & MDAPE & \textbf{\textit{13.48}} & \textbf{\underline{8.93}} & 23.92 & 71.06 & 88.36 & 71.33 & 89.55 & 161.99 & 124.93 & 57.19 & 36.48 & 34.37 & 96.66 & 102.27 & 16.05 \\
 &     & ${\rm TU}_1$ & \textbf{\underline{0.07}} & \textbf{\textit{0.07}} & 0.11 & 0.34 & 0.32 & 0.27 & 0.36 & 0.46 & 0.38 & 0.22 & 0.16 & 0.16 & 0.33 & 0.33 & 0.08 \\
 &     & MASE & \textbf{\textit{2.49}} & \textbf{\underline{1.99}} & 4.47 & 16.01 & 16.51 & 13.30 & 17.62 & 30.00 & 21.66 & 10.52 & 6.18 & 6.29 & 16.20 & 14.20 & 2.66 \\
\cline{3-18}
 & 24M & RMSE & \textbf{\underline{0.42}} & 1.69 & \textbf{\textit{0.99}} & 5.04 & 1.70 & 2.20 & 3.07 & 2.60 & 2.15 & 2.15 & 5.01 & 2.67 & 3.30 & 1.77 & 1.31 \\
 &     & SMAPE & \textbf{\underline{0.09}} & 0.56 & \textbf{\textit{0.23}} & 0.80 & 0.35 & 0.45 & 0.51 & 0.51 & 0.43 & 0.46 & 0.77 & 0.53 & 0.57 & 0.33 & 0.29 \\
\bottomrule
\end{tabular}
}
\end{adjustbox}
\end{table*}
\FloatBarrier

\begin{table*}[!ht]
\caption{Evaluation of the SZBVARx model’s performance relative to 14 baseline forecasters across all forecast horizons (12-month-ahead \& 24-month-ahead) for France (\textbf{\underline{best}} and \textbf{\textit{second-best}} results are highlighted).}
\label{tab:fra_perf_eval_G7}
\centering
\begin{adjustbox}{width=1\textwidth, height=4.5cm}
{\tiny
\renewcommand{\arraystretch}{0.85}
\begin{tabular}{cclccccccccccccccc}
\toprule
Variables & Horizons & Metrics & \textbf{VARx} & \textbf{TVAR} & \textbf{VES} & \textbf{NBEATSx} & \textbf{LGBMx} & \textbf{DTS\_CBx} & \textbf{DTS\_XGBx} & \textbf{NHITSx} & \textbf{TSMIXERx} & \textbf{BRNNx} & \textbf{XGBx} & \textbf{CatBoostx} & \textbf{TiDEx} & \textbf{TFTx} & \textbf{\textcolor{blue}{SZBVARx}} \\
\midrule
\multirow{10}{*}{Unemployment Rate} & 12M & RMSE & 0.52 & \textbf{\underline{0.30}} & \textbf{\textit{0.30}} & 2.30 & 1.31 & 0.91 & 0.91 & 1.67 & 0.81 & 2.04 & 0.50 & 0.95 & 2.50 & 1.29 & 0.48 \\
 &     & SMAPE & 0.06 & \textbf{\textit{0.03}} & \textbf{\underline{0.03}} & 0.27 & 0.16 & 0.11 & 0.11 & 0.18 & 0.08 & 0.24 & 0.06 & 0.11 & 0.33 & 0.14 & 0.05 \\
 &     & MDAPE & 6.42 & \textbf{\textit{3.57}} & \textbf{\underline{3.02}} & 16.95 & 15.63 & 11.37 & 11.73 & 21.01 & 5.86 & 24.75 & 5.80 & 11.55 & 31.17 & 16.97 & 6.12 \\
 &     & ${\rm TU}_1$ & 0.04 & \textbf{\underline{0.02}} & \textbf{\textit{0.02}} & 0.16 & 0.08 & 0.06 & 0.06 & 0.10 & 0.05 & 0.12 & 0.03 & 0.06 & 0.18 & 0.08 & 0.03 \\
 &     & MASE & 2.24 & \textbf{\textit{1.36}} & \textbf{\underline{1.32}} & 8.81 & 6.62 & 4.68 & 4.46 & 7.96 & 3.37 & 10.53 & 2.34 & 4.70 & 10.31 & 5.96 & 2.02 \\
\cline{3-18}
 & 24M & RMSE & \textbf{\textit{0.45}} & 1.25 & 0.88 & 2.08 & 1.11 & 1.41 & 1.58 & 1.20 & 2.24 & 2.18 & 1.41 & 1.56 & 1.93 & 1.97 & \textbf{\underline{0.38}} \\
 &     & SMAPE & \textbf{\textit{0.05}} & 0.16 & 0.12 & 0.20 & 0.13 & 0.17 & 0.16 & 0.12 & 0.26 & 0.26 & 0.15 & 0.19 & 0.21 & 0.21 & \textbf{\underline{0.04}} \\
\bottomrule
\end{tabular}
}
\end{adjustbox}
\end{table*}
\FloatBarrier

\begin{table*}[!ht]
\caption{Evaluation of the SZBVARx model’s performance relative to 14 baseline forecasters across all forecast horizons (12-month-ahead \& 24-month-ahead) for Germany (\textbf{\underline{best}} and \textbf{\textit{second-best}} results are highlighted).}
\label{tab:ger_perf_eval_G7}
\centering
\begin{adjustbox}{width=1\textwidth, height=4.5cm}
{\tiny
\renewcommand{\arraystretch}{0.85}
\begin{tabular}{cclccccccccccccccc}
\toprule
Variables & Horizons & Metrics & \textbf{VARx} & \textbf{TVAR} & \textbf{VES} & \textbf{NBEATSx} & \textbf{LGBMx} & \textbf{DTS\_CBx} & \textbf{DTS\_XGBx} & \textbf{NHITSx} & \textbf{TSMIXERx} & \textbf{BRNNx} & \textbf{XGBx} & \textbf{CatBoostx} & \textbf{TiDEx} & \textbf{TFTx} & \textbf{\textcolor{blue}{SZBVARx}} \\
\midrule
\multirow{10}{*}{Unemployment Rate} & 12M & RMSE & 0.28 & 0.37 & \textbf{\textit{0.26}} & 2.59 & 1.65 & 2.05 & 2.00 & 5.46 & 1.48 & 3.61 & 2.73 & 2.39 & 2.27 & 3.08 & \textbf{\underline{0.22}} \\
 &     & SMAPE & 0.07 & 0.08 & \textbf{\underline{0.06}} & 0.62 & 0.37 & 0.49 & 0.46 & 0.92 & 0.37 & 0.73 & 0.57 & 0.55 & 0.46 & 0.64 & \textbf{\textit{0.06}} \\
 &     & MDAPE & 5.84 & 6.34 & \textbf{\textit{4.91}} & 65.68 & 41.26 & 66.25 & 61.67 & 166.07 & 42.19 & 118.44 & 93.71 & 76.44 & 35.57 & 100.79 & \textbf{\underline{3.76}} \\
 &     & ${\rm TU}_1$ & 0.05 & 0.06 & \textbf{\textit{0.04}} & 0.32 & 0.21 & 0.25 & 0.24 & 0.47 & 0.19 & 0.37 & 0.31 & 0.28 & 0.29 & 0.33 & \textbf{\underline{0.04}} \\
 &     & MASE & 1.15 & 1.43 & \textbf{\textit{0.93}} & 11.04 & 7.53 & 10.17 & 9.59 & 26.94 & 7.10 & 18.00 & 13.00 & 11.91 & 8.12 & 14.89 & \textbf{\underline{0.93}} \\
\cline{3-18}
 & 24M & RMSE & 0.76 & \textbf{\underline{0.36}} & 0.67 & 2.63 & \textbf{\textit{0.51}} & 1.20 & 0.90 & 1.42 & 1.69 & 3.86 & 0.73 & 1.24 & 1.62 & 0.69 & 0.69 \\
 &     & SMAPE & 0.24 & \textbf{\underline{0.10}} & 0.19 & 0.52 & \textbf{\textit{0.14}} & 0.32 & 0.24 & 0.34 & 0.42 & 0.77 & 0.19 & 0.33 & 0.34 & 0.17 & 0.21 \\
\bottomrule
\end{tabular}
}
\end{adjustbox}
\end{table*}
\FloatBarrier

\begin{table*}[!ht]
\caption{Evaluation of the SZBVARx model’s performance relative to 14 baseline forecasters across all forecast horizons (12-month-ahead \& 24-month-ahead) for Japan (\textbf{\underline{best}} and \textbf{\textit{second-best}} results are highlighted).}
\label{tab:jap_perf_eval_G7}
\centering
\begin{adjustbox}{width=1\textwidth, height=4.5cm}
{\tiny
\renewcommand{\arraystretch}{0.85}
\begin{tabular}{cclccccccccccccccc}
\toprule
Variables & Horizons & Metrics & \textbf{VARx} & \textbf{TVAR} & \textbf{VES} & \textbf{NBEATSx} & \textbf{LGBMx} & \textbf{DTS\_CBx} & \textbf{DTS\_XGBx} & \textbf{NHITSx} & \textbf{TSMIXERx} & \textbf{BRNNx} & \textbf{XGBx} & \textbf{CatBoostx} & \textbf{TiDEx} & \textbf{TFTx} & \textbf{\textcolor{blue}{SZBVARx}} \\
\midrule
\multirow{10}{*}{Unemployment Rate} & 12M & RMSE & 0.28 & 0.30 & \textbf{\textit{0.14}} & 2.73 & 0.24 & 0.34 & \textbf{\underline{0.11}} & 2.76 & 0.65 & 1.43 & 0.41 & 0.38 & 3.19 & 0.67 & 0.39 \\
 &     & SMAPE & 0.09 & 0.10 & \textbf{\textit{0.04}} & 1.03 & 0.08 & 0.12 & \textbf{\underline{0.03}} & 0.53 & 0.22 & 0.43 & 0.13 & 0.12 & 0.70 & 0.23 & 0.13 \\
 &     & MDAPE & 7.55 & 8.59 & \textbf{\textit{3.37}} & 85.17 & 9.57 & 14.28 & \textbf{\underline{1.94}} & 75.05 & 24.12 & 53.68 & 17.63 & 13.31 & 114.46 & 8.44 & 11.98 \\
 &     & ${\rm TU}_1$ & 0.05 & 0.06 & \textbf{\textit{0.03}} & 0.44 & 0.04 & 0.06 & \textbf{\underline{0.02}} & 0.36 & 0.11 & 0.22 & 0.08 & 0.07 & 0.40 & 0.14 & 0.07 \\
 &     & MASE & 2.67 & 2.89 & \textbf{\textit{1.18}} & 25.93 & 2.44 & 3.48 & \textbf{\underline{0.87}} & 24.25 & 7.01 & 15.63 & 3.96 & 3.66 & 31.56 & 5.21 & 3.82 \\
\cline{3-18}
 & 24M & RMSE & 0.19 & \textbf{\textit{0.16}} & \textbf{\underline{0.15}} & 2.52 & 0.27 & 0.38 & 0.32 & 2.03 & 1.93 & 1.45 & 0.41 & 0.46 & 1.21 & 0.69 & 0.22 \\
 &     & SMAPE & 0.06 & \textbf{\textit{0.05}} & \textbf{\underline{0.05}} & 0.58 & 0.08 & 0.13 & 0.10 & 0.44 & 0.53 & 0.44 & 0.13 & 0.15 & 0.31 & 0.23 & 0.07 \\
\bottomrule
\end{tabular}
}
\end{adjustbox}
\end{table*}
\FloatBarrier

\begin{table*}[!ht]
\caption{Evaluation of the SZBVARx model’s performance relative to 14 baseline forecasters across all forecast horizons (12-month-ahead \& 24-month-ahead) for the UK (\textbf{\underline{best}} and \textbf{\textit{second-best}} results are highlighted).}
\label{tab:uk_perf_eval_G7}
\centering
\begin{adjustbox}{width=1\textwidth, height=4.5cm}
{\tiny
\renewcommand{\arraystretch}{0.85}
\begin{tabular}{cclccccccccccccccc}
\toprule
Variables & Horizons & Metrics & \textbf{VARx} & \textbf{TVAR} & \textbf{VES} & \textbf{NBEATSx} & \textbf{LGBMx} & \textbf{DTS\_CBx} & \textbf{DTS\_XGBx} & \textbf{NHITSx} & \textbf{TSMIXERx} & \textbf{BRNNx} & \textbf{XGBx} & \textbf{CatBoostx} & \textbf{TiDEx} & \textbf{TFTx} & \textbf{\textcolor{blue}{SZBVARx}} \\
\midrule
\multirow{10}{*}{Unemployment Rate} & 12M & RMSE & 0.68 & \textbf{\underline{0.22}} & 0.69 & 1.75 & 0.71 & 0.58 & \textbf{\textit{0.39}} & 3.89 & 0.69 & 1.72 & 0.63 & 0.42 & 1.93 & 3.04 & 0.47 \\
 &     & SMAPE & 0.17 & \textbf{\underline{0.05}} & 0.17 & 0.35 & 0.13 & 0.11 & \textbf{\textit{0.08}} & 0.63 & 0.13 & 0.34 & 0.11 & 0.09 & 0.31 & 0.47 & 0.11 \\
 &     & MDAPE & 16.33 & \textbf{\underline{4.77}} & 16.18 & 31.68 & 12.98 & 10.26 & \textbf{\textit{6.68}} & 94.56 & 12.62 & 40.92 & 9.53 & 7.64 & 34.38 & 75.49 & 10.97 \\
 &     & ${\rm TU}_1$ & 0.09 & \textbf{\underline{0.03}} & 0.09 & 0.19 & 0.08 & 0.07 & \textbf{\textit{0.05}} & 0.32 & 0.08 & 0.17 & 0.07 & 0.05 & 0.19 & 0.27 & 0.06 \\
 &     & MASE & 4.20 & \textbf{\underline{1.30}} & 4.27 & 9.86 & 3.81 & 3.13 & \textbf{\textit{2.09}} & 24.87 & 3.84 & 11.08 & 3.21 & 2.36 & 10.49 & 17.64 & 2.80 \\
\cline{3-18}
 & 24M & RMSE & 0.77 & 1.56 & \textbf{\textit{0.34}} & 3.14 & 1.42 & 1.49 & 1.48 & 2.94 & 2.37 & 1.88 & 1.82 & 1.72 & 2.06 & 2.77 & \textbf{\underline{0.22}} \\
 &     & SMAPE & 0.18 & 0.35 & \textbf{\textit{0.07}} & 0.49 & 0.29 & 0.32 & 0.26 & 0.53 & 0.45 & 0.38 & 0.35 & 0.35 & 0.34 & 0.47 & \textbf{\underline{0.05}} \\
\bottomrule
\end{tabular}
}
\end{adjustbox}
\end{table*}
\FloatBarrier

\begin{table*}[!ht]
\caption{Evaluation of the SZBVARx model’s performance relative to 14 baseline forecasters across all forecast horizons (12-month-ahead \& 24-month-ahead) for Italy (\textbf{\underline{best}} and \textbf{\textit{second-best}} results are highlighted).}
\label{tab:ita_perf_eval_G7}
\centering
\begin{adjustbox}{width=1\textwidth, height=4.5cm}
{\tiny
\renewcommand{\arraystretch}{0.85}
\begin{tabular}{cclccccccccccccccc}
\toprule
Variables & Horizons & Metrics & \textbf{VARx} & \textbf{TVAR} & \textbf{VES} & \textbf{NBEATSx} & \textbf{LGBMx} & \textbf{DTS\_CBx} & \textbf{DTS\_XGBx} & \textbf{NHITSx} & \textbf{TSMIXERx} & \textbf{BRNNx} & \textbf{XGBx} & \textbf{CatBoostx} & \textbf{TiDEx} & \textbf{TFTx} & \textbf{\textcolor{blue}{SZBVARx}} \\
\midrule
\multirow{10}{*}{Unemployment Rate} & 12M & RMSE & 1.31 & 2.45 & 1.52 & 2.11 & 1.94 & 1.63 & 1.50 & 3.38 & 1.98 & 2.16 & 1.33 & 1.42 & \textbf{\textit{1.27}} & 2.70 & \textbf{\underline{0.55}} \\
 &     & SMAPE & 0.15 & 0.26 & 0.18 & 0.22 & 0.22 & 0.19 & 0.16 & 0.36 & 0.28 & 0.25 & \textbf{\textit{0.15}} & 0.16 & 0.16 & 0.25 & \textbf{\underline{0.06}} \\
 &     & MDAPE & 16.21 & 28.31 & 19.53 & 15.60 & 23.55 & 20.81 & 17.02 & 40.61 & 21.48 & 28.57 & 15.09 & 17.51 & \textbf{\textit{12.77}} & 22.89 & \textbf{\underline{5.15}} \\
 &     & ${\rm TU}_1$ & \textbf{\textit{0.08}} & 0.14 & 0.09 & 0.14 & 0.11 & 0.10 & 0.09 & 0.18 & 0.15 & 0.13 & 0.08 & 0.09 & 0.09 & 0.16 & \textbf{\underline{0.04}} \\
 &     & MASE & 2.16 & 4.02 & 2.59 & 2.88 & 3.28 & 2.76 & 2.30 & 5.79 & 3.18 & 3.77 & 2.17 & 2.26 & \textbf{\textit{1.93}} & 3.88 & \textbf{\underline{0.82}} \\
\cline{3-18}
 & 24M & RMSE & 1.28 & 4.77 & \textbf{\textit{1.21}} & 2.00 & 1.86 & 2.54 & 2.75 & 1.77 & 2.88 & 1.95 & 3.51 & 2.71 & 4.30 & 2.98 & \textbf{\underline{0.96}} \\
 &     & SMAPE & 0.14 & 0.42 & \textbf{\textit{0.13}} & 0.19 & 0.20 & 0.27 & 0.27 & 0.19 & 0.30 & 0.22 & 0.32 & 0.28 & 0.41 & 0.27 & \textbf{\underline{0.11}} \\
\bottomrule
\end{tabular}
}
\end{adjustbox}
\end{table*}
\FloatBarrier

Taken together, our results document the enhanced forecast performance of SZBVARx in complex, shock-prone macroeconomic environments. SZBVARx outperforms the classic VARx and VES in the nonstationary regimes and, in particular, provides much lower forecast variance than the machine learning and deep learning approaches when the sample sizes are small or moderately large. In particular, CatBoostx and VARx appear to be sensible baselines for the semi-long-term forecast (exchange rate and interest rate series), but quickly deteriorate in volatile and persistent environments. The broad evidence suggests the benefits of Bayesian shrinkage, explicitly modeling uncertainty, and carefully calibrating the prior for multivariate macroeconomic forecasting. The model’s performance evaluation endorses SZBVARx as a principled framework that provides a robust, stable, and interpretable platform for policy analysis across different G7 regimes.

\subsection{Statistical significance of the results}\label{Sec_stat_signif}
As a further robustness check on the stability of our findings, we also conduct the multiple comparisons with the best (MCB) procedure \citep{lee2018proper}. The MCB is a non-parametric multiple testing procedure that simultaneously compares multiple algorithms and settings with regard to their predictive accuracy. In the MCB test, we rank all algorithms according to the RMSE metric across all five target macroeconomic variables, and two forecast horizons (``semi-long-term'' and ``long-term'') for all G7 economies. The MCB test is based on the Nemenyi post-hoc procedure that corrects for the increase in Type I error arising from the multiple simultaneous comparisons \citep{koning2005m3,svetunkov2023forecasting}. For each algorithm $i$ evaluated across $D$ datasets, the mean rank $\bar{R}_i$ is computed as:
\begin{equation}
\bar{R}_i = \frac{1}{D}\sum_{j=1}^{D} R_{ij}
\end{equation}
where $R_{ij}$ denotes the rank of algorithm $i$ on dataset $j$. The critical distance at significance level $\alpha$ follows the Tukey distribution:

\begin{equation}
CD_{\alpha} = q_{\alpha}\sqrt{\frac{\mathcal{A}(\mathcal{A}+1)}{6D}}
\end{equation}
\begin{figure}[!htbp]
 \centering
  \includegraphics[width=\textwidth]
  {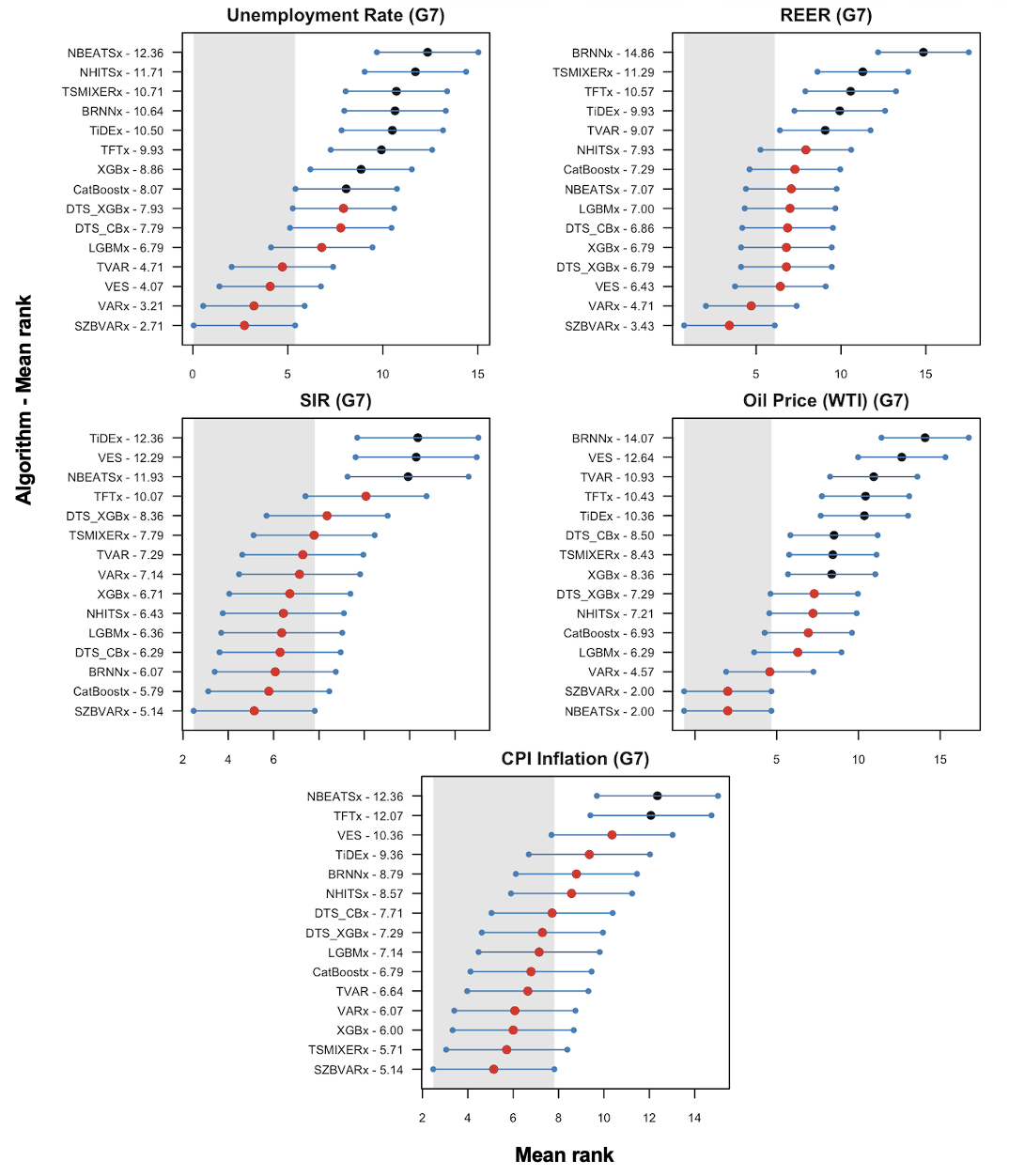}
   \caption{Multiple comparisons with the best (MCB) plots for G7 economies across 12-month (semi-long-term) ahead and 24-month (long-term) ahead forecasting horizons, ranking algorithms by RMSE metric. Key indicators: Unemployment Rate, REER, SIR, Oil Price (WTI), and CPI Inflation. Labels (e.g., SZBVARx–2.71) denote model identifiers and their mean ranks, where lower values indicate superior out-of-sample predictive accuracy.}
   \label{fig:MCB_all_G7}
\end{figure}

In this expression, $q_{\alpha}$ is the critical value from the Tukey distribution at level $\alpha$, $\mathcal{A}$ is the number of algorithms, and $D$ is the evaluation datasets. Two algorithms are statistically significantly different if $|R_i-R_j|> CD_{\alpha}$. As shown in Figure \ref{fig:MCB_all_G7}, the MCB analysis reveals a non-negligible degree of performance heterogeneity across the forecasting algorithms for the five macroeconomic indicators. SZBVARx consistently achieves the lowest mean ranks and its dominance in terms of RMSE performance over the full set of key variables is statistically confirmed. SZBVARx has mean ranks of 2.71, 3.43, 5.14, 2.00, and 5.14 for the unemployment rate, REER, SIR, oil price (WTI), and CPI inflation, respectively. The statistical significance is derived from the critical intervals resulting from the MCB procedure with the critical distance value of 5.34 (the upper part of the critical distance for the SZBVARx model, represented by the grey shaded regions in Figure~\ref{fig:MCB_all_G7}, is used as the reference value for the test and is compared to other competing algorithms), where a number of top methods such as NBEATSx, NHITSx, and TSMIXERx have mean ranks that are significantly higher with non-overlapping critical intervals. In the case of deep learning and transformer-based models, TiDEx and TFTx exhibit a moderate level of performance for most of the variables, whereas BRNNx is conspicuously placed at the bottom for REER and oil price (WTI), with mean ranks that are above 10. At the same time, there also exists a group of competitive alternatives that occupy a second-tier position across several variables. In particular, VARx often is on par with top-performing models, and CatBoostx, outperforms most of the machine learning alternatives. CatBoostx demonstrates competitive performance for specific macroeconomic variables such as SIR, where it is the second-best (5.79) method only after SZBVARx, thus supporting its potential for practical applicability to monetary policy-related time series. NBEATSx is also able to show promising performance for oil price forecasting, with a mean rank that is the same as SZBVARx (2.00). The MCB test also suggests that the results are robust across the two forecast horizons. The fact that this pattern of algorithmic performance is observed across variables with fundamentally different statistical characteristics and volatility patterns further reinforces the evidence in favor of SZBVARx and its ability to provide consistently accurate forecasts for a range of multivariate macroeconomic time series for the G7 economies. The most strategically relevant baseline algorithms for comparison with SZBVARx across G7 countries, according to the results of the MCB test, are VARx and CatBoostx. The former is able to demonstrate competitive performance for unemployment rate and REER forecasting, where it consistently achieves second place in these two indicators. The latter exhibits strong performance for SIR, where it ranks second after SZBVARx. This aspect can be seen as a sign of its specialized applicability to time series that are directly relevant for monetary policy. Thus, their combination as complementary baselines is an ideal way to assess the relative performance of SZBVARx across the full set of variables and forecasting horizons.

Expanding on the MCB test described above, we apply a Murphy diagram (MD) difference analysis to further stress-test the statistical significance of the forecasting performances across G7 nations. The MD difference approach considered in this section operationalizes the extremal score framework of \citet{ehm2016quantiles} for rigorous pairwise comparison of forecasting methods on a large class of consistent scoring functions. The MD difference approach specifies extremal scoring functions that evaluate forecast performance on account of threshold-dependent decision problems \citep{ziegel2017murphy}. For two competing forecasts from different forecasting methods with corresponding observations, where forecasts are point predictions, the extremal score at threshold parameter $\theta$ is defined as:
\begin{equation}
\mathcal{S}_\theta(x, y) = \begin{cases}
(\alpha - 1)(y - \theta) & \text{if } y \leq \theta < x \\
\alpha(y - \theta) & \text{if } x \leq \theta < y \\
0 & \text{otherwise}
\end{cases}
\label{eq:extremal_score}
\end{equation}
where $x$ represents the forecast value, $y$ represents the actual observed value at time $t$, $\alpha \in (0,1)$ is the expectile level parameter controlling asymmetric penalization (with $\alpha = 0.5$ providing symmetric scoring), and $\theta \in \mathbb{R}$ is the threshold parameter \citep{ehm2016quantiles}. This formulation implements the elementary scoring function for expectiles, representing a special case of the consistent scoring function family for expectile functionals as established in \citet{gneiting2011making}. Following the scoring convention, lower extremal scores indicate superior forecasting performance. The Murphy diagram difference function computes the performance differential between two forecasting methods across the spectrum of threshold values:
\begin{equation}
D(\theta) = \overline{\mathcal{S}}_{\theta}(\hat{y}_{\text{SZBVARx}}, y) - \overline{\mathcal{S}}_{\theta}(\hat{y}_{\text{baseline}}, y) = \frac{1}{N}\sum_{t=1}^{N}[\mathcal{S}_\theta(\hat{y}_{\text{SZBVARx},t}, y_t) - \mathcal{S}_\theta(\hat{y}_{\text{baseline},t}, y_t)]
\label{eq:murphy_diff}
\end{equation}
where $\overline{\mathcal{S}}_{\theta}(\hat{y}_j, y)$ represents the mean extremal score for forecasting method $j$ at threshold $\theta$, and $N$ is the total number of forecast-observation pairs. Since lower extremal scores indicate better performance, negative values of $D(\theta)$ indicate superior performance of SZBVARx, while positive values suggest the baseline method (VARx or CatBoost) performs better at the given threshold. To account for potential heteroskedasticity and autocorrelation in the score differences, we employ the Newey-West heteroskedasticity- and autocorrelation-consistent (HAC) variance estimator with Bartlett kernel and lag truncation parameter $\ell$ \citep{newey1987simple}:
\begin{equation}
\widehat{\operatorname{Var}}[D(\theta)] = \frac{1}{N}\Big(\hat{\gamma}_0 + 2\sum_{j=1}^{\ell} w_j \hat{\gamma}_j\Big)
\label{eq:hac_variance}
\end{equation}
where $w_j = 1 - \frac{j}{\ell+1}$ are Bartlett kernel weights that ensure positive semi-definiteness of the covariance matrix estimator, and $\hat{\gamma}_j$ are the sample autocovariances of the mean-demeaned score differences:
\begin{equation}
\hat{\gamma}_j = \frac{1}{N}\sum_{t=j+1}^{N} \big(d_t(\theta)-\bar{d}(\theta)\big)\,\big(d_{t-j}(\theta)-\bar{d}(\theta)\big)
\label{eq:autocovariance}
\end{equation}
where $d_t(\theta) = \mathcal{S}_\theta(\hat{y}_{\text{SZBVARx},t}, y_t) - \mathcal{S}_\theta(\hat{y}_{\text{baseline},t}, y_t)$ denotes the score difference at time $t$, and $\bar{d}(\theta) = \frac{1}{N}\sum_{t=1}^N d_t(\theta)$ is the sample mean of score differences \citep{andrews1991heteroskedasticity}. The lag truncation parameter is chosen as $\ell = \lfloor 4(N/100)^{2/9} \rfloor$ following the automatic bandwidth selection rule of \citet{newey1994automatic}. Pointwise confidence bands are constructed as:
\begin{equation}
D(\theta) \pm z_{\alpha/2} \sqrt{\widehat{\operatorname{Var}}[D(\theta)]}
\label{eq:confidence_bands}
\end{equation}
where $z_{\alpha/2}$ is the critical value from the standard normal distribution for the desired confidence level (90\% in our implementation). The threshold parameter $\theta$ orchestrates forecast evaluation by selectively emphasizing distinct dimensions of prediction error asymmetry \citep{holzmann2014expectile}. Lower $\theta$ values amplify sensitivity to systematic underestimation, while elevated values intensify penalties for consistent overforecasting. This parametric flexibility empowers Murphy difference diagrams to deliver distribution-agnostic performance benchmarking across competing forecasting architectures, revealing error distribution patterns and establishing dominance hierarchies under heterogeneous scoring regimes \citep{jordan2019evaluating}.

\begin{figure}[!htbp]
 \centering
  \includegraphics[width=\textwidth]
  {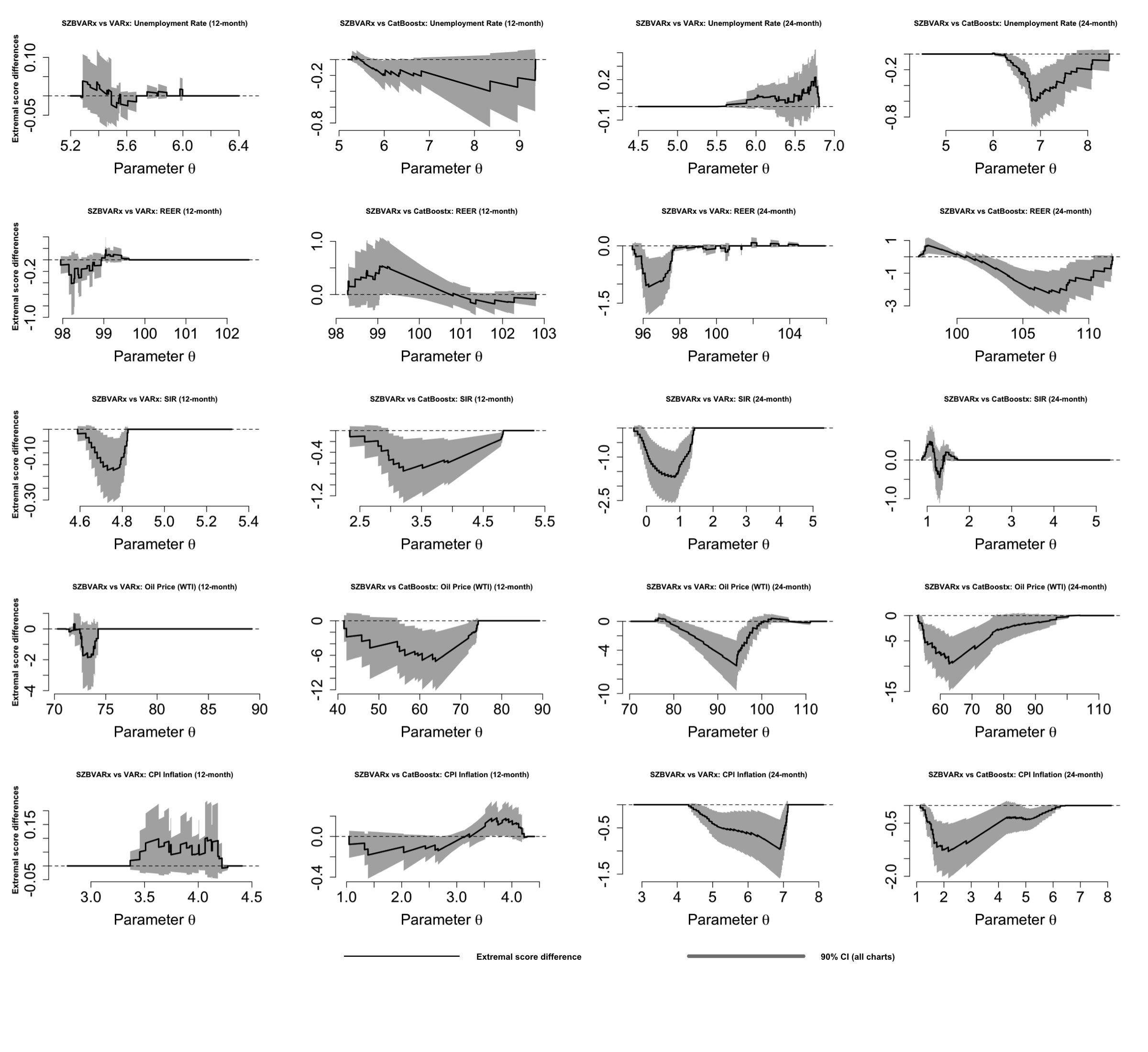}
   \caption{Murphy diagram difference plots comparing SZBVARx forecasting performance against VARx and CatBoostx baselines across 12-month and 24-month-ahead forecast horizons for Canada. Panels in the first and second columns correspond to the 12-month horizon, while panels in the third and fourth columns display the results for the 24-month horizon. The analysis covers five key macroeconomic indicators: Unemployment Rate, REER, SIR, Oil Price (WTI), and CPI Inflation. Each subplot displays extremal score differences with 90\% HAC-based confidence bands (gray shaded areas) across threshold parameter values. Negative differences indicate superior SZBVARx performance, while the magnitude reflects the strength of the performance advantage at each decision threshold.}
   \label{fig:MD_canada_12M_24M}
\end{figure}
This threshold-adaptive methodology complements our MCB analysis by revealing performance heterogeneity across diverse decision-theoretic contexts, providing a more nuanced understanding of forecasting dominance patterns beyond aggregate error metrics\footnote{Implementation uses the \texttt{murphydiagram} \textbf{R} package's \texttt{murphydiagram\_diff} function \citep{krueger2019murphydiagram} with HAC-based confidence bands and automatic lag truncation.}.

Figure \ref{fig:MD_canada_12M_24M} presents the Murphy difference diagram for Canada. For the unemployment rate, SZBVARx outperforms the alternatives in most of the threshold ranges between 5.2 and 6.4, while in certain decision criteria, VARx also becomes competitive. For REER, the positive difference areas are generally assigned to CatBoostx in threshold ranges between 98 and 100, while the negative differences dominate overall, suggesting the superiority of SZBVARx. The oil price also displays negative differences almost everywhere, although there are certain decision criteria where the challenger models prevail. In the case of CPI inflation, the positive difference areas are mainly related to VARx ($\theta \approx 3.4-4.3$) and CatBoostx ($\theta \approx 3.5-4.1$) at the 12-month horizon, which may be interpreted as the superiority of the baseline models at those threshold values. At 24 months, the differences are negative for all decision criteria, which in turn implies the dominance of SZBVARx. In summary, there is no methodology that is clearly superior to the others for all of the considered threshold parameters and forecast horizons. However, each method displays its excellence in certain decision regimes. This complementarity across approaches and decision criteria for Canada paves the way to further our analysis with the other G7 countries.

A more nuanced view of the competitive relationships emerges when looking at the remaining G7 plots available in Appendix~\ref{app:Appendix_stat_significance} (Figures \ref{fig:MD_usa} through \ref{fig:MD_italy}). These plots verify the general supremacy of SZBVARx but also feature zones where either the standard VARx or CatBoostx forecasts dominate across different threshold intervals. This result indicates that forecasting dominance is not unidirectional and that different methodologies present their own advantages depending on the economic variable and the decision-theoretic framework in focus. When considering the competitive landscape across countries, there are significant variations. In the United States, SZBVARx shows dominance in the unemployment indicator ($\theta \approx 3.5-7$) and in oil price, while baseline models are more competitive in CPI inflation decision regimes ($\theta \approx 3.5-5.0$ in the case of 12-month horizon, $\theta \approx 5.5-8.0$ in the case of 24-month horizon). The case of France presents a more marked supremacy of SZBVARx, mainly in unemployment ($\theta \approx 6.8-9.0$) and SIR ($\theta \approx 2.8-4.0$). The traditional models remain competitive in some specific regimes for REER ($\theta \approx 96.0-97.0$). The German case can be considered the one that most strongly confirms the superiority of SZBVARx, with negative differences across the whole spectrum of selected variables and decision thresholds of interest: unemployment rate ($\theta \approx 2.8-3.2$), REER ($\theta \approx 98.0-99.0$), SIR ($\theta \approx 3.3-4.0$), and CPI inflation ($\theta \approx 6.0-8.0$). In contrast, Japan shows the most even competitive situation, with the baseline approaches showing superiority in unemployment rate ($\theta \approx 2.7-2.9$) and in a few CPI inflation regimes ($\theta \approx 3.4-4.0$). SZBVARx models dominate the forecasting task in REER ($\theta \approx 75-85$, $\theta \approx 90-105$). The case of the United Kingdom illustrates well the complementarity of the methodological approaches. On the one hand, SZBVARx presents a clear superiority in the REER forecasting task ($\theta \approx 95-105$). On the other hand, the traditional approaches are more competitive in oil price (WTI) and in CPI inflation ($\theta \approx 70-85$, $\theta \approx 4.5-10$). The Italian situation is similar to the one just described, with baseline superiority in SIR ($\theta \approx 0-1$) and CPI inflation ($\theta \approx 2-8$), and SZBVARx advantage in unemployment rate and REER.

Augmenting the detailed findings of the Murphy difference analysis, we complement our evaluation of forecast performance differences with statistical inference on the (multivariate) Diebold-Mariano (DM) test statistic of \cite{mariano2012statistical}\footnote{The multivariate Diebold-Mariano test is implemented using the \texttt{multDM} \textbf{R} package by Drachal (2025). The R document can be found here: \url{https://cran.r-project.org/web/packages/multDM/multDM.pdf.}}. The DM test is a Wald-type test that generalizes the original one model-versus-one to testing for equal predictive accuracy of two or more non-nested models over a set of target series. The multivariate framework allows accounting for serial correlation and heteroskedasticity of loss differentials while jointly testing the null hypothesis that the competing models are all equally accurate against the alternative hypothesis that they are not. The resulting finite-sample corrected test statistic $S_c$ has an asymptotic chi-squared ($\chi^2$) distribution, and the finite-sample correction is shown to improve size control in moderate samples. The test is invariant to the ordering of the models, and the testing procedure can be flexibly applied using various loss functions. We follow \cite{mariano2012statistical} and use absolute scaled error loss, block length $q=1$ (standard choice), and finite-sample correction in our implementation. Table~\ref{tab:multDM_forecast_performance_g7} reports the p-values from pairwise comparison of SZBVARx versus VARx and CatBoostx across the G7 economies and 12-month and 24-month forecast horizons. Results that are statistically significant at 10\% (reported in bold) indicate that, overall, SZBVARx exhibits superior predictive accuracy, especially at longer horizons and relative to CatBoostx. Notably, for Canada, the United States, France, Germany, Japan and the United Kingdom at 24-month horizon, the SZBVARx outperforms both baselines and exhibits several cases of significance at 12-month horizon. VARx and CatBoostx occasionally show parity or local advantage, but the multivariate DM test results largely confirm the relative superiority and robustness of the SZBVARx approach for multivariate macroeconomic forecasting in the presence of exogenous uncertainty shocks. 

To augment the multivariate Diebold-Mariano test results further, we also employ the unconditional multivariate Giacomini-White (GW) test \citep{borup2024predicting}. The GW test framework has three main advantages: (i) HAC covariance estimation with Parzen kernel to account for heteroskedasticity and serial correlation in forecast errors \footnote{Multivariate GW test has been implemented using \texttt{feval} Python package (\url{https://github.com/ogrnz/feval}).}, (ii) an explicit accounting for estimation uncertainty in nested model comparisons relevant for rolling window evaluation exercises, and (iii) Wald-type of the loss differential vector, allowing for joint hypothesis testing and more powerful statistical inference \citep{giacomini2006tests, mariano2012statistical, hansen2011model}. The GW test evaluates $H_0: \mathbb{E}[\Delta \mathcal{L}_{t+1}] = 0$ (``unconditional case" since no conditioning variables used) using a statistic $\mathcal{S} = T \bar{d}' \hat{\Sigma}^{-1} \bar{d}$, where $\bar{d}$ is the mean loss differential vector, $T$ is the number of forecast observations and $\hat{\Sigma}$ is the HAC covariance matrix. Rejection of the null hypothesis suggests that at least one of the models is superior in predictive performance.

Table~\ref{tab:mult_GW_g7} in Appendix~\ref{app:Appendix_stat_significance} reports p-values from the unconditional multivariate GW test for SZBVARx versus VARx and CatBoostx, spanning G7 economies, five macroeconomic indicators, and both 12- and 24-month forecast horizons. The results provide compelling evidence of SZBVARx’s forecasting edge, with 101 out of 140 tests (72.14\%) significant at the 10\% level. Significance rates are uniformly high across countries—ranging from 65\% (Canada), 70\% (France, United States) to 75\% (United Kingdom, Germany, Italy, Japan)—and across variables, peaking at 78.57\% for both unemployment rate and SIR, and remaining robust for REER (75.00\%), oil price (WTI) (71.43\%), and CPI inflation (57.14\%). Horizon-specific results further underscore SZBVARx’s dominance at longer horizons: the model outperforms CatBoostx in 80.00\% and VARx in 77.14\% of 24-month tests, compared to 65.71\% (for both VARx and CatBoostx) at 12-month horizon, respectively. Notably, robust significance is observed for CPI inflation and SIR forecasts in Canada, the United States, France, Japan, and the United Kingdom at 24-month horizon, with several significant results at 12-month horizon as well. Although our implementation is unconditional, the remarkable consistency and breadth of significant results across countries, variables, and horizons provides robust statistical evidence for the forecasting supremacy of SZBVARx over both statistical and machine learning benchmarks, especially at extended horizons and under macroeconomic uncertainty.
\begin{table*}[!ht]
\centering
\caption{Forecasting performance comparison using the Multivariate Diebold-Mariano Test for SZBVARx, VARx, and CatBoostx across G7 countries at 12-month-ahead and 24-month-ahead forecast horizons. \textbf{Bold with *} values denote significance at the 10\% level}
\label{tab:multDM_forecast_performance_g7}
\begin{adjustbox}{width=0.7\textwidth}
\begin{tabular}{lccc} \hline
\toprule
\textbf{Country} & 
\begin{tabular}{@{}c@{}}\textbf{Forecast} \\ \textbf{Horizon}\end{tabular} & 
\begin{tabular}{@{}c@{}}\textbf{SZBVARx Vs VARx} \\ \textbf{(p-value)}\end{tabular} & 
\begin{tabular}{@{}c@{}}\textbf{SZBVARx Vs CatBoostx} \\ \textbf{(p-value)}\end{tabular} \\
\midrule 

Canada  & 12 & 0.446 & \textbf{0.015*} \\
        & 24 & \textbf{0.002*} & \textbf{0.000*} \\
\hline
US      & 12 & \textbf{0.073*} & \textbf{0.010*} \\
        & 24 & \textbf{0.010*} & \textbf{0.000*} \\
\hline
France  & 12 & 0.464 & \textbf{0.002*} \\
        & 24 & \textbf{0.019*} & \textbf{0.000*} \\
\hline
Germany & 12 & 0.114 & \textbf{0.068*} \\
        & 24 & \textbf{0.005*} & \textbf{0.030*} \\
\hline
Japan   & 12 & \textbf{0.031*} & \textbf{0.002*} \\
        & 24 & \textbf{0.044*} & \textbf{0.000*} \\
\hline
UK      & 12 & \textbf{0.004*} & 0.686 \\
        & 24 & \textbf{0.011*} & \textbf{0.073*} \\
\hline
Italy   & 12 & 0.123 & 0.606 \\
        & 24 & \textbf{0.001*} & 0.378 \\
\bottomrule
\end{tabular}
\end{adjustbox}
\end{table*}
\FloatBarrier

\subsection{Uncertainty Quantification through Predictive Credible Intervals}\label{CPPI_coverage_G7}
In addition to point forecasts, characterizing the predictive uncertainty is also of great interest for economic decision-making and risk management. Bayesian forecasting yields the \emph{predictive} distribution for future observations \emph{conditional on posterior-mean parameters}. We report credible prediction intervals that reflect \emph{future shock uncertainty} conditional on posterior-mean parameters $(\bar{\mu},\bar{\Phi}_1,\ldots,\bar{\Phi}_p,\bar{\Gamma},\bar{\Sigma})$ after hyperparameter tuning \citep{koop2010bayesian, geweke2006bayesian} from Section~\ref{szbvarx_forecast}\footnote{This ensures intervals are centered on optimal point forecasts from hyperparameter tuning rather than re-simulating from different specifications. The snap-centering adjustment aligns the simulated distribution with the deterministic forecast path while preserving stochastic uncertainty. Intervals are constructed component-wise (marginal); joint credible regions are not reported.}. A $\gamma$-level credible interval $[L_{T+h}, U_{T+h}]$ satisfies $\mathbb{P}(L_{T+h} \leq y_{T+h} \leq U_{T+h} \mid y_{1:T}) = \gamma$. We simulate future shocks for each draw $s$ and horizon $h = 1, \ldots, H$ as:\footnote{Here, $s = 1, \ldots, S$ indexes Monte Carlo shock draws, with parameters fixed at posterior means.}
\begin{equation}\label{eq:forecast_shock_sim}
\varepsilon_{T+h}^{(s)} \sim \mathcal{N}_m(0, \bar{\Sigma}),
\end{equation}
and compute the forecast trajectory stepwise:
\begin{equation}\label{eq:forecast_trajectory}
\tilde{y}_{T+h \mid y_{T}}^{(s)} = \bar{\mu} + \sum_{i=1}^{p} \bar{\Phi}_i \tilde{y}_{T+h-i \mid y_{T}}^{(s)} + \bar{\Gamma} x_{T+h} + \varepsilon_{T+h}^{(s)},
\end{equation}
initialized with $\tilde{y}_{T+1-j \mid y_{T}}^{(s)} = y_{T+1-j}$ for $j = 1, \ldots, p$.\footnote{The exogenous path uses the last $H$ training observations: $x_{T+h} = x_{T-H+h}$ for $h = 1, \ldots, H$, ensuring no data leakage.} To ensure forecast consistency, we apply snap-centering: $\delta_h = \hat{y}_{T+h \mid y_{T}}^{\text{PF}} - \hat{y}_{T+h \mid y_{T}}$, where $\hat{y}_{T+h \mid y_{T}}$ is the deterministic forecast (with $\varepsilon_{T+h} \equiv 0$) and $\hat{y}_{T+h \mid y_{T}}^{\text{PF}}$ is the tuned point forecast,\footnote{The adjustment $\delta_h$ is a vector of length $m$.} yielding:
\begin{equation}\label{eq:centered_trajectory}
\hat{y}_{T+h \mid y_{T}}^{(s)} = \tilde{y}_{T+h \mid y_{T}}^{(s)} + \delta_h.
\end{equation} 
We impose admissibility constraints by truncating draws to economically meaningful support bounds $[a_\ell, b_\ell]$ for each variable $\ell$, where $a_\ell$ and $b_\ell$ are determined by the natural constraints of each macroeconomic variable.\footnote{E.g., unemployment rate $\in [0, 100]$, real exchange rate $\in [0, \infty)$. Acceptance rate $\rho_{\ell,h}$ measures the proportion of admissible draws; low rates indicate potential misspecification.} Let $\mathcal{S}_{\ell,h} = \{s : a_\ell \leq \hat{y}_{\ell,T+h \mid y_{T}}^{(s)} \leq b_\ell\}$ denote admissible draws. The $\gamma$-level credible interval is the shortest interval containing the point forecast with at least $\gamma$ probability mass:\footnote{When the point forecast lies outside the truncated empirical support, we anchor at the point forecast and extend to the nearest $\gamma$-mass window, ensuring $\hat{y}_{\ell,T+h \mid y_{T}}^{\text{PF}} \in [L_{\ell,T+h}, U_{\ell,T+h}]$.}
\begin{equation}\label{eq:credible_interval}
[L_{\ell,T+h}, U_{\ell,T+h}] = \underset{[L, U]: \hat{y}_{\ell,T+h \mid y_{T}}^{\text{PF}} \in [L,U]}{\arg\min} \left\{ U - L : \frac{1}{|\mathcal{S}_{\ell,h}|} \sum_{s \in \mathcal{S}_{\ell,h}} \mathbb{1}\{L \leq \hat{y}_{\ell,T+h \mid y_{T}}^{(s)} \leq U\} \geq \gamma \right\}.
\end{equation}
When truncation binds, the effective mass covered is $\gamma_{\text{eff}} = \gamma \cdot \rho_{\ell,h}$. These PF-anchored shortest-mass (Highest Posterior Density-style) intervals guarantee $\hat{y}^{\text{PF}}_{\ell,T+h}\in[L_{\ell,T+h},U_{\ell,T+h}]$ and are appropriate under truncation-induced skewness. We use $S = 1000$ draws and $\gamma = 0.50$.

\begin{figure}[!htbp]
\centering
\includegraphics[width=\textwidth,height=0.9\textheight,keepaspectratio]
{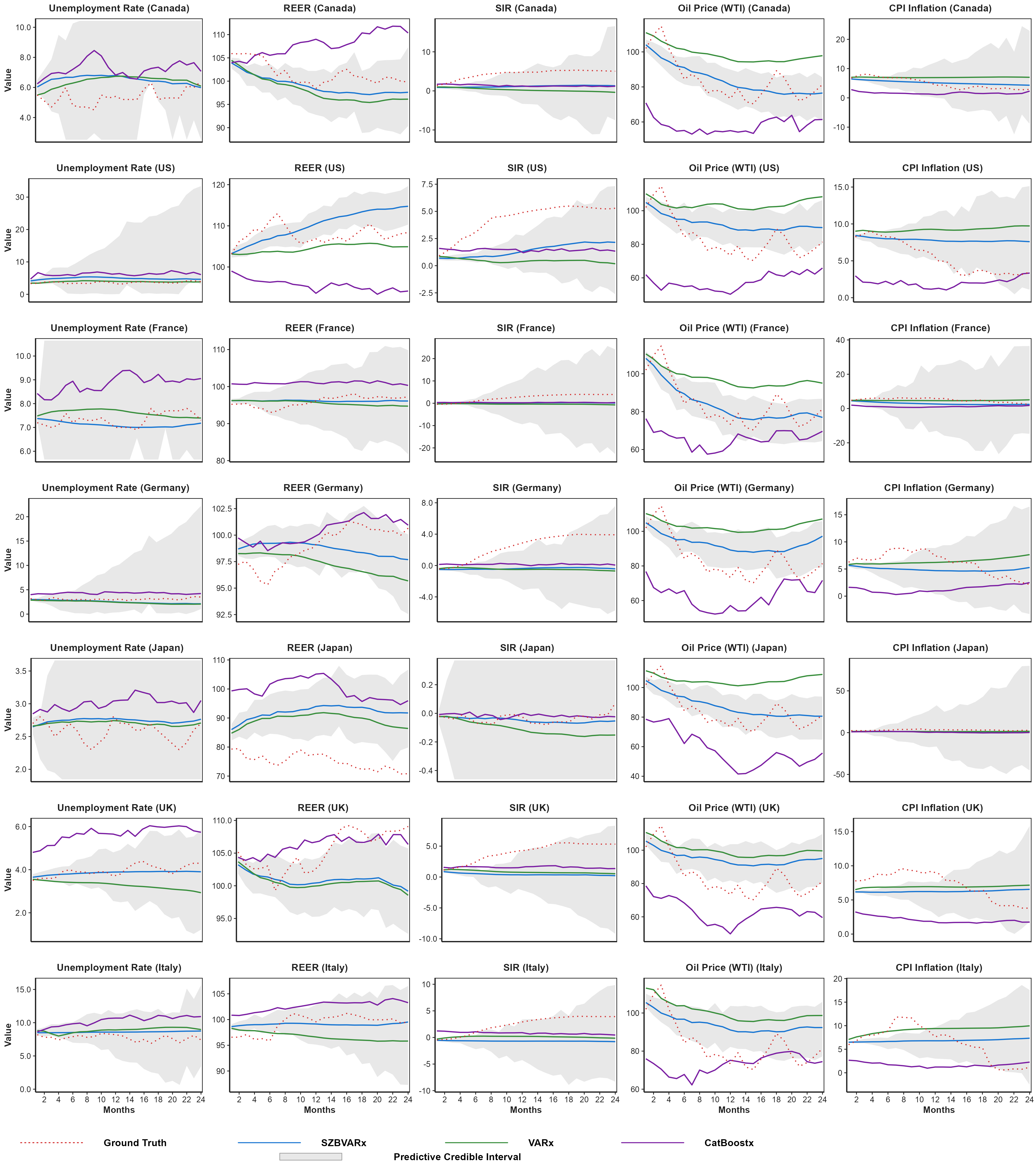}
\caption{Visualization of 24-month-ahead forecasts for five macroeconomic indicators across G7 countries. Each panel displays the ground truth (red dotted line), point forecasts from SZBVARx (blue), VARx (green), and CatBoostx (purple) models, along with predictive credible intervals from SZBVARx (grey shaded regions) quantifying forecast uncertainty. Variables shown are: Unemployment Rate, REER, SIR, Oil Price (WTI), and CPI Inflation Rate. Rows represent countries (Canada, US, France, Germany, Japan, UK, Italy from top to bottom), while columns represent the five macroeconomic variables.}
\label{fig:PPI_24M_G7}
\end{figure}
\begin{figure}[!htbp]
\centering
\includegraphics[width=\textwidth]{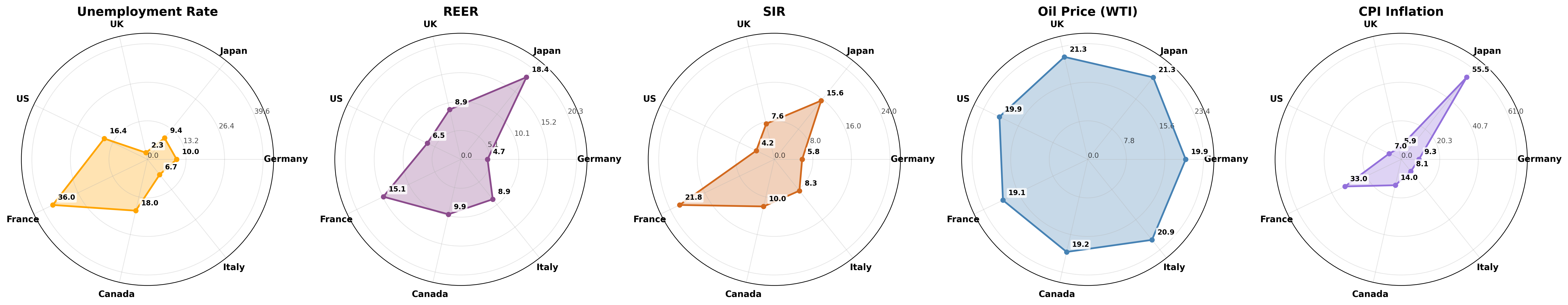}
\caption{Visualization of average predictive credible interval widths for 24-month-ahead forecasts across five G7 macroeconomic indicators using SZBVARx model. Each subplot displays interval width distributions across G7 economies for: Unemployment Rate, REER, SIR, Oil Price (WTI), and CPI Inflation. Lower interval widths indicate superior prediction precision and enhanced forecast reliability.}
\label{fig:PPI_avg_interval_width_24M_G7}
\end{figure}
Figure~\ref{fig:PPI_24M_G7} reports the out-of-sample forecast performance of 24-month-ahead forecast of all G7 countries. We observe that the heterogeneity across the countries and across the variables in terms of the predictive uncertainty is remarkable. The SZBVARx model outperforms the VARx and CatBoostx baselines in comparison to the ground truth. We find that the credible intervals include the realized values in a majority of cases across countries and variables, providing empirical evidence that the uncertainty quantification procedure described in Equations~\eqref{eq:forecast_shock_sim}--\eqref{eq:credible_interval} yields well-calibrated predictive distributions for the 24-month forecast horizon. It is worth noting that in all countries and for all variables, the credible intervals are small in the first forecast horizons and they widen at longer horizons. This occurs because forecast uncertainty accumulates over longer horizons as shock uncertainty compounds over time in Bayesian predictive intervals.
We see relatively narrow intervals for unemployment rate forecasts in the United Kingdom and Italy, moderate uncertainty bands in Japan and Germany, and substantially wider uncertainty in Canada, France, and the United States. For REER predictions, we observe the tightest intervals in Germany and the United States, followed by the United Kingdom, France, and Italy, while Canada and Japan feature broader uncertainty. Tight intervals are also observed in the United States and Germany for SIR forecasts, while the uncertainty is moderate in the United Kingdom, Italy, and Japan, and fairly wide in France and Canada, in line with elevated monetary policy uncertainty in those countries. In contrast, the intervals for oil price predictions are consistently wide but relatively homogeneous across all G7 countries, highlighting the high volatility and geopolitical risk sensitivity in energy markets. CPI inflation forecasts stand out with the most cross-country heterogeneity: the United Kingdom, the United States, and Germany maintain relatively narrow intervals, while Italy and Canada exhibit moderate to wide uncertainty, and France and Japan display exceptionally wide intervals, reflecting high inflation volatility and policy uncertainty in the forecast period. 
The 12-month ahead forecast results, presented in Appendix~\ref{app:Appendix_PPI_12M} and Figure~\ref{fig:PPI_12M_G7}, exhibit qualitatively similar patterns with narrower credible intervals overall, reflecting reduced forecast uncertainty at shorter horizons.

To better understand the cross-country heterogeneity in the forecast uncertainty, Figure~\ref{fig:PPI_avg_interval_width_24M_G7} plots the average width of the 50\% predictive credible intervals across the five macroeconomic variables in each of the G7 countries\footnote{We report the average width of 50\% predictive credible intervals to provide a standardized measure of forecast uncertainty that is directly comparable across countries and variables. Mathematically, the interval width $W_{0.50} = Q_{0.75} - Q_{0.25}$ captures the interquartile range of the predictive distribution, where $Q_p$ denotes the $p$-th quantile. This choice is consistent with $\gamma = 0.50$ used in constructing the predictive credible intervals and provides a robust measure of central dispersion that is less sensitive to extreme tail behavior than wider intervals, making it particularly suitable for cross-country comparisons.}. The radar chart shows that the unemployment rate intervals have the smallest range in the United Kingdom (2.3) and the widest range in France (36.0); the REER intervals are the narrowest in Germany (4.7) and the widest in Japan (18.4); the SIR intervals range from the smallest in the United States (4.2) to the largest in France (21.8) reflecting the cross-country heterogeneity in the monetary policy uncertainty; the oil price intervals are the most homogeneous across the G7 countries (19--21) with the narrowest range in Germany; and the CPI inflation intervals are the most dispersed ranging from the United Kingdom (5.9) to Japan (55.5), with France (33.0) and Japan exhibiting abnormally high inflation uncertainty during the forecast horizon. These point-wise summary measures are in line with the visual patterns in Figure~\ref{fig:PPI_24M_G7} and highlight the importance of quantifying the forecast uncertainty on both the country and the variable dimensions.

\section{Policy Implications}\label{Sec_Policy_Implications}
The empirical results of this paper point to several policy-relevant facts. First, uncertainty shocks are economically meaningful and transmit through identifiable channels. Geopolitical risk and equity-market volatility primarily operate via the external sector—appreciations (depreciations) of the REER and movements in oil prices; while economic policy uncertainty and monetary policy uncertainty affect the SIR, inflation, and unemployment. Second, transmission is state dependent: impulse responses are typically larger and faster in high-rate (tight) regimes and more persistent but diminished in low-rate (easy) regimes, indicating an interaction between uncertainty, financial frictions, and available policy space \citep{caldara2016macroeconomic,lhuissier2021regime}. Third, the proposed SZBVARx delivers superior accuracy at 12–24 months, consistent with the advantages of Bayesian shrinkage and stochastic-volatility priors in high-dimensional settings \citep{banbura2010large,carriero2019large}. 
Finally, the Bayesian predictive credible intervals demonstrate reasonable empirical coverage and appropriately scale with forecast horizon, providing credible uncertainty quantification that is essential for policy communication, risk assessment, and scenario planning under macroeconomic uncertainty \citep{giannone2015prior,clark2011real,knuppel2015evaluating}.

These results have several policy implications. For monetary policy, the decision framework should condition explicitly on the rate regime and contemporaneous uncertainty indicators. As wavelet coherence indicates that the four uncertainty series are generally leading the macro variables, and the macro block is endogenous, it is reasonable to consider economic policy uncertainty, geopolitical risk, equity-market volatility, and monetary policy uncertainty as exogenous shocks for the purpose of scenario design. In practice, this means that central banks should generate regime-conditioned (high-rate vs. low-rate) parallel projections and evaluate the degree to which unemployment and inflation paths start to materially diverge; where divergence is sizeable, they should lean against amplification in tight conditions and tolerate slower real-side adjustment, recognizing asymmetric interactions between uncertainty and financial conditions \citep{caldara2016macroeconomic,lhuissier2021regime}. Given its medium-horizon advantage, the SZBVARx is also well suited to forward guidance, where coherence across variables over 1–2 years matters for credibility \citep{bernanke2020new,yellen2017inflation}.

For exchange rate and external sector management, the high prominence of the external channel suggests that geopolitical risk and financial-market volatility may push risk premia and capital flows in ways that induce safe-haven appreciations for some currencies and depreciation for others, with first-round effects on competitiveness and imported inflation. In particular, small, open, and energy-intensive economies should accompany monetary policy moves with Forex liquidity backstops, reserve management, and transparent communication about how uncertainty shocks translate into exchange-rate movements and the inflation target \citep{kilian2009not,caldara2022measuring}. In the euro area, cross-country heterogeneity in exchange-rate sensitivity further argues for the provision of national macro-stabilization tools to accompany area-wide policy in case of common shocks with asymmetric external impacts.

For energy and commodity risk, the forecast evaluation results indicate that oil-price projections exhibit the widest prediction intervals—particularly during global stress periods—suggesting that fiscal and regulatory authorities should incorporate energy-price stress tests into budget planning and sectoral oversight while supporting hedging strategies that reduce exposure to abrupt commodity swings \citep{kilian2009not}. Where energy import dependence is high, pre-specified contingencies (such as temporary energy-levy adjustments or targeted transfers) can be activated when prediction intervals for oil-price pass-through indicate elevated near-term inflation risk.

For fiscal policy, uncertainty shocks in low-rate regimes tend to raise unemployment while dampening inflation, strengthening the case for automatic stabilizers that are most active when monetary space is constrained \citep{auerbach2012measuring}. Predictive credible intervals provide a basis for sizing fiscal buffers ex ante: wider unemployment or inflation intervals warrant larger precautionary space and more conservative debt-sustainability margins. The model indicates particularly wide prediction intervals for CPI inflation or short-term interest rates; thus, budget authorities should broaden the revenue and expenditure scenario ranges in their frameworks.

For macroprudential policy, equity-market volatility and monetary policy uncertainty propagate via financial conditions and rate expectations. The width and skew of credible intervals around interest rate and exchange rate forecasts can therefore guide countercyclical capital buffer calibration and liquidity coverage targets. When the model signals widening short rate and real effective exchange rate uncertainty under tight regimes, earlier activation and slower release of buffers can mitigate credit-channel amplification; when regimes are stable and credible intervals are narrow, more targeted borrower-based measures may be preferred to broad capital tightening \citep{carstens2021central}.

Finally, for forecast governance and communication, institutions should leverage the point-forecast gains of SZBVARx. Publishing fan charts based on predictive credible intervals can shift attention from single central paths to risk-weighted ranges, consistent with best practice in inflation-report communication and with the need to avoid overconfidence from parametric intervals that may be misspecified \citep{britton1998inflation}. A practical workflow is to (i) nowcast and monitor the exogenous uncertainty block with thresholds that trigger scenario refreshes; (ii) run and disclose parallel regime-conditioned projections, documenting policy choices against the implied amplification-versus-persistence trade-off; and (iii) communicate credible intervals to anchor expectations and guide contingency planning. While judgment is indispensable in rare regime shifts or institutional changes, the architecture developed here offers a practical and evidence-based template for resilient policy under uncertainty in the G7.

\section{Conclusion and Discussion}\label{Section_Conclusion}
This paper develops and evaluates an uncertainty-resilient Bayesian VAR framework with exogenous drivers (SZBVARx) for multivariate macroeconomic forecasting in the G7. The framework integrates a compact block of external uncertainty indicators - EPU, GPR, USEMV, and USMPU - with five core macroeconomic variables: unemployment rate, REER, SIR, oil price (WTI), and CPI inflation. The central premise is that policy and geopolitical uncertainty, financial-market volatility, and monetary policy uncertainty are systematic determinants of macroeconomic outcomes. Embedding these drivers directly in the forecasting system improves point predictions and clarifies the mechanisms through which external risks propagate across G7 economies.

Two features set the proposed method apart on a methodological level. First, non-linear local projections \citep{jorda2005estimation} are used to estimate impulse responses, which are state-dependent and path-dependent, in contrast to tightly specified structural VARs that require strong parametric restrictions.
Second, Bayesian point forecasts are combined with credible intervals derived from the posterior predictive distribution to quantify forecast uncertainty across horizons \citep{giannone2015prior,clark2011real}. In large-scale benchmarks against 14 different methods, from classical time series to machine learning and deep learning, at multiple horizons, the SZBVARx systematically improves accuracy across a range of metrics. A suite of statistical tests confirms that improvements are economically meaningful and not driven by any single metric. Taken together, these results provide a new standard for macroeconomic forecasting in advanced economies that explicitly models uncertainty and regime dependence. Although the previous section already provides a detailed discussion of policy implications, it is worth emphasizing the immediate relevance of the framework for policymakers. The SZBVARx produces coherent medium-term projections across variables, transparent regime-conditioned impulse responses that make the amplification–persistence trade-off explicit, and Bayesian credible intervals suitable for fan-chart communication and contingency planning. In an environment characterized by recurrent global shocks and evolving monetary regimes, these features constitute a practical decision-support toolkit and a sound methodological foundation for future, uncertainty-aware forecasting.

The findings in this paper open up several avenues for interesting extensions. Newspaper-based measures of uncertainty are noisy and cross-country comparable. Market-implied indicators, sentiment-based gauges, or measures of supply-chain and climate-policy risk, in particular, could be used to strengthen the robustness of our findings. Regime classification is based on the short-term policy rate. Composite state variables that aggregate information from the policy rate and financial-conditions indices or term-premium measures may improve classification. The parsimonious forecasting block could be expanded to more sectoral disaggregates (e.g., tradables vs. non-tradables inflation; vacancy–unemployment dynamics) to better isolate policy-relevant margins. On the uncertainty quantification side, while this paper uses Bayesian predictive credible intervals obtained from the posterior predictive distribution, an important avenue for future research is the use of multivariate conformal prediction intervals \citep{vovk2005algorithmic,romano2019conformalized}. This would be a particularly promising direction for macroeconomic forecasting, where model uncertainty is pervasive and coverage validity is critical for policy communication. A fully real-time assessment, in which the mapping from forecasts to nowcasts would account for publication lags and data revisions, would make the framework even more operationally amenable. We finally point out that the approach could be easily extended to emerging markets and developing economies, many of which are affected by different transmission channels, subject to higher volatility, and with typically less policy space to maneuver than G7 economies. We leave the exploration of the above directions for future work.


\section*{Competing interest}
There are no competing interests to be declared.
\section*{Data and Code Availability Statement}
The macroeconomic time series data for the G7 countries analyzed in this study were sourced from the Federal Reserve Economic Data (FRED) repository (\url{https://fred.stlouisfed.org/}). Measures of economic policy uncertainty, geopolitical risk, and related uncertainty indices were obtained from the Policy Uncertainty database (\url{https://www.policyuncertainty.com/}). The code and compiled datasets used to reproduce the results of this study are available at: \url{https://github.com/ctanujit/Macrocasting}. 


\FloatBarrier

\bibliography{Bibliography}

\newpage

\appendix

\renewcommand{\thesection}{A.\arabic{section}}
\counterwithin{figure}{section}
\counterwithin{table}{section}
\renewcommand{\thefigure}{A.\arabic{section}.\arabic{figure}}
\renewcommand{\thetable}{A.\arabic{section}.\arabic{table}}

\setcounter{section}{0}
\setcounter{figure}{0}
\setcounter{table}{0}

\section{Appendix}\label{app:Appendix}
\subsection{Trend Plots}\label{app:Appendix_trend_plots_G7}
\begin{figure}[!htbp]
\centering
\includegraphics[width=\textwidth]
{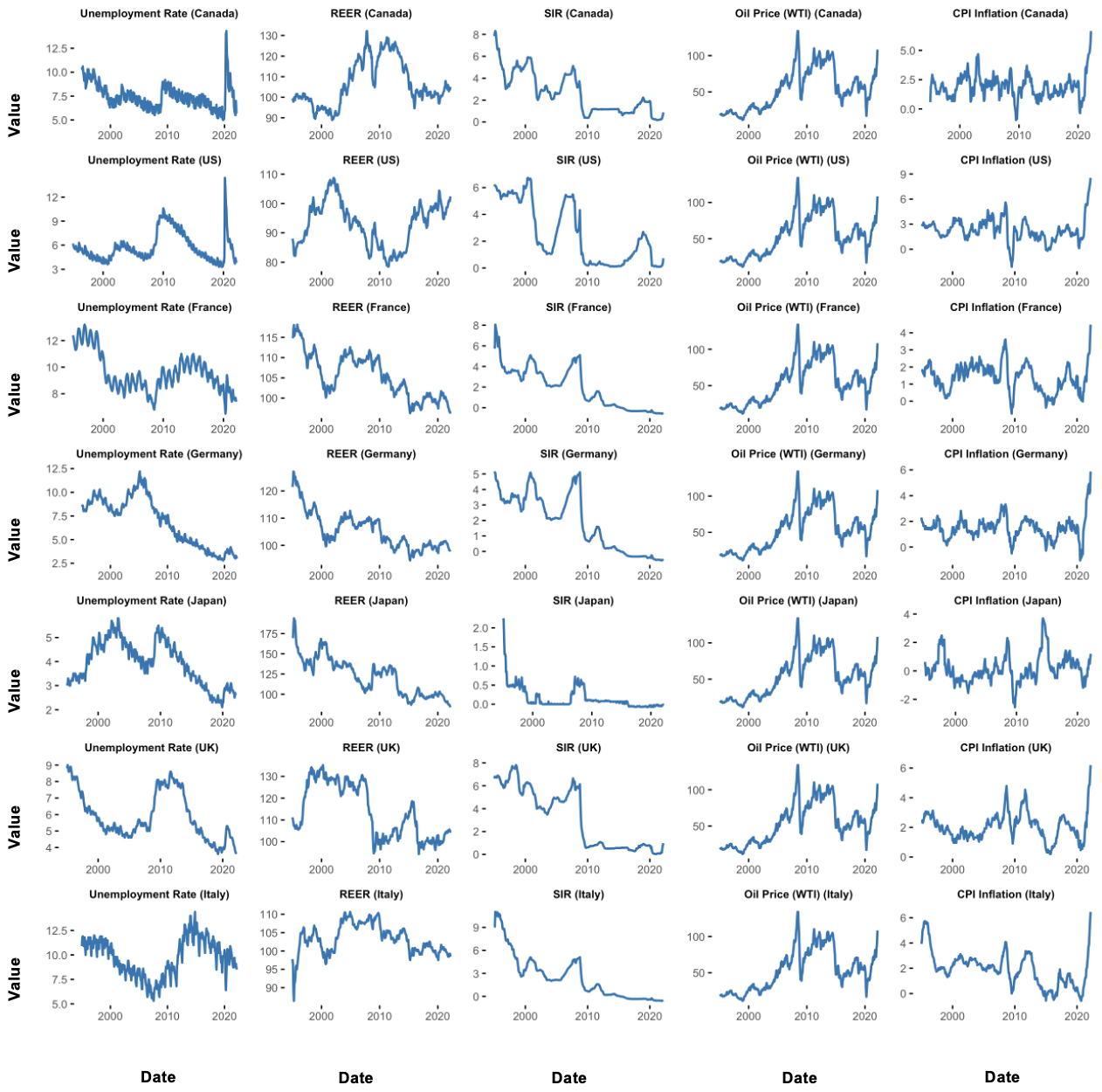}
\caption{Trend plots for Unemployment Rate, REER, SIR, Oil Price (WTI), and CPI Inflation across G7 countries. Each subplot visualizes the time series evolution of a specific macroeconomic variable for a given country over the training period (Jan, 1995 -- Mar, 2022).}
\label{fig:trend_G7}
\end{figure}
\FloatBarrier

\subsection{ACF Plots}\label{app:Appendix_acf_pacf}
\begin{figure}[!htbp]
\centering
\includegraphics[width=\textwidth]
{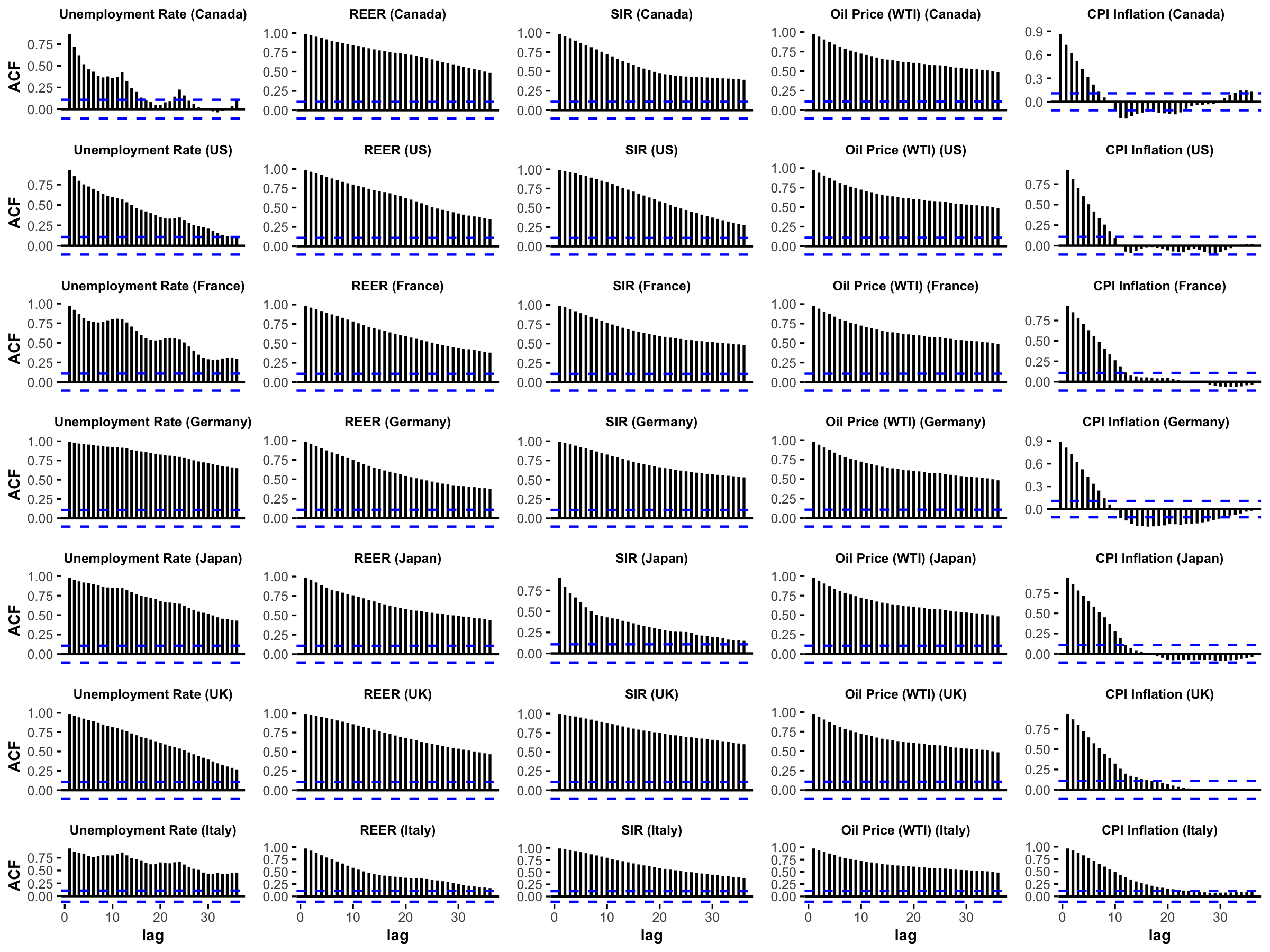}
\caption{ACF plots for Unemployment Rate, REER, SIR, Oil Price (WTI), and CPI Inflation across G7 countries. Each subplot displays the autocorrelation structure for a specific macroeconomic variable and country over the training period (Jan, 1995 -- Mar, 2022).}
\label{fig:acf_G7}
\end{figure}
\FloatBarrier
\FloatBarrier
\subsection{OLS-CUSUM test for structural breakpoint analysis}\label{app:Appendix_structural_breakpoints}
\begin{figure}[!htbp]
\centering
\includegraphics[width=0.92\textwidth]
{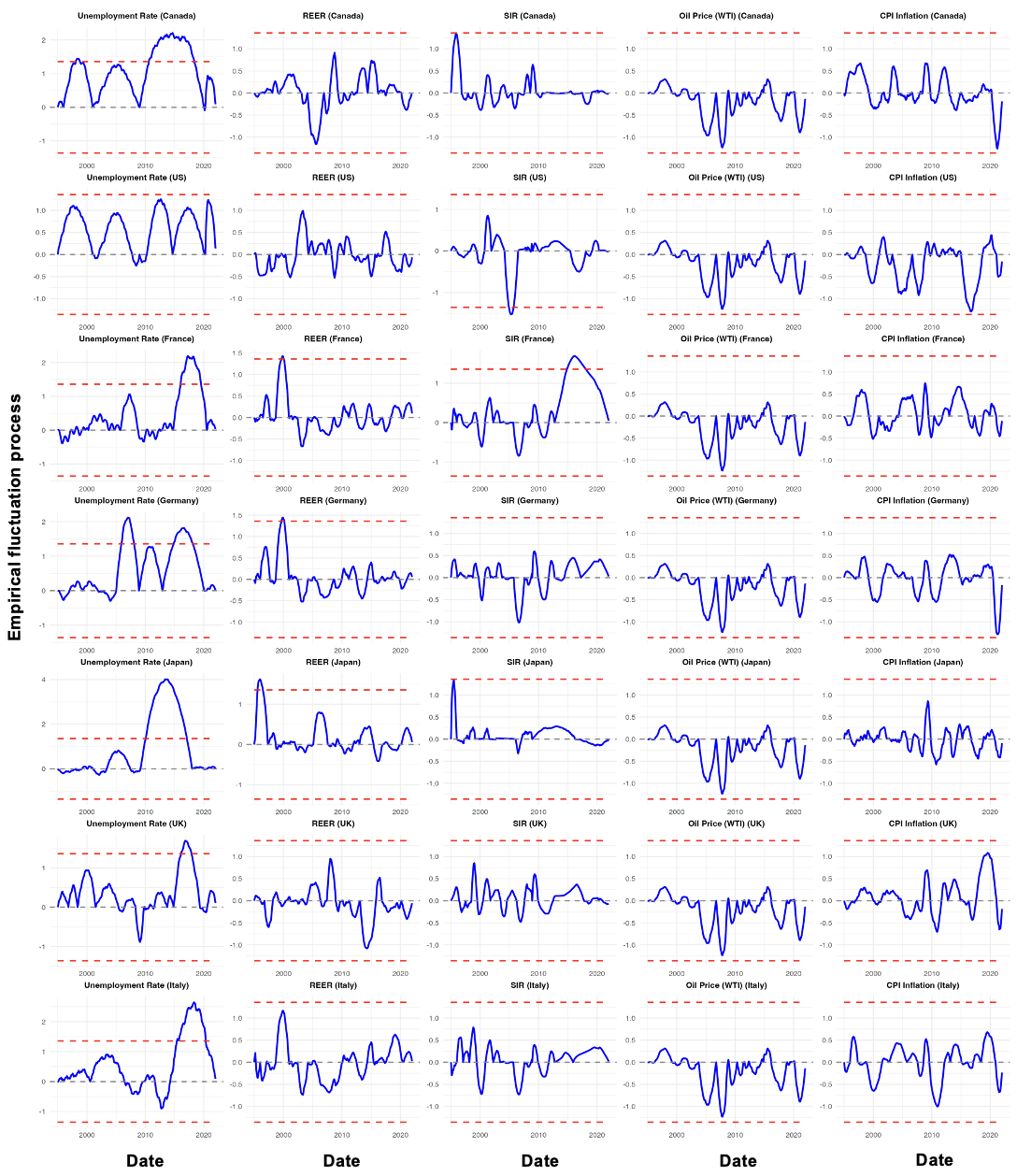}
\caption{OLS-CUSUM test plots for structural breakpoint detection in Unemployment Rate, REER, SIR, Oil Price (WTI), and CPI Inflation across G7 countries. Blue lines show the CUSUM process; red dashed lines indicate 95\% confidence boundaries. Excursions beyond the boundaries signal structural breaks in the respective macroeconomic series.}
\label{fig:OLS_CUSUM_SB_G7}
\end{figure}
\FloatBarrier

\subsection{Causal Analysis}\label{app:Appendix_causal_analysis}
This appendix presents the formal mathematical details of the False Discovery Rate (FDR) corrected Wavelet Coherence Analysis (WCA) workflow used for finding time-frequency-localized associations and lead-lag relations between five key macroeconomic variables and four exogenous uncertainty shocks over the G7 sample. The implementation via the \texttt{biwavelet} package in \textbf{R} leverages the most recent advances in signal decomposition, bias correction, and valid multiple testing control to guide valid discovery of statistically significant findings.

Let $x(t)$ and $y(t)$ be two discrete-time processes sampled at interval $dt$, with $t$ indexing observation times. The continuous wavelet transform (CWT) of $x(t)$, using a complex Morlet mother wavelet $\psi(\cdot)$ with $\omega_0 = 6$ and unit-energy normalization, is given by:
\begin{equation*}
\mathcal{W}_{x}(s, \tau) = \sqrt{\frac{dt}{s}} \sum_{t'} x(t') \overline{\psi}\left( \frac{(t' - \tau)dt}{s} \right),
\end{equation*}
where $s$ is the scale (inverse frequency), $\tau$ the translation (in time), and $\overline{\psi}$ the complex conjugate of the mother wavelet \citep{torrence1998practical}. The cross-wavelet transform between $x(t)$ and $y(t)$ is:
\begin{equation*}
\mathcal{W}_{xy}(s, \tau) = \mathcal{W}_{x}(s, \tau) \cdot \overline{\mathcal{W}_{y}(s, \tau)},
\end{equation*}
where $\overline{\mathcal{W}_{y}(s, \tau)}$ is the complex conjugate of the CWT of $y(t)$. 

The localized wavelet coherence, bounded by $\mathcal{R}^2(s, \tau) \in [0,1]$, is computed as:
\begin{equation*}
\mathcal{R}^2(s, \tau) =
\frac{ \left| \mathcal{S} \left (s^{-1} \mathcal{W}_{xy}(s, \tau) \right ) \right|^2 }
{ \mathcal{S} \left ( s^{-1} |\mathcal{W}_{x}(s, \tau)|^2 \right ) \cdot \mathcal{S} \left ( s^{-1} |\mathcal{W}_{y}(s, \tau)|^2 \right ) },
\end{equation*}
where $\mathcal{S}(\cdot)$ represents a two-dimensional smoothing operator applied in both time and scale dimensions using a Morlet-consistent time-scale smoothing kernel \citep{torrence1998practical, grinsted2004application}. The phase difference, providing lead-lag information, is given by:
\begin{equation*}
\phi_{xy}(s, \tau) = \arg\left(\mathcal{S}\left(s^{-1}\mathcal{W}_{xy}(s, \tau)\right)\right),
\end{equation*}
where $\arg(\cdot)$ denotes the argument of the complex number. Lead-lag is interpreted via the phase-arrow convention: right-pointing arrows indicate in-phase relationships, left-pointing arrows indicate anti-phase relationships; arrows tilting upward indicate $x$ leads $y$, and downward-tilting arrows indicate $y$ leads $x$; the angle magnitude encodes the lag duration.

To address low-frequency bias in cross-power, we apply the scale normalization of \citet{veleda2012cross}, premultiplying each transform by $\sqrt{s}$ before forming cross-power. This correction chiefly affects cross-power estimates, while the normalized coherence $\mathcal{R}^2(s, \tau)$ remains essentially invariant after smoothing operations. For auto-power bias background, see \citet{liu2007rectification}.

Statistical significance is assessed via Monte Carlo simulation using AR(1) surrogate time series that preserve the lag-1 autocorrelation structure of the original data, generating empirical null distributions for $\mathcal{R}^2(s, \tau)$ at each $(s, \tau)$ point with $B = 1000$ realizations and significance level $\alpha = 0.05$. All testing excludes regions within the cone of influence (COI) where edge effects contaminate the wavelet transform \citep{torrence1998practical}.

As the method involves multiple simultaneous hypothesis tests across the time-frequency grid, control for false discoveries is ensured through a two-stage approach. First, the Benjamini-Hochberg FDR procedure \citep{benjamini1995controlling} is applied at each scale under positive regression dependency on subsets conditions:
\begin{itemize}
  \item For a given scale $s$, let $m$ denote the number of time points under consideration outside the COI. Compute a set of p-values $\{ p_{s, \tau_1}, p_{s, \tau_2}, ..., p_{s, \tau_m} \}$ corresponding to the test for statistical significance of coherence at each temporal location $\tau_j$.
  \item Sort the p-values in ascending order, denoted as $p_{(1)} \leq p_{(2)} \leq \ldots \leq p_{(m)}$, where $p_{(k)}$ is the $k^{th}$ smallest p-value among the $m$ hypotheses at scale $s$.
  \item For a chosen false discovery rate (FDR) level $\alpha$, determine the largest integer $k^*$ such that $p_{(k)} \leq \frac{k}{m} \alpha$.
  \item All points for which the p-value $p_{s,\tau} \leq p_{(k^*)}$ are then declared statistically significant at FDR level $\alpha$.
\end{itemize}

Second, to conservatively perform a global sensitivity analysis that allows for possible dependence in the time-frequency domain, we compute pooled Benjamini-Yekutieli q-values \citep{benjamini2001control} over all $(s, \tau)$ points that control FDR at any pre-specified level under any possible dependence structure. Cluster-wise contiguity on the time-frequency plane can also be optionally used to respect the geometry of spatial dependence. Here $k$ indexes through the hypotheses ranked at a particular scale. $s$, $m$ is the number of time points tested at that scale outside of the COI, and $\alpha$ is the chosen FDR level (e.g., $0.10$).

The proposed correction method rigorously limits the anticipated rate of false positives amongst all detected time-frequency points, thereby boosting the confidence in findings from high-resolution WCA investigations in macroeconomic frameworks. The findings denote correlation trends and lead-lag interconnections, as opposed to structural causation, with phase variations delivering directional clues concerning the temporal order between shocks in uncertainty and macroeconomic metrics.
\FloatBarrier

\begin{figure}[!htbp]
\centering
\includegraphics[width=\textwidth]
{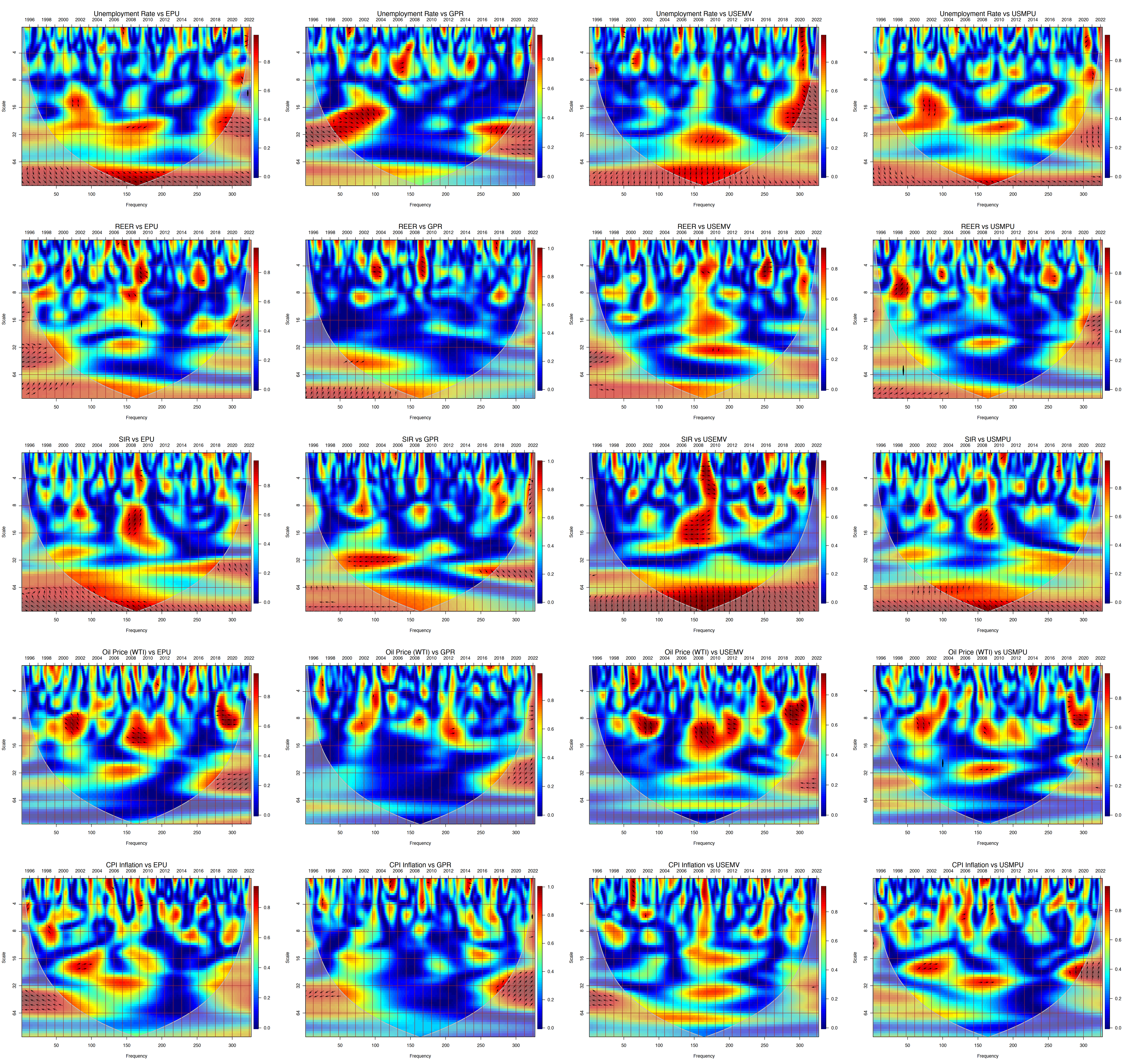}
\caption{Scale-wise FDR-corrected wavelet coherence spectra for the US (Jan, 1995 – Mar, 2022), displaying Unemployment Rate, REER, SIR, Oil Price (WTI), and CPI Inflation against EPU, GPR, USEMV, and USMPU. Each subplot visualizes the time-frequency coherence between a macroeconomic variable (y-axis: Scale; x-axis: Frequency) and an uncertainty index, with warm colors (red/yellow) indicating intervals of high, statistically significant co-movement after FDR correction. The direction of phase arrows indicates lead-lag relationships: rightward arrows mean the series move in phase; leftward, in anti-phase; upward arrows signify the uncertainty index leads; downward, that it lags. Shaded regions inside the cone of influence reflect reliable coherence, while grid consistency enables robust cross-variable and cross-period comparison. The level of significance ($\alpha$) for the FDR adjustment is set at 10\%.}
\label{fig:WCA_usa}
\end{figure}
\FloatBarrier

\begin{figure}[!htbp]
\centering
\includegraphics[width=\textwidth]
{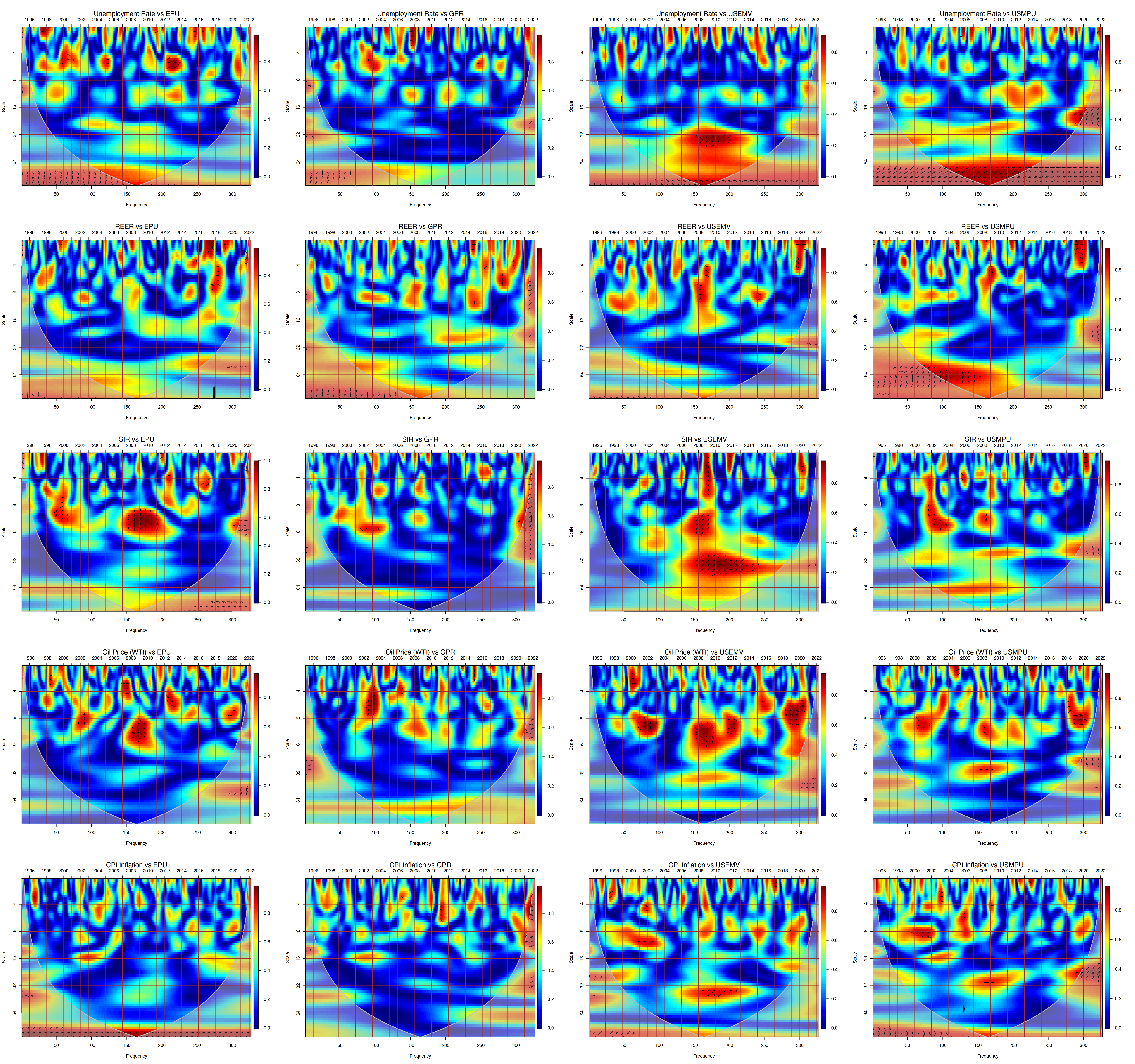}
\caption{Scale-wise FDR-corrected wavelet coherence spectra for France (Jan, 1995 – Mar, 2022), displaying Unemployment Rate, REER, SIR, Oil Price (WTI), and CPI Inflation against EPU, GPR, USEMV, and USMPU. Each subplot visualizes the time-frequency coherence between a macroeconomic variable (y-axis: Scale; x-axis: Frequency) and an uncertainty index, with warm colors (red/yellow) indicating intervals of high, statistically significant co-movement after FDR correction. The direction of phase arrows indicates lead-lag relationships: rightward arrows mean the series move in phase; leftward, in anti-phase; upward arrows signify the uncertainty index leads; downward, that it lags. Shaded regions inside the cone of influence reflect reliable coherence, while grid consistency enables robust cross-variable and cross-period comparison. The level of significance ($\alpha$) for the FDR adjustment is set at 10\%.}
\label{fig:WCA_france}
\end{figure}
\FloatBarrier

\begin{figure}[!htbp]
\centering
\includegraphics[width=\textwidth]
{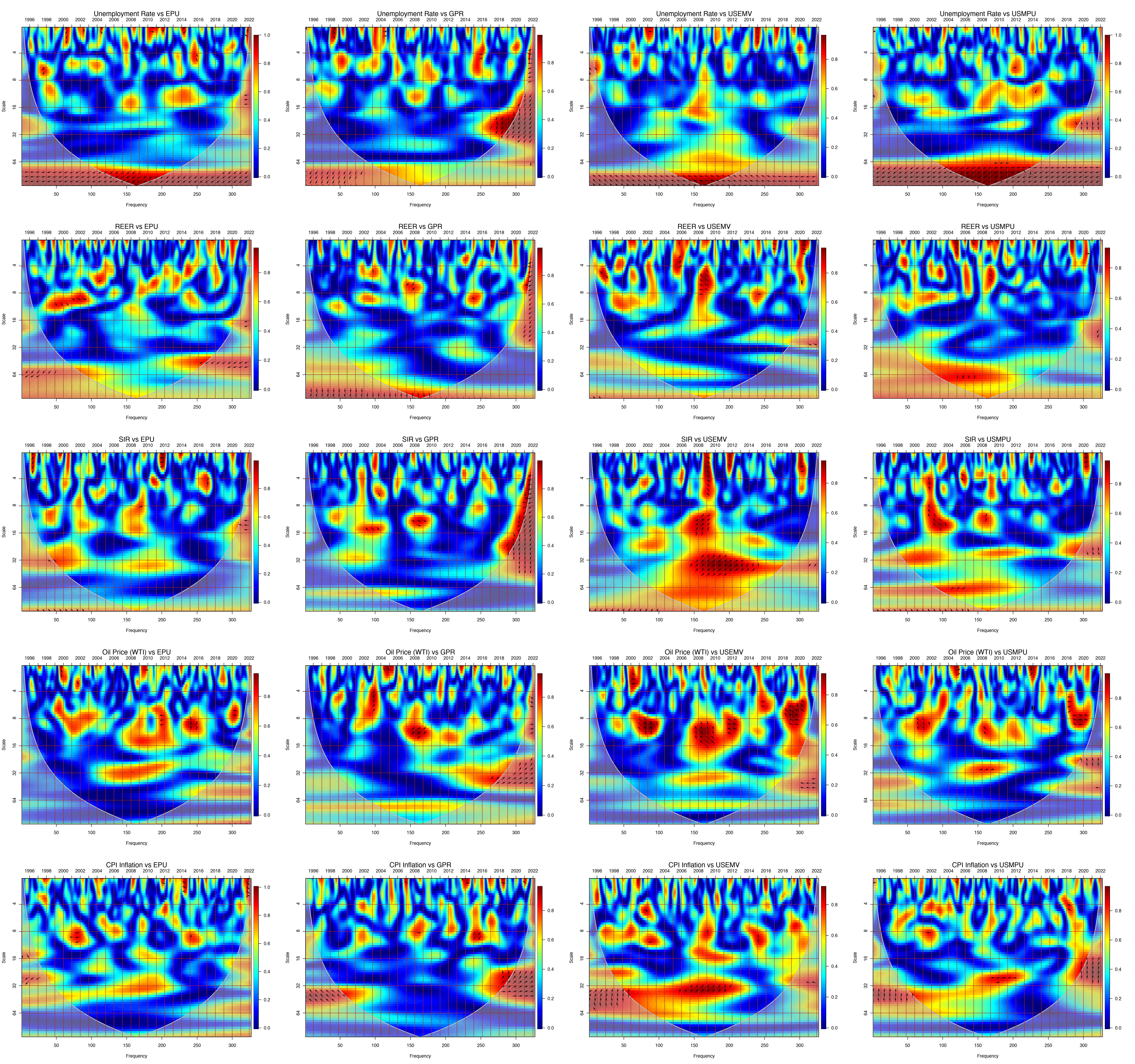}
\caption{Scale-wise FDR-corrected wavelet coherence spectra for Germany (Jan, 1995 – Mar, 2022), displaying Unemployment Rate, REER, SIR, Oil Price (WTI), and CPI Inflation against EPU, GPR, USEMV, and USMPU. Each subplot visualizes the time-frequency coherence between a macroeconomic variable (y-axis: Scale; x-axis: Frequency) and an uncertainty index, with warm colors (red/yellow) indicating intervals of high, statistically significant co-movement after FDR correction. The direction of phase arrows indicates lead-lag relationships: rightward arrows mean the series move in phase; leftward, in anti-phase; upward arrows signify the uncertainty index leads; downward, that it lags. Shaded regions inside the cone of influence reflect reliable coherence, while grid consistency enables robust cross-variable and cross-period comparison. The level of significance ($\alpha$) for the FDR adjustment is set at 10\%.}
\label{fig:WCA_germany}
\end{figure}
\FloatBarrier

\begin{figure}[!htbp]
\centering
\includegraphics[width=\textwidth]
{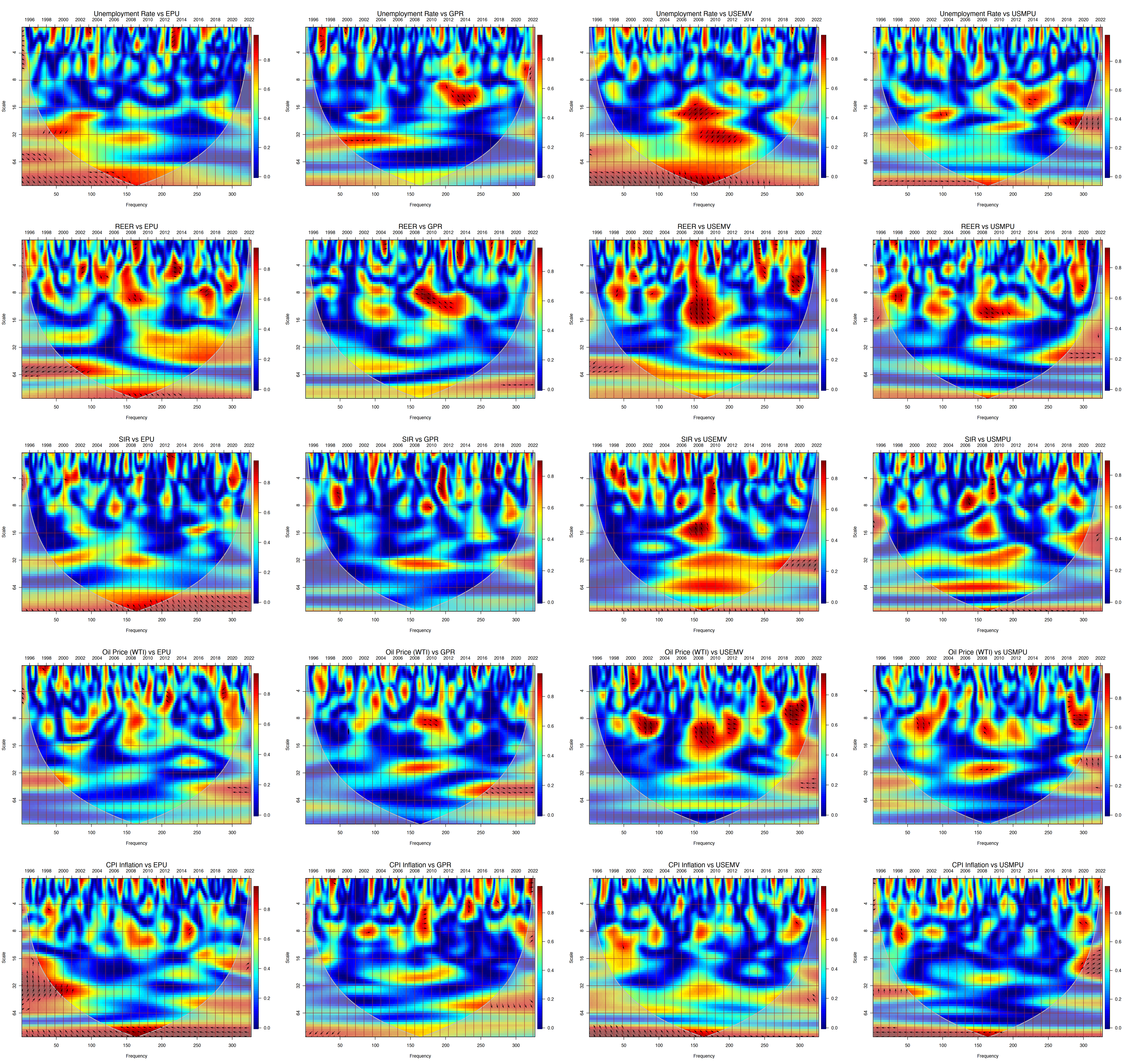}
\caption{Scale-wise FDR-corrected wavelet coherence spectra for Japan (Jan, 1995 – Mar, 2022), displaying Unemployment Rate, REER, SIR, Oil Price (WTI), and CPI Inflation against EPU, GPR, USEMV, and USMPU. Each subplot visualizes the time-frequency coherence between a macroeconomic variable (y-axis: Scale; x-axis: Frequency) and an uncertainty index, with warm colors (red/yellow) indicating intervals of high, statistically significant co-movement after FDR correction. The direction of phase arrows indicates lead-lag relationships: rightward arrows mean the series move in phase; leftward, in anti-phase; upward arrows signify the uncertainty index leads; downward, that it lags. Shaded regions inside the cone of influence reflect reliable coherence, while grid consistency enables robust cross-variable and cross-period comparison. The level of significance ($\alpha$) for the FDR adjustment is set at 10\%.}
\label{fig:WCA_japan}
\end{figure}
\FloatBarrier

\begin{figure}[!htbp]
\centering
\includegraphics[width=\textwidth]
{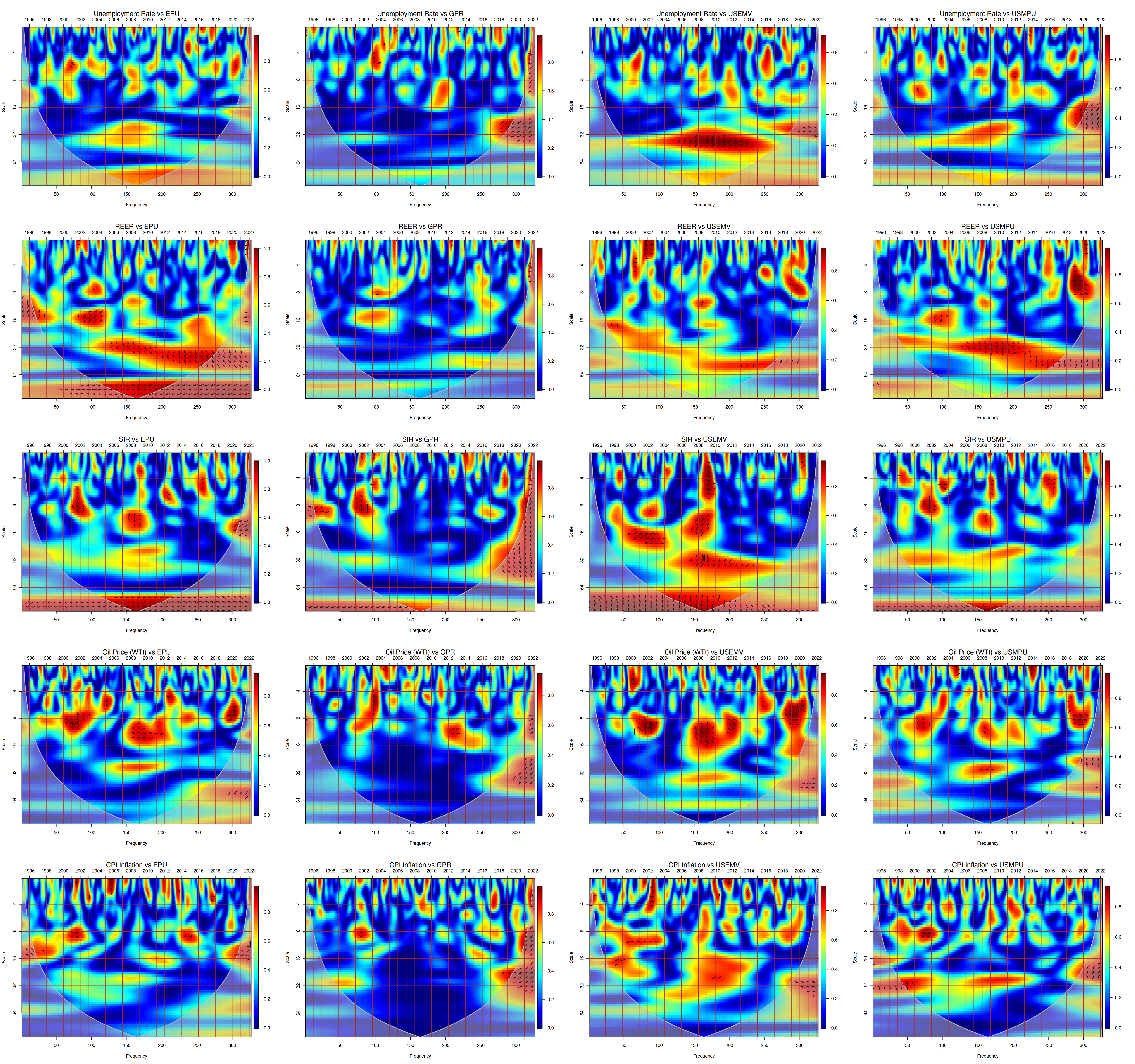}
\caption{Scale-wise FDR-corrected wavelet coherence spectra for the UK (Jan, 1995 – Mar, 2022), displaying Unemployment Rate, REER, SIR, Oil Price (WTI), and CPI Inflation against EPU, GPR, USEMV, and USMPU. Each subplot visualizes the time-frequency coherence between a macroeconomic variable (y-axis: Scale; x-axis: Frequency) and an uncertainty index, with warm colors (red/yellow) indicating intervals of high, statistically significant co-movement after FDR correction. The direction of phase arrows indicates lead-lag relationships: rightward arrows mean the series move in phase; leftward, in anti-phase; upward arrows signify the uncertainty index leads; downward, that it lags. Shaded regions inside the cone of influence reflect reliable coherence, while grid consistency enables robust cross-variable and cross-period comparison. The level of significance ($\alpha$) for the FDR adjustment is set at 10\%.}
\label{fig:WCA_uk}
\end{figure}
\FloatBarrier

\begin{figure}[!htbp]
\centering
\includegraphics[width=\textwidth]
{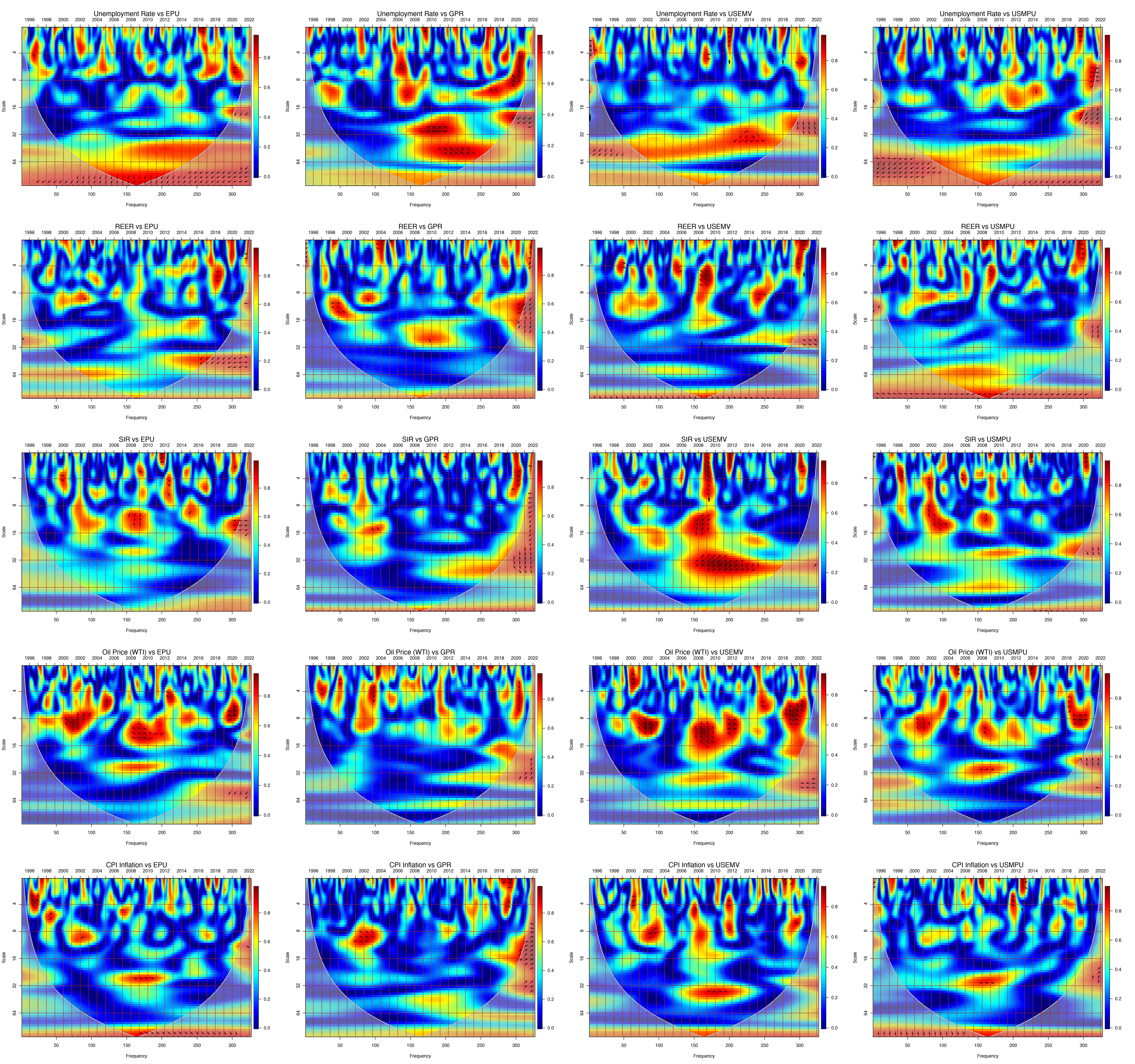}
\caption{Scale-wise FDR-corrected wavelet coherence spectra for Italy (Jan, 1995 – Mar, 2022), displaying Unemployment Rate, REER, SIR, Oil Price (WTI), and CPI Inflation against EPU, GPR, USEMV, and USMPU. Each subplot visualizes the time-frequency coherence between a macroeconomic variable (y-axis: Scale; x-axis: Frequency) and an uncertainty index, with warm colors (red/yellow) indicating intervals of high, statistically significant co-movement after FDR correction. The direction of phase arrows indicates lead-lag relationships: rightward arrows mean the series move in phase; leftward, in anti-phase; upward arrows signify the uncertainty index leads; downward, that it lags. Shaded regions inside the cone of influence reflect reliable coherence, while grid consistency enables robust cross-variable and cross-period comparison. The level of significance ($\alpha$) for the FDR adjustment is set at 10\%.}
\label{fig:WCA_italy}
\end{figure}
\FloatBarrier
\subsection{Methodological framework for non-linear local projections}\label{app:Appendix_IRFS_NLP}
Impulse responses are estimated horizon-wise using nonlinear local projections as implemented in the \texttt{lpirfs} package \citep{RJ-2019-052}, extending \citet{jorda2005estimation} to smoothly switch regimes via a logistic function. For each G7 country $c$ and forecast horizon $h$, the system
\begin{equation}\label{eq:nllp}
\begin{aligned}
\mathbf{y}_{c,t+h}
&= \boldsymbol{\alpha}_c^{\,h}
+ \sum_{j=1}^{p} \mathbf{B}_{j,1,c}^{\,h}\Big(\mathbf{y}_{c,t-j}\,[1 - F(z_{c,t-1})]\Big)
+ \sum_{j=1}^{p} \mathbf{B}_{j,2,c}^{\,h}\Big(\mathbf{y}_{c,t-j}\,F(z_{c,t-1})\Big) \\
&\quad + \sum_{m=1}^{M} \boldsymbol{\Gamma}_{m,1,c}^{\,h}\,\big(u_{m,c,t}\,[1 - F(z_{c,t-1})]\big)
+ \sum_{m=1}^{M} \boldsymbol{\Gamma}_{m,2,c}^{\,h}\,\big(u_{m,c,t}\,F(z_{c,t-1})\big) \\
&\quad + \delta_c^{\,h}\,\mathrm{trend}_{t}
+ \boldsymbol{\varepsilon}_{c,t+h}^{\,h},
\end{aligned}
\end{equation}
is estimated by OLS for each $h=0,\dots,H$, where $\mathbf{y}_{c,t}$ is a $K \times 1$ vector of macro variables (unemployment rate, REER, SIR, Oil Price (WTI), and, CPI inflation; $K=5$), $u_{m,c,t}$ denotes uncertainty shocks (EPU, GPR, USEMV, USMPU; $M=4$), and $F(z_{c,t-1}) = [1+\exp(\gamma z_{c,t-1})]^{-1}$ is the transition function of the standardized short-term rate $z_{c,t-1}$. Lag order $p$ is selected via Bayesian Information Criterion (BIC) and held fixed across $h$. The regime index is driven by the standardized short-term interest rate (SIR)
\[
z_{c,t-1} \;=\; \frac{\mathrm{SIR}_{c,t-1}-\mu_c}{\sigma_c},
\]
where $\mu_c$ and $\sigma_c$ are the within-country sample mean and standard deviation computed once and held fixed across horizons. The (decreasing) logistic transition function is
\[
F\!\left(z_{c,t-1}\right) \;=\; \frac{1}{1+\exp\!\big(\gamma\,z_{c,t-1}\big)}, \qquad \gamma>0,
\]
so that higher values of $z_{c,t-1}$ (tight/high-rate periods) imply $F(z_{c,t-1})\approx 0$ and the $(1-F)$-weighted regime (regime 1, “high IR”) dominates, whereas lower values of $z_{c,t-1}$ (loose/low-rate periods) imply $F(z_{c,t-1})\approx 1$ and the $F$-weighted regime (regime 2, “low IR”) dominates. We lag the transition function to avoid contemporaneous feedback. We consider two complementary implementations consistent with \citet{RJ-2019-052}:
(i) Identified-shock nonlinear LPs (\texttt{lp\_nl\_iv}). In practice, we estimate a separate model for each uncertainty proxy $m$, treating $u_{m,c,t}$ as the identified shock series. In this case, the regime-$r$ impulse response of variable $i$ at horizon $h$ to a unit innovation in $u_m$ is the $i$th element of $\boldsymbol{\Gamma}_{m,r,c}^{\,h}$, extracted using the selection vector $\mathbf{e}_i'$, where $\mathbf{e}_i$ is a $K \times 1$ unit vector with unity in the $i$th position and zeros elsewhere, enabling isolation of the specific variable's response from the multivariate coefficient vector:
\[
\mathrm{IRF}_{i,m,r}^{\,h} \;=\; \mathbf{e}_i'\,\boldsymbol{\Gamma}_{m,r,c}^{\,h}, \qquad r\in\{1,2\}.
\]
(ii) VAR-based nonlinear IRFs (\texttt{lp\_nl}). When no external shock series is supplied, \texttt{lpirfs} estimates a reduced VAR to obtain the residual covariance $\boldsymbol{\Sigma}$, applies a Cholesky factorization to construct orthogonal shocks, and uses the local projection coefficients in \eqref{eq:nllp} to compute state-dependent IRFs by horizon, as detailed in \citet{RJ-2019-052}. HAC Newey--West standard errors yield confidence intervals:
\[
\mathrm{CI}_{95\%} = [\hat{\theta} - 1.96\,\hat{\sigma}_{NW}(\hat{\theta}),\, \hat{\theta} + 1.96\,\hat{\sigma}_{NW}(\hat{\theta})].
\]
Model reliability requires sufficient variation in $F(z_{c,t-1})$ so both regimes are well-represented. This framework enables flexible, state-dependent IRF estimation while accounting for serial correlation and smooth monetary regime shifts, following \citep{RJ-2019-052}.

\subsection{IRF Plots for G7}\label{app:Appendix_IRFS_USA_ITA}
\begin{figure}[!htbp]
 \centering
\includegraphics[width=\textwidth,height=0.85\textheight,keepaspectratio]
{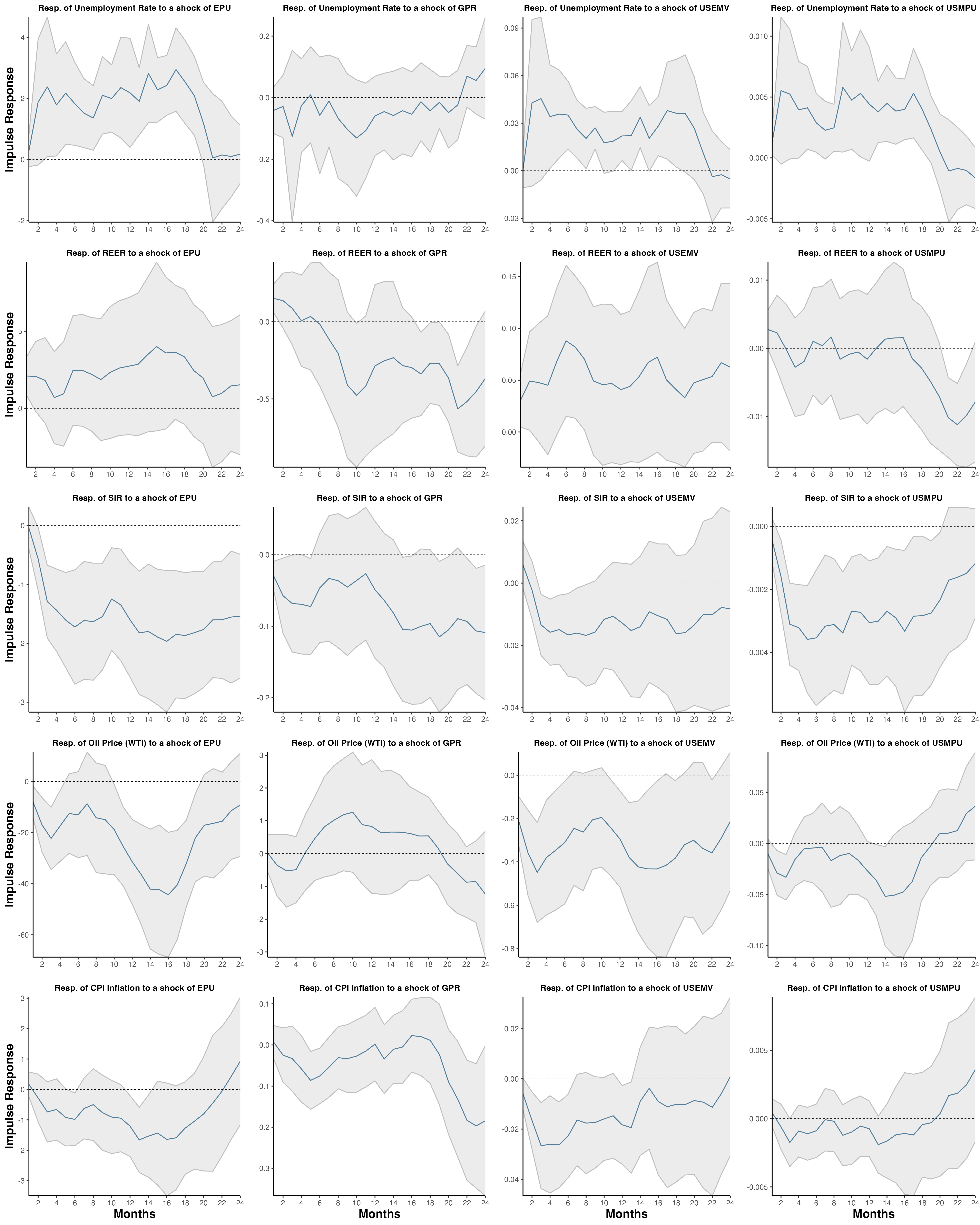}
\caption{Impulse response functions (nonlinear local projections) of Unemployment Rate, REER, SIR, Oil Price (WTI), and CPI Inflation to EPU, GPR, USEMV, and USMPU shocks for the US (high-rate regime). Each row shows responses of a macro variable; each column, a distinct uncertainty shock. Solid \textbf{\textcolor{blue}{blue}} lines depict IRFs with 95\% confidence bands (grey). The x-axis is horizon (months), the y-axis the response magnitude. Compare responses by reading across rows (by variable) or down columns (by shock); zero response is marked by the dotted line.} 
\label{fig:irf_figure1_usa}
\end{figure}
\FloatBarrier
\begin{figure}[!htbp]
 \centering
\includegraphics[width=\textwidth,height=0.85\textheight,keepaspectratio]
{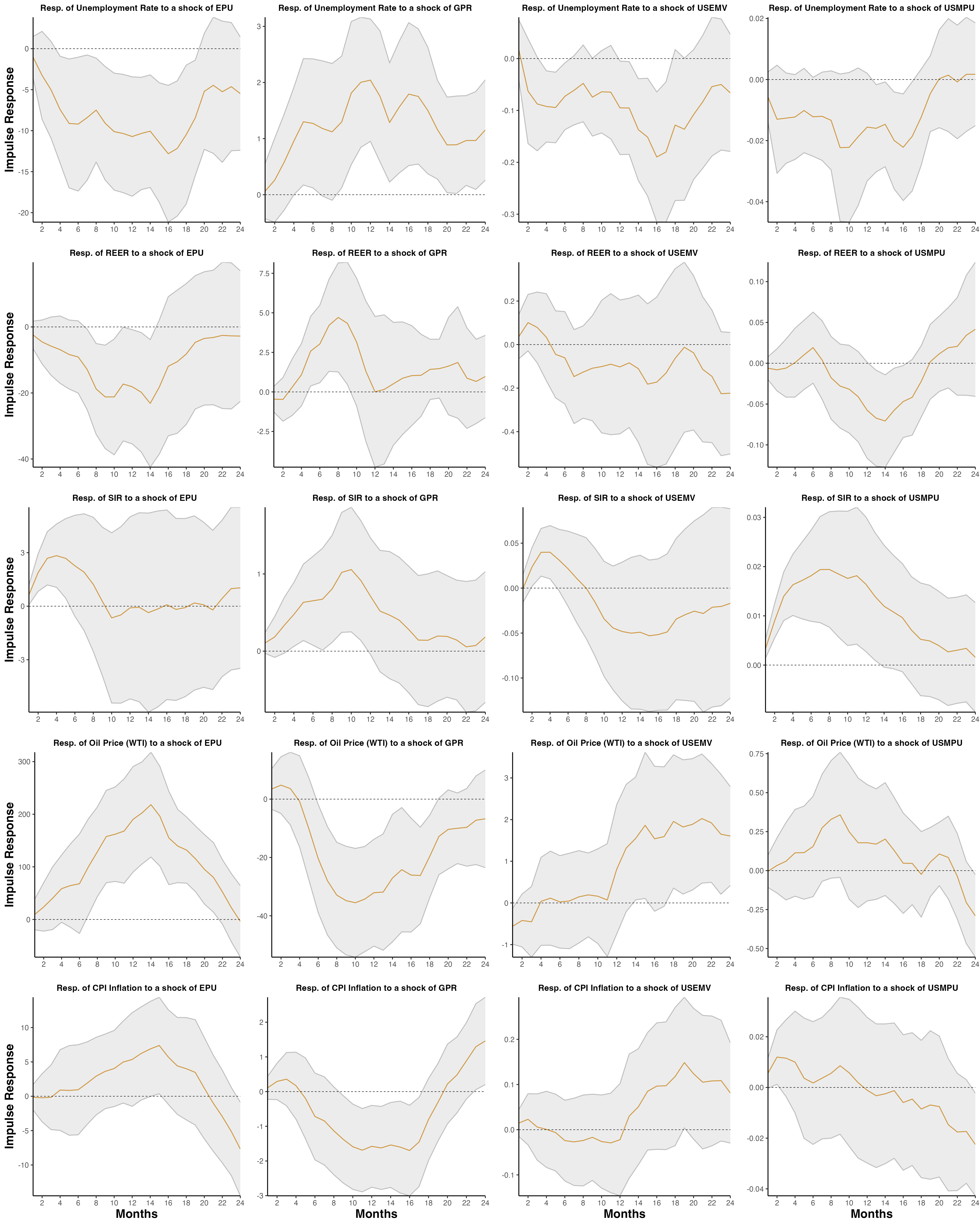}
   \caption{Impulse response functions (nonlinear local projections) of Unemployment Rate, REER, SIR, Oil Price (WTI), and CPI Inflation to EPU, GPR, USEMV, and USMPU shocks for the US (low-rate regime). Each row shows responses of a macro variable; each column, a distinct uncertainty shock. Solid \textbf{\textcolor{orange}{orange}} lines depict IRFs with 95\% confidence bands (grey). The x-axis is horizon (months), the y-axis the response magnitude. Compare responses by reading across rows (by variable) or down columns (by shock); zero response is marked by the dotted line.} 
   \label{fig:irf_figure2_usa}
\end{figure}
\FloatBarrier
\begin{figure}[!htbp]
 \centering
\includegraphics[width=\textwidth,height=0.85\textheight,keepaspectratio]
{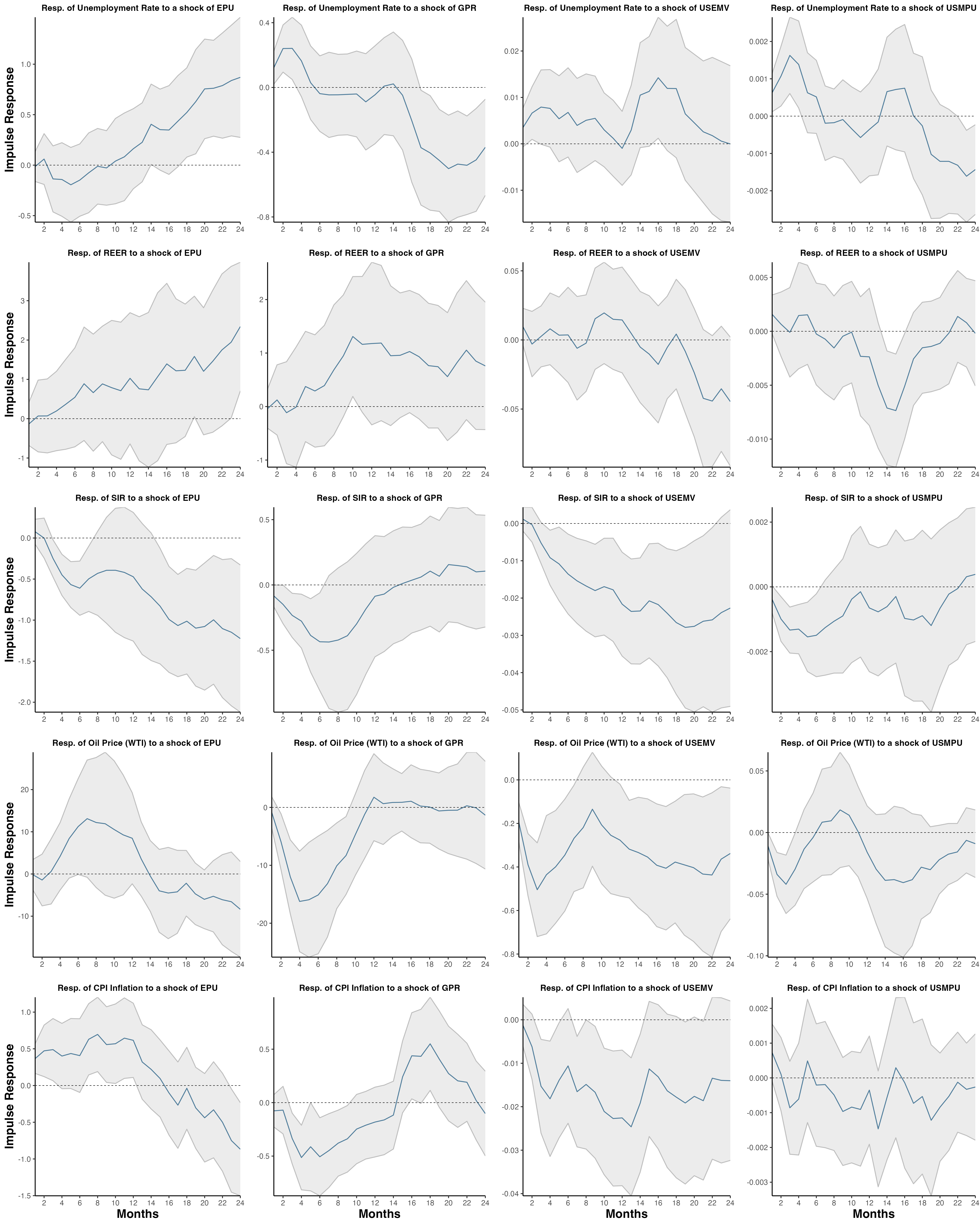}
   \caption{Impulse response functions (nonlinear local projections) of Unemployment Rate, REER, SIR, Oil Price (WTI), and CPI Inflation to EPU, GPR, USEMV, and USMPU shocks for France (high-rate regime). Each row shows responses of a macro variable; each column, a distinct uncertainty shock. Solid \textbf{\textcolor{blue}{blue}} lines depict IRFs with 95\% confidence bands (grey). The x-axis is horizon (months), the y-axis the response magnitude. Compare responses by reading across rows (by variable) or down columns (by shock); zero response is marked by the dotted line.} 
   \label{fig:irf_figure1_fra}
\end{figure}
\FloatBarrier
\begin{figure}[!htbp]
 \centering
\includegraphics[width=\textwidth,height=0.85\textheight,keepaspectratio]
{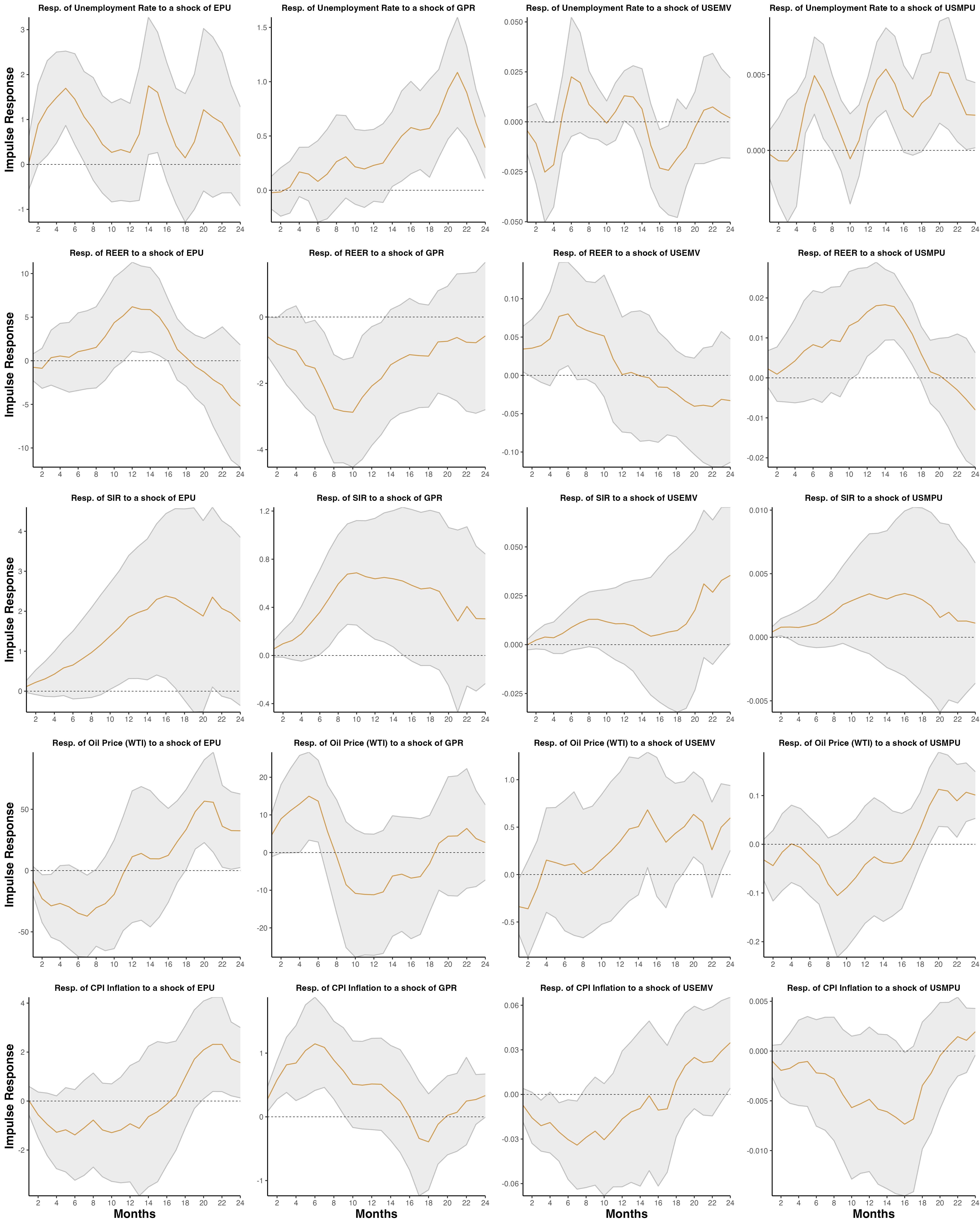}
   \caption{Impulse response functions (nonlinear local projections) of Unemployment Rate, REER, SIR, Oil Price (WTI), and CPI Inflation to EPU, GPR, USEMV, and USMPU shocks for France (low-rate regime). Each row shows responses of a macro variable; each column, a distinct uncertainty shock. Solid \textbf{\textcolor{orange}{orange}} lines depict IRFs with 95\% confidence bands (grey). The x-axis is horizon (months), the y-axis the response magnitude. Compare responses by reading across rows (by variable) or down columns (by shock); zero response is marked by the dotted line.}
   \label{fig:irf_figure2_fra}
\end{figure}
\FloatBarrier
\begin{figure}[!htbp]
 \centering
\includegraphics[width=\textwidth,height=0.85\textheight,keepaspectratio]
{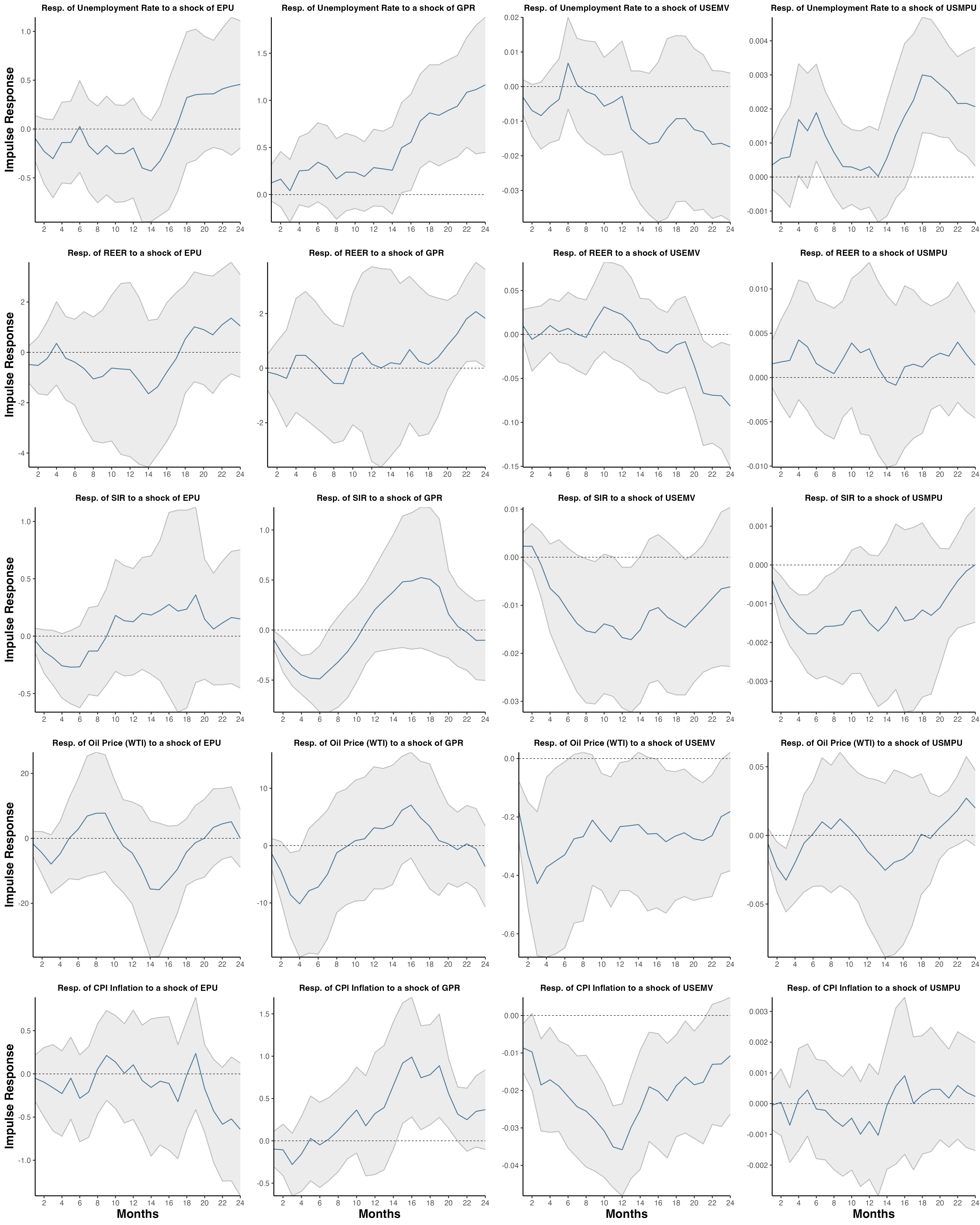}
   \caption{Impulse response functions (nonlinear local projections) of Unemployment Rate, REER, SIR, Oil Price (WTI), and CPI Inflation to EPU, GPR, USEMV, and USMPU shocks for Germany (high-rate regime). Each row shows responses of a macro variable; each column, a distinct uncertainty shock. Solid \textbf{\textcolor{blue}{blue}} lines depict IRFs with 95\% confidence bands (grey). The x-axis is horizon (months), the y-axis the response magnitude. Compare responses by reading across rows (by variable) or down columns (by shock); zero response is marked by the dotted line.}
   \label{fig:irf_figure1_ger}
\end{figure}
\FloatBarrier
\begin{figure}[!htbp]
 \centering
\includegraphics[width=\textwidth,height=0.85\textheight,keepaspectratio]
{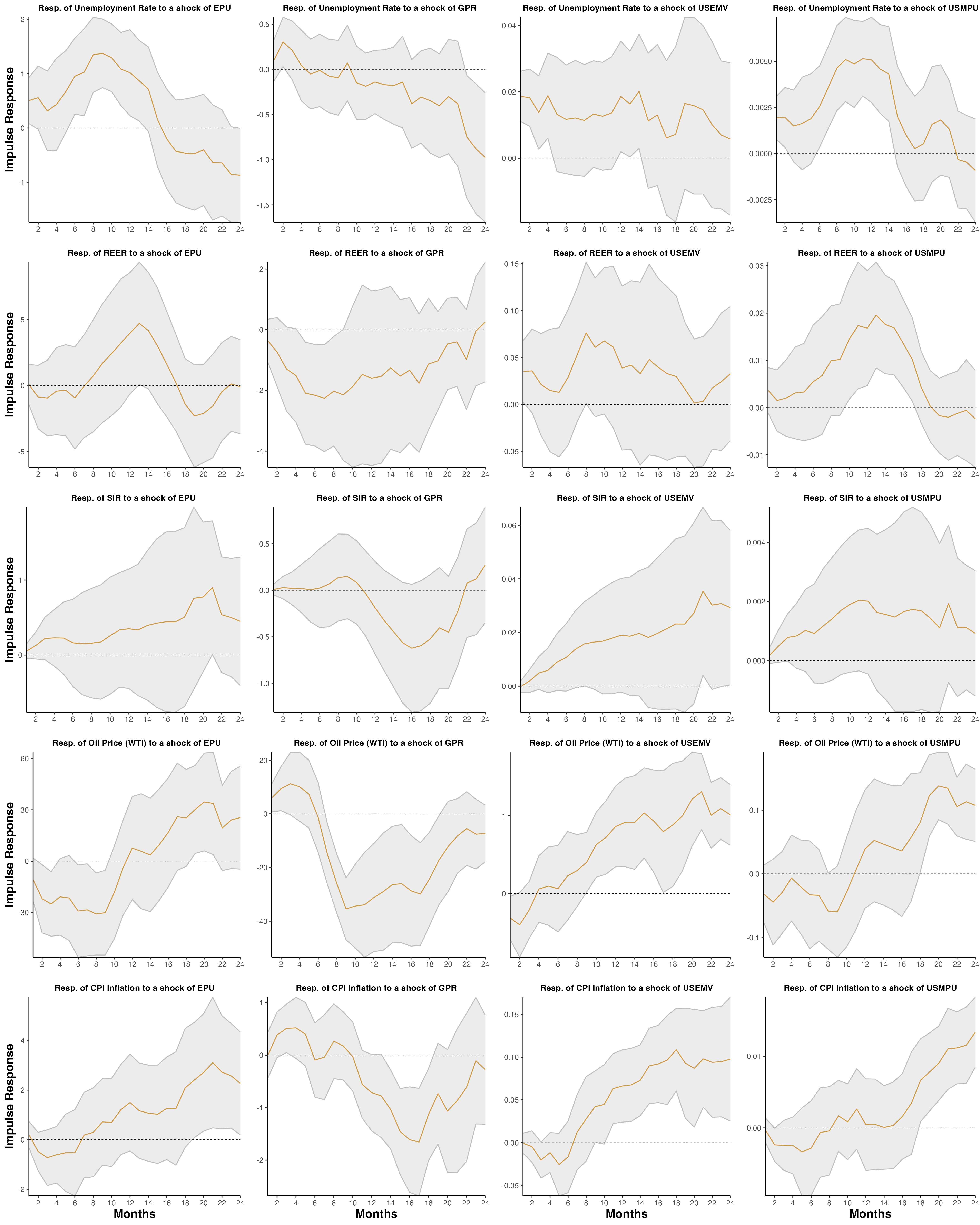}
   \caption{Impulse response functions (nonlinear local projections) of Unemployment Rate, REER, SIR, Oil Price (WTI), and CPI Inflation to EPU, GPR, USEMV, and USMPU shocks for Germany (low-rate regime). Each row shows responses of a macro variable; each column, a distinct uncertainty shock. Solid \textbf{\textcolor{orange}{orange}} lines depict IRFs with 95\% confidence bands (grey). The x-axis is horizon (months), the y-axis the response magnitude. Compare responses by reading across rows (by variable) or down columns (by shock); zero response is marked by the dotted line.}
   \label{fig:irf_figure2_ger}
\end{figure}
\FloatBarrier
\begin{figure}[!htbp]
 \centering
\includegraphics[width=\textwidth,height=0.85\textheight,keepaspectratio]
{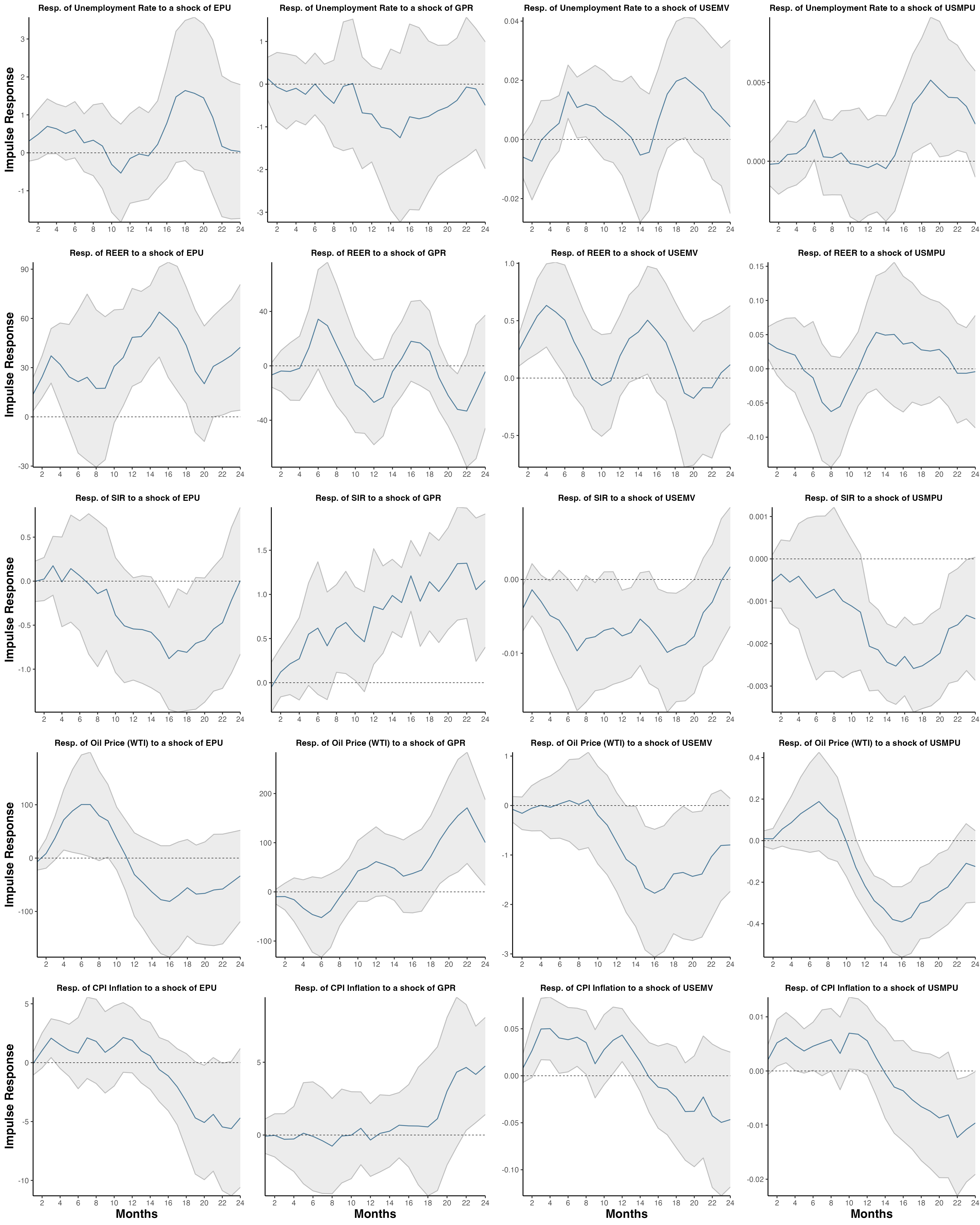}
   \caption{Impulse response functions (nonlinear local projections) of Unemployment Rate, REER, SIR, Oil Price (WTI), and CPI Inflation to EPU, GPR, USEMV, and USMPU shocks for Japan (high-rate regime). Each row shows responses of a macro variable; each column, a distinct uncertainty shock. Solid \textbf{\textcolor{blue}{blue}} lines depict IRFs with 95\% confidence bands (grey). The x-axis is horizon (months), the y-axis the response magnitude. Compare responses by reading across rows (by variable) or down columns (by shock); zero response is marked by the dotted line.}
   \label{fig:irf_figure1_jap}
\end{figure}
\FloatBarrier
\begin{figure}[!htbp]
 \centering
\includegraphics[width=\textwidth,height=0.85\textheight,keepaspectratio]
{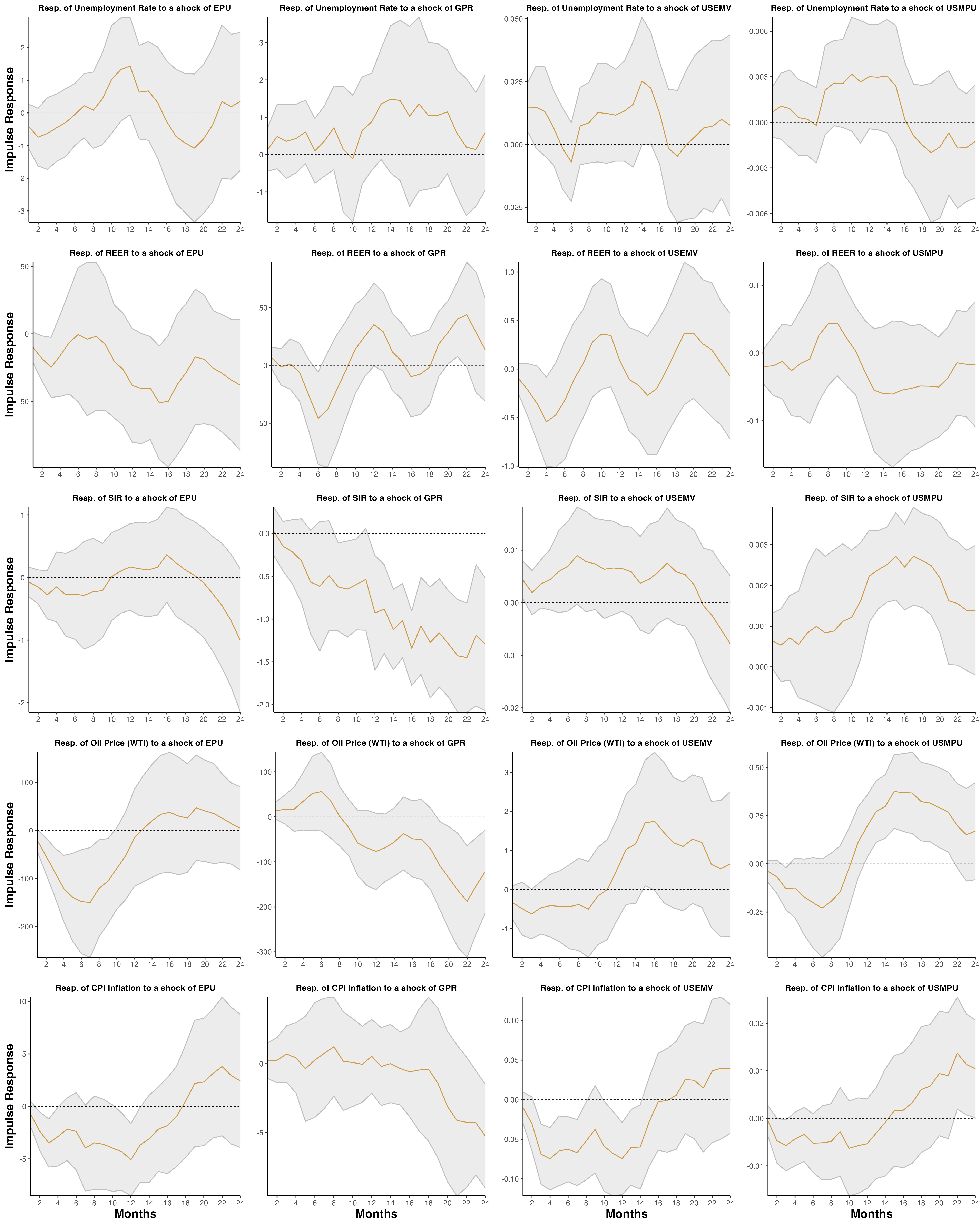}
   \caption{Impulse response functions (nonlinear local projections) of Unemployment Rate, REER, SIR, Oil Price (WTI), and CPI Inflation to EPU, GPR, USEMV, and USMPU shocks for Japan (low-rate regime). Each row shows responses of a macro variable; each column, a distinct uncertainty shock. Solid \textbf{\textcolor{orange}{orange}} lines depict IRFs with 95\% confidence bands (grey). The x-axis is horizon (months), the y-axis the response magnitude. Compare responses by reading across rows (by variable) or down columns (by shock); zero response is marked by the dotted line.}
   \label{fig:irf_figure2_jap}
\end{figure}
\FloatBarrier
\begin{figure}[!htbp]
 \centering
\includegraphics[width=\textwidth,height=0.85\textheight,keepaspectratio]
{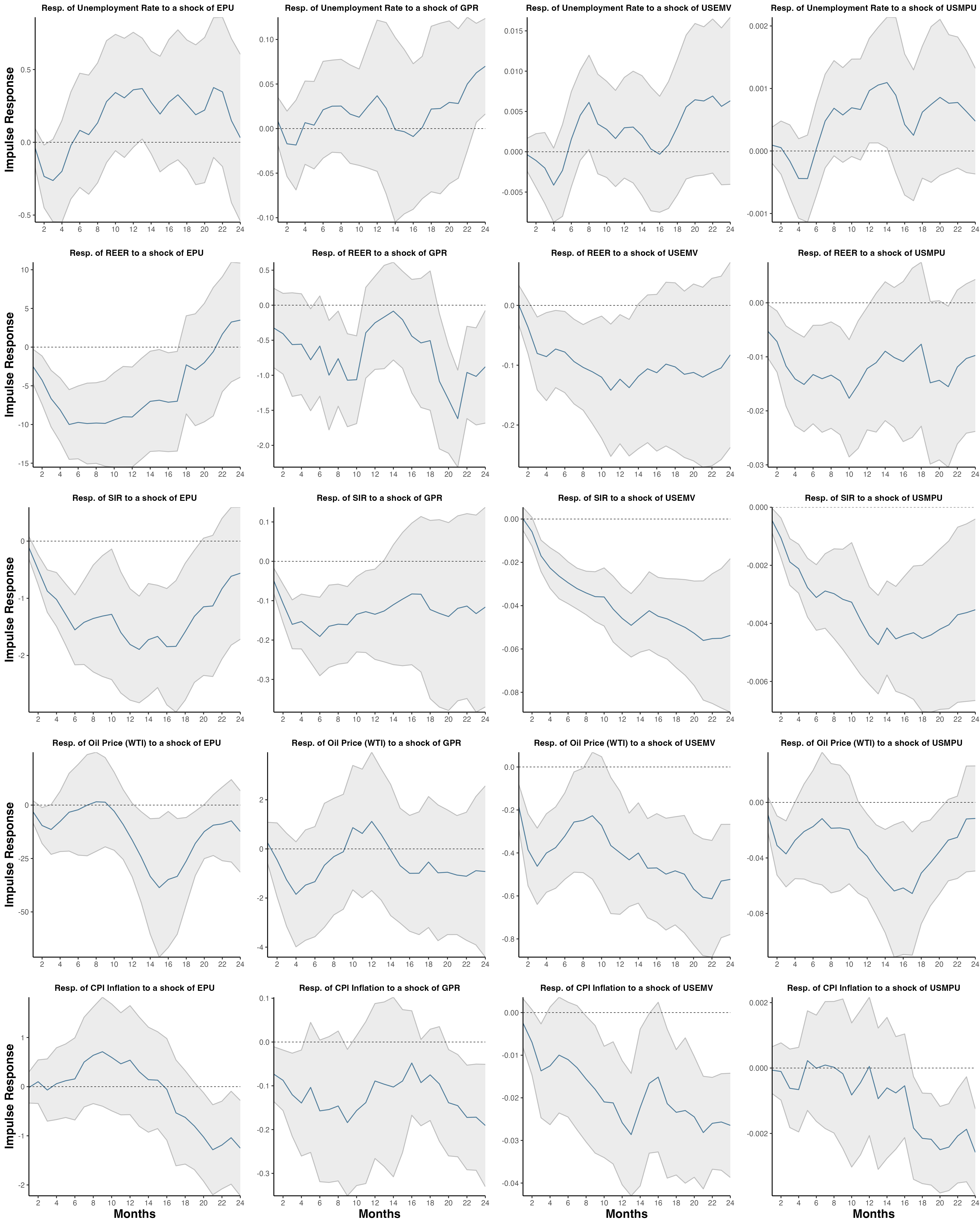}
   \caption{Impulse response functions (nonlinear local projections) of Unemployment Rate, REER, SIR, Oil Price (WTI), and CPI Inflation to EPU, GPR, USEMV, and USMPU shocks for the UK (high-rate regime). Each row shows responses of a macro variable; each column, a distinct uncertainty shock. Solid \textbf{\textcolor{blue}{blue}} lines depict IRFs with 95\% confidence bands (grey). The x-axis is horizon (months), the y-axis the response magnitude. Compare responses by reading across rows (by variable) or down columns (by shock); zero response is marked by the dotted line.}
   \label{fig:irf_figure1_uk}
\end{figure}
\FloatBarrier
\begin{figure}[!htbp]
 \centering
\includegraphics[width=\textwidth,height=0.85\textheight,keepaspectratio]
{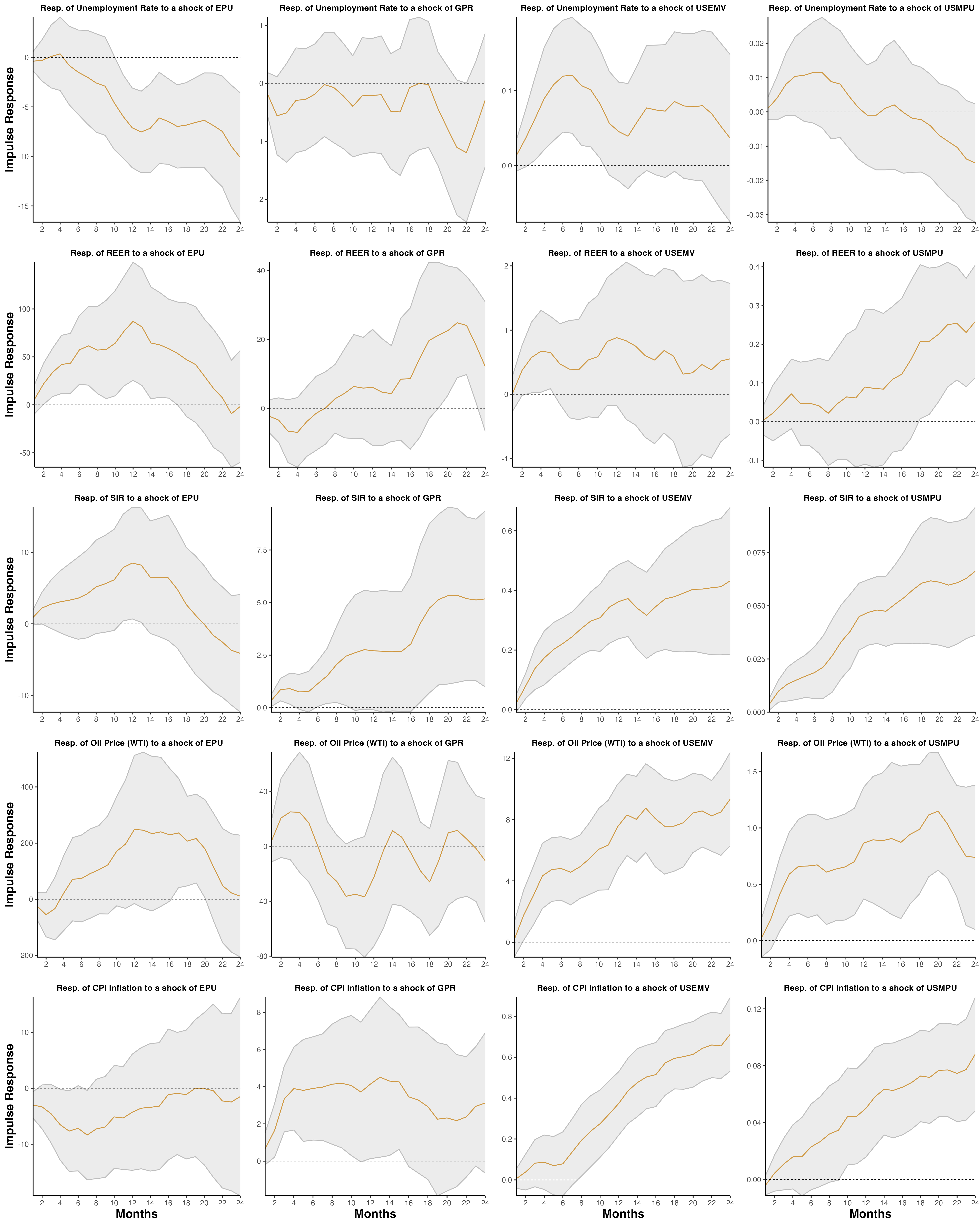}
   \caption{Impulse response functions (nonlinear local projections) of Unemployment Rate, REER, SIR, Oil Price (WTI), and CPI Inflation to EPU, GPR, USEMV, and USMPU shocks for the UK (low-rate regime). Each row shows responses of a macro variable; each column, a distinct uncertainty shock. Solid \textbf{\textcolor{orange}{orange}} lines depict IRFs with 95\% confidence bands (grey). The x-axis is horizon (months), the y-axis the response magnitude. Compare responses by reading across rows (by variable) or down columns (by shock); zero response is marked by the dotted line.}
   \label{fig:irf_figure2_uk}
\end{figure}
\FloatBarrier
\begin{figure}[!htbp]
 \centering
\includegraphics[width=\textwidth,height=0.85\textheight,keepaspectratio]
{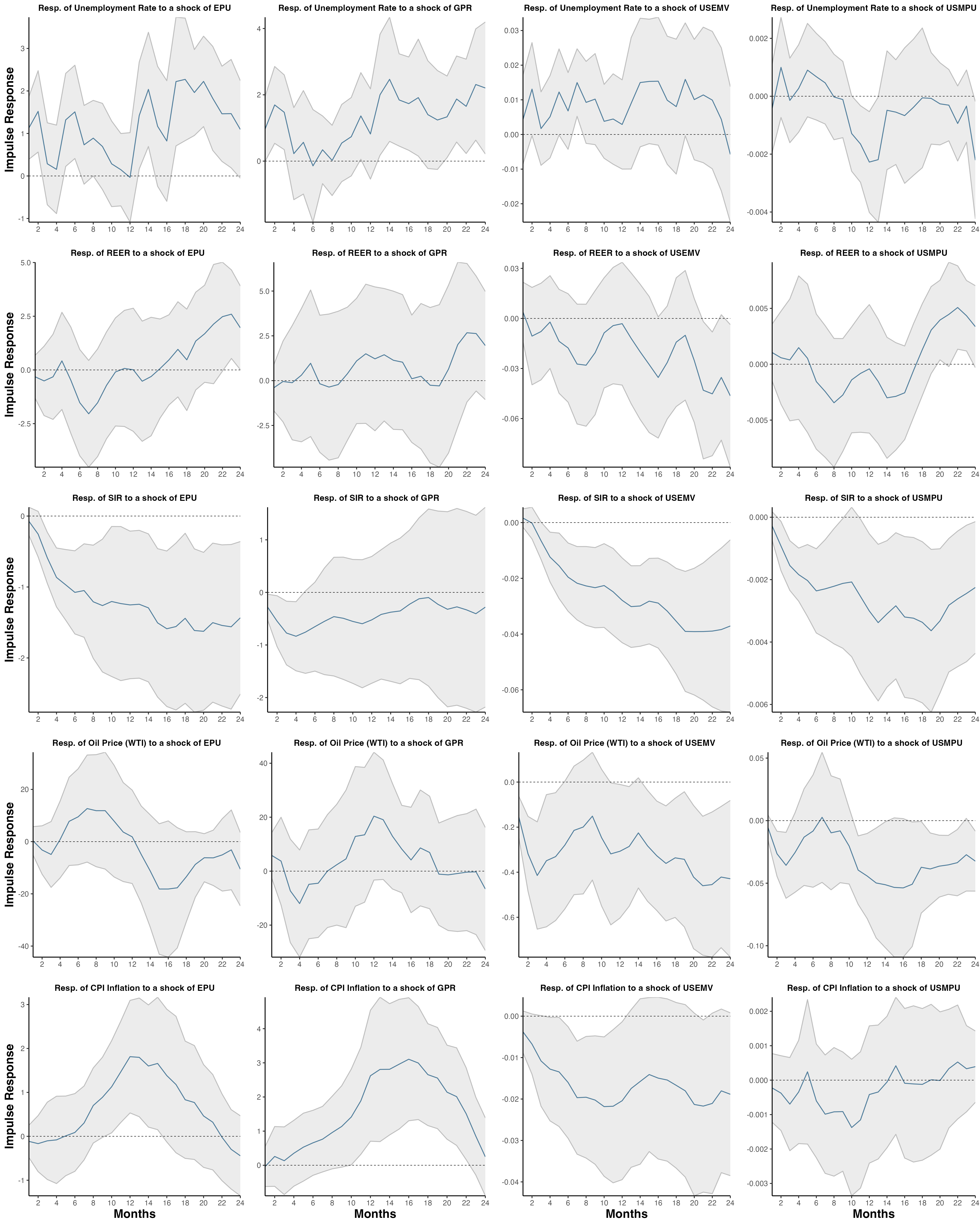}
   \caption{Impulse response functions (nonlinear local projections) of Unemployment Rate, REER, SIR, Oil Price (WTI), and CPI Inflation to EPU, GPR, USEMV, and USMPU shocks for Italy (high-rate regime). Each row shows responses of a macro variable; each column, a distinct uncertainty shock. Solid \textbf{\textcolor{blue}{blue}} lines depict IRFs with 95\% confidence bands (grey). The x-axis is horizon (months), the y-axis the response magnitude. Compare responses by reading across rows (by variable) or down columns (by shock); zero response is marked by the dotted line.}
   \label{fig:irf_figure1_ita}
\end{figure}
\FloatBarrier
\begin{figure}[!htbp]
 \centering
\includegraphics[width=\textwidth,height=0.85\textheight,keepaspectratio]
{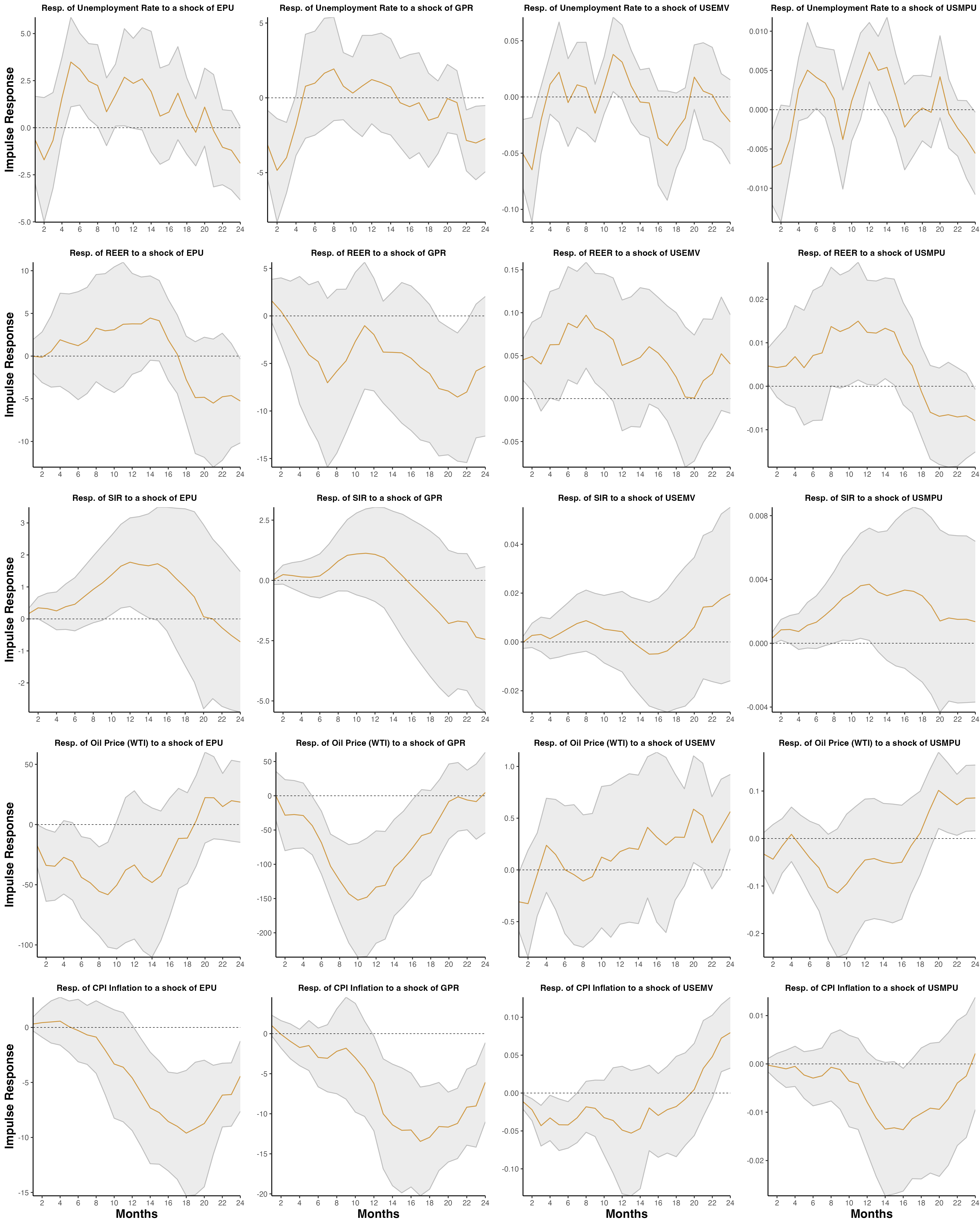}
   \caption{Impulse response functions (nonlinear local projections) of Unemployment Rate, REER, SIR, Oil Price (WTI), and CPI Inflation to EPU, GPR, USEMV, and USMPU shocks for Italy (low-rate regime). Each row shows responses of a macro variable; each column, a distinct uncertainty shock. Solid \textbf{\textcolor{orange}{orange}} lines depict IRFs with 95\% confidence bands (grey). The x-axis is horizon (months), the y-axis the response magnitude. Compare responses by reading across rows (by variable) or down columns (by shock); zero response is marked by the dotted line.} 
   \label{fig:irf_figure2_ita}
\end{figure}
\FloatBarrier

\subsection{Baseline Models: An overview}\label{app:Appendix_Baseline}
Our baseline models include:
\begin{itemize}
    \item \textit{Vector Autoregressive model with exogenous covariates (VARx):} The Vector Autoregression with exogenous variables model is a system of equations describing the joint dynamics of a set of time series with external predictors, and thereby facilitates rich cross-variable and shock transmission channels\citep{sims1980macroeconomics, kilian2006new, hamilton2020time, tsay2005analysis}.
    \item \textit{Threshold Vector Autoregressive model (TVAR):} TVAR models permit the coefficients to switch between different regimes specified by the value of a threshold variable, and so can be used to capture nonlinearities and regime shifts in time series data \citep{tsay1998testing, lo2001threshold, hansen2011threshold}. A downside to TVAR models is that they do not allow for exogenous covariates. In practice, a $p$-lag TVAR model with two regimes can be specified as: \begin{equation*} \mathbf{y}_t = \begin{cases} \mu^{(1)} + \sum_{i=1}^{p} \mathbf{A}_i^{(1)} \mathbf{y}_{t-i} + \varepsilon^{(1)}_t, & \text{if } q_{t-d} \leq r, \\ \mu^{(2)} + \sum_{i=1}^{p} \mathbf{A}_i^{(2)} \mathbf{y}_{t-i} + \varepsilon^{(2)}_t, & \text{if } q_{t-d} > r, \end{cases} \end{equation*} where $q_{t-d}$ is the threshold variable (typically a lagged endogenous series), $r$ is the estimated threshold, $d$ is the delay, and the $\mathbf{A}_i^{(j)}$'s are regime-specific coefficients. The threshold is selected by searching for the optimal value in the data. Forecasts are then generated by fitting a TVAR over the training data, and then predicting forward to obtain the required horizons.
    \item \textit{Vector Exponential Smoothing (VES):} VES jointly smooths and forecasts multiple time series, capturing trend and seasonality \citep{svetunkov2023new, hyndman2008forecasting, de2010vector}. VES does not natively incorporate exogenous variables.
    \item \textit{Neural Basis Expansion Analysis for Time Series with exogenous variables (NBEATSx): }A deep neural model using stacked fully-connected blocks and backward/forward basis expansions, for flexible and highly accurate forecasting, that allows for exogenous regressors, excels at pure time series signal extraction \citep{oreshkin2019n}.
    \item \textit{Neural Hierarchical Interpolation for Time Series Forecasting with exogenous covariates \\ (NHITSx): }An extended version of NBEATSx that factorizes a time series with a set of hierarchical interpolation modules. Hierarchical interpolation can better model short- and long-term patterns simultaneously. NHITSx can also effectively support long input and output sequences by directly learning the hierarchical basis and interpolation weights. It improves model accuracy and computational efficiency \citep{Challu2023NHITS}.
    \item \textit{Block Recurrent Neural Network with exogenous covariates (BRNNx):} BlockRNN refers to recurrent neural network models (including LSTM and GRU blocks) structured for multivariate forecasting. These models are adept at learning sequential dependencies, providing strong performance for time series with significant temporal structure or autocorrelation, and can also handle exogenous covariates \citep{herzen2022darts}.
    \item \textit{Time-Series Mixer with exogenous covariates (TSMIXERx):} TSMIXERx is an all-MLP (multi-layer perceptron) neural architecture without recurrent or convolutional components. It models time series by mixing information both along the time axis (temporal mixing) and variable axis (feature mixing) through specialized mixer layers, offering memory and computational scalability. TSMIXERx is adept at capturing cross-variable dependencies and can process exogenous variables efficiently \citep{chen2023tsmixer, ekambaram2023tsmixer}.
    \item \textit{CatBoostx, Light Gradient Boosting Machine (LGBMx), \& Xtreme Gradient Boosting Machine (XGBx) with exogenous covairates:} These are tree ensemble methods that accept both categorical and numerical features, can model nonlinearities, and can ingest exogenous variables as additional features (\citealp{prokhorenkova2018catboost, ke2017lightgbm, chen2016xgboost, herzen2022darts}).
    \item \textit{Temporal Fusion Transformer with exogenous covariates (TFTx): } TFTx incorporates attention and gating mechanisms to balance variable selection, interpretability, and model expressiveness for multi-horizon time series forecasting with exogenous data \citep{lim2021temporal}.
    \item \textit{Time-Series Dense Encoder with exogenous covariates (TiDEx): } A recently developed deep neural architecture with dense encoders and decoders, designed for long-term time series prediction, offering a good trade-off between model expressivity and efficiency\citep{das2023long}.
    \item \textit{DishTS with CatBoostx ($DTS\_CBx$) \& DishTS with XGBx ($DTS\_XGBx$) with exogenous covariates:} These two hybrid models use DishTS (Distribution Shift in Time Series Forecasting approach as proposed by \cite{fan2023dishts}) to normalize the time series and extract robust feature representations to tackle nonstationarity and distribution shifts. The time series transformed with DishTS, along with the covariates, are fed to CatBoostx or XGBoostx ensemble regressors. These models combine the bias-correction and normalization from DishTS with modern gradient boosting ensembles for forecasting strength and regularization, resulting in an improved robustness to structural shifts for macroeconomic data \citep{fan2023dishts}.
\end{itemize}
It is worth emphasizing that, as in their univariate versions, TVAR and VES models in their multivariate forms do not allow for exogenous variables to enter the model specification. This rules out the possibility to fully account for the presence of external shocks in a multivariate setting. Overall, this very comprehensive set of baselines allow us to robustly benchmark SZBVARx against both conventional econometric approaches and the frontiers of machine learning.

\subsection{Statistical significance of the results}\label{app:Appendix_stat_significance}
\begin{figure}[htbp!]
 \centering
  \includegraphics[width=0.94\textwidth]
  {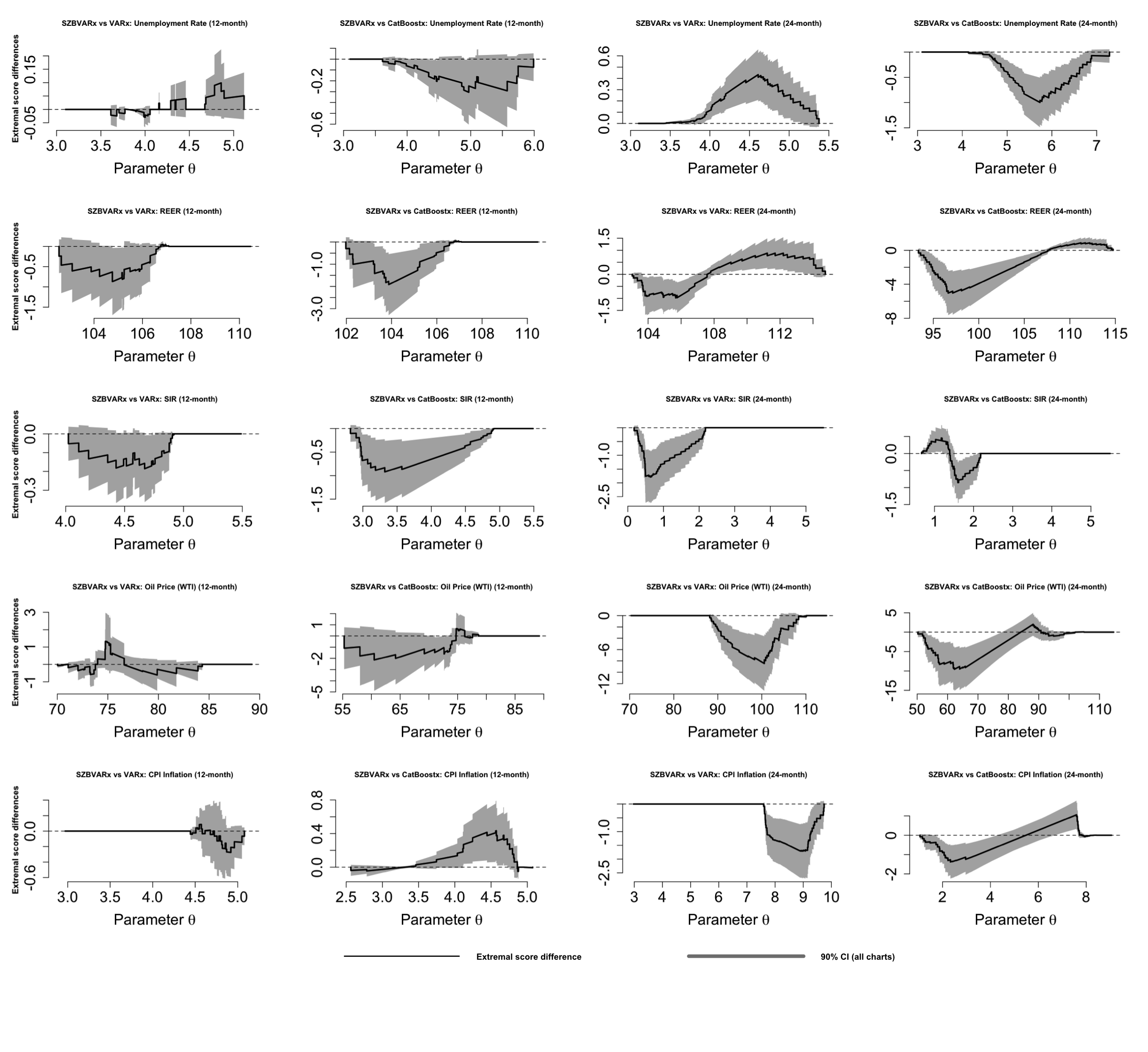}
   \caption{Murphy diagram difference plots comparing SZBVARx forecasting performance against VARx and CatBoostx baselines across 12-month-and 24-month-ahead forecast horizons for the US. Panels in the first and second columns correspond to the 12-month horizon, while panels in the third and fourth columns display the results for the 24-month horizon. The analysis covers five key macroeconomic indicators: Unemployment Rate, REER, SIR, Oil Price (WTI), and CPI Inflation. Each subplot displays extremal score differences with 90\% HAC-based confidence bands (gray shaded areas) across threshold parameter values. Negative differences indicate superior SZBVARx performance, while the magnitude reflects the strength of performance advantage at each decision threshold.}
   \label{fig:MD_usa}
\end{figure}
\FloatBarrier
\begin{figure}[htbp!]
 \centering
  \includegraphics[width=\textwidth]
  {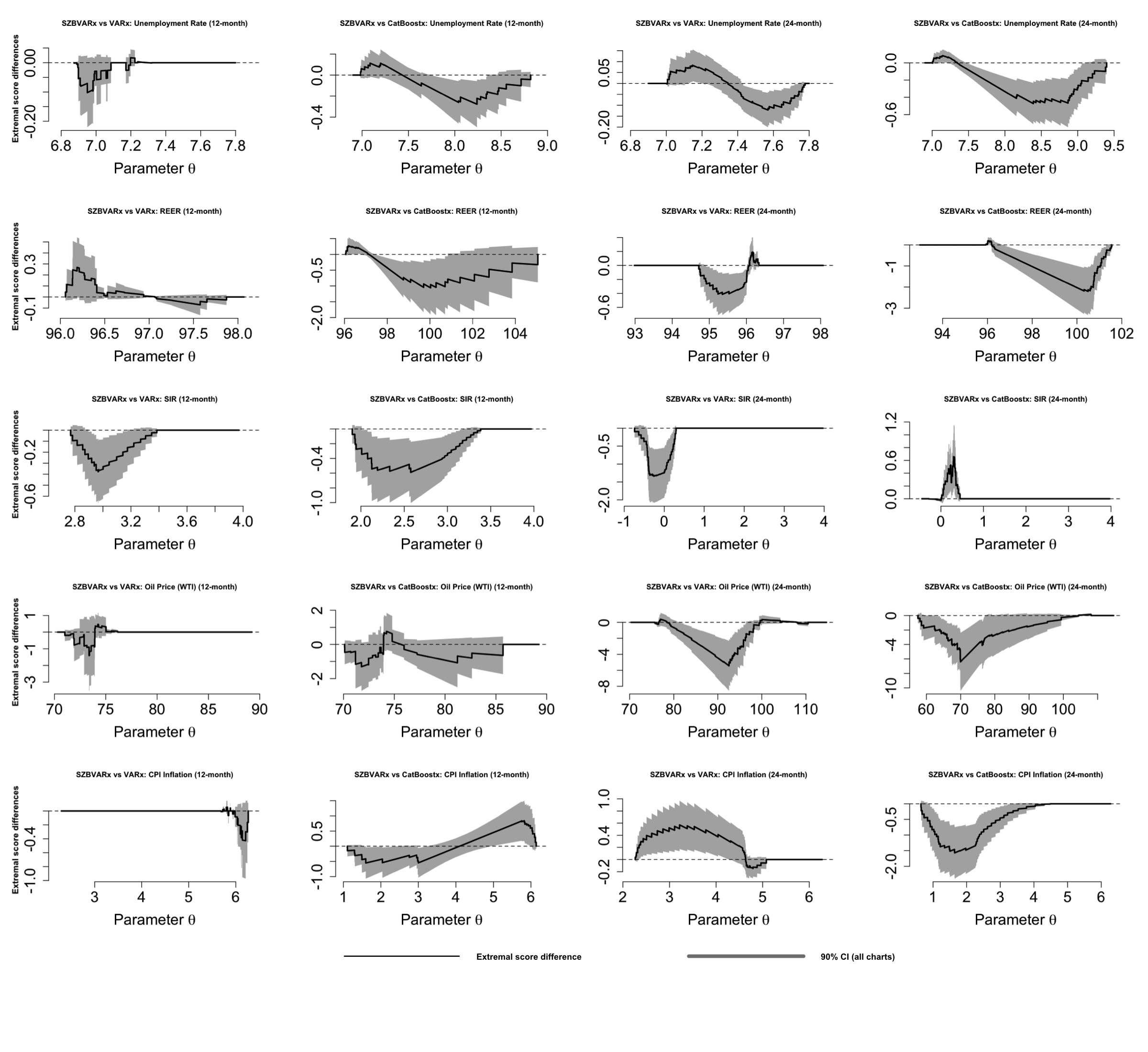}
   \caption{Murphy diagram difference plots comparing SZBVARx forecasting performance against VARx and CatBoostx baselines across 12-month-and 24-month-ahead forecast horizons for France. Panels in the first and second columns correspond to the 12-month horizon, while panels in the third and fourth columns display the results for the 24-month horizon. The analysis covers five key macroeconomic indicators: Unemployment Rate, REER, SIR, Oil Price (WTI), and CPI Inflation. Each subplot displays extremal score differences with 90\% HAC-based confidence bands (gray shaded areas) across threshold parameter values. Negative differences indicate superior SZBVARx performance, while the magnitude reflects the strength of the performance advantage at each decision threshold.}
   \label{fig:MD_france}
\end{figure}
\FloatBarrier
\begin{figure}[htbp!]
 \centering
  \includegraphics[width=\textwidth]
  {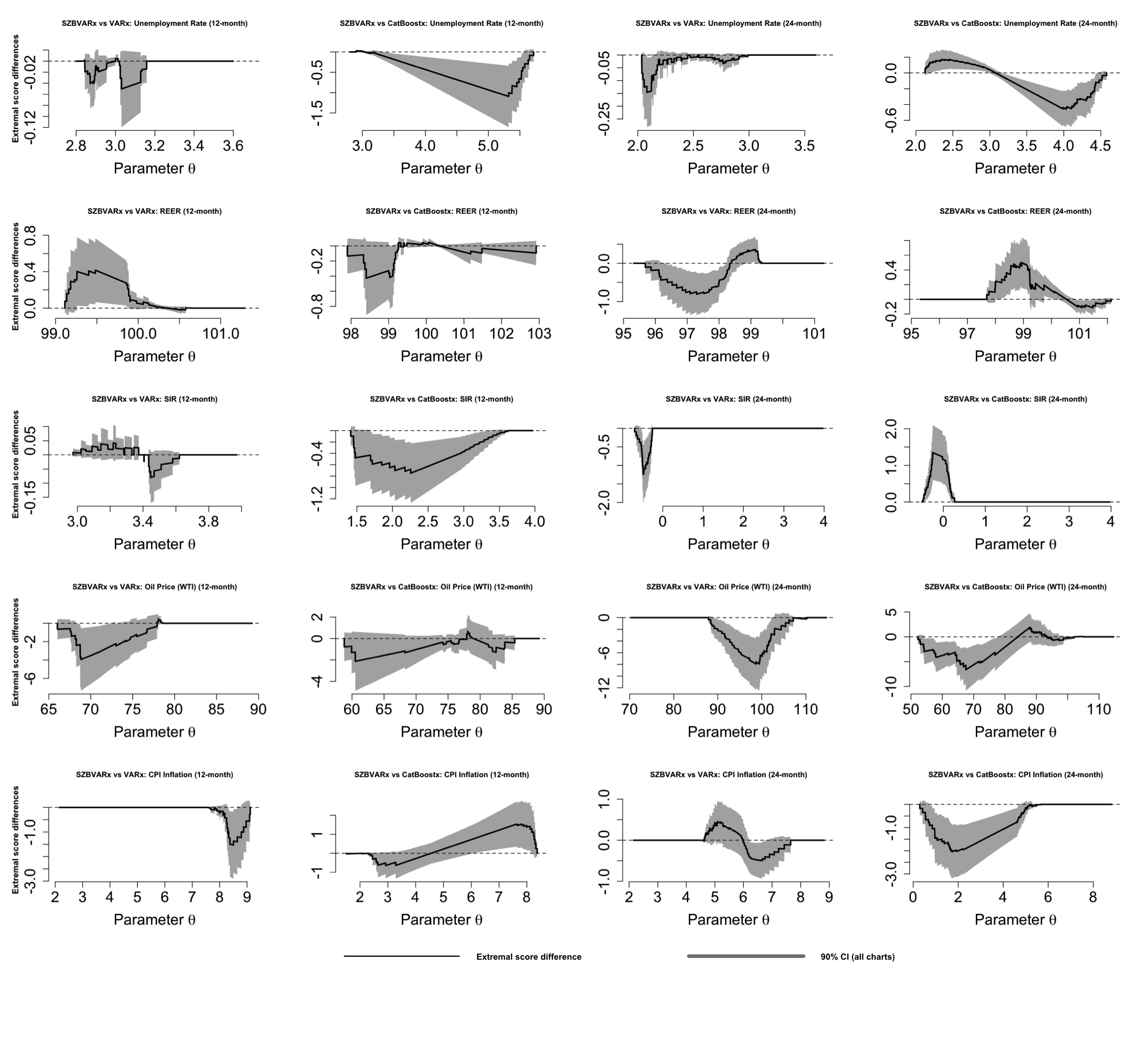}
   \caption{Murphy diagram difference plots comparing SZBVARx forecasting performance against VARx and CatBoostx baselines across 12-month-and 24-month-ahead forecast horizons for Germany. Panels in the first and second columns correspond to the 12-month horizon, while panels in the third and fourth columns display the results for the 24-month horizon. The analysis covers five key macroeconomic indicators: Unemployment Rate, REER, SIR, Oil Price (WTI), and CPI Inflation. Each subplot displays extremal score differences with 90\% HAC-based confidence bands (gray shaded areas) across threshold parameter values. Negative differences indicate superior SZBVARx performance, while the magnitude reflects the strength of the performance advantage at each decision threshold.}
   \label{fig:MD_germany}
\end{figure}
\FloatBarrier
\begin{figure}[htbp!]
 \centering
  \includegraphics[width=\textwidth]
  {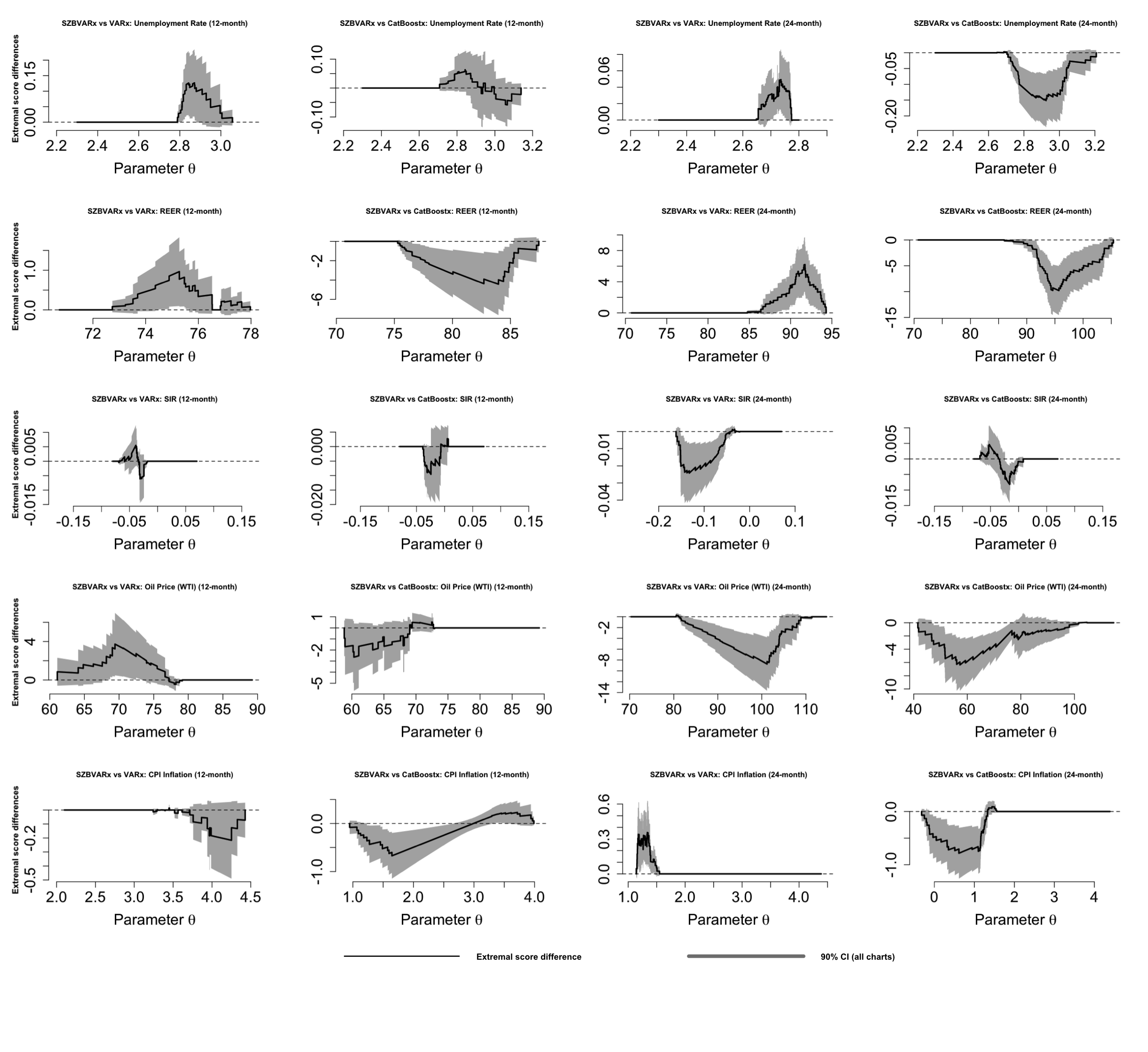}
   \caption{Murphy diagram difference plots comparing SZBVARx forecasting performance against VARx and CatBoostx baselines across 12-month-and 24-month-ahead forecast horizons for Japan. Panels in the first and second columns correspond to the 12-month horizon, while panels in the third and fourth columns display the results for the 24-month horizon. The analysis covers five key macroeconomic indicators: Unemployment Rate, REER, SIR, Oil Price (WTI), and CPI Inflation. Each subplot displays extremal score differences with 90\% HAC-based confidence bands (gray shaded areas) across threshold parameter values. Negative differences indicate superior SZBVARx performance, while the magnitude reflects the strength of the performance advantage at each decision threshold.}
   \label{fig:MD_japan}
\end{figure}
\FloatBarrier
\begin{figure}[htbp!]
 \centering
  \includegraphics[width=\textwidth]
  {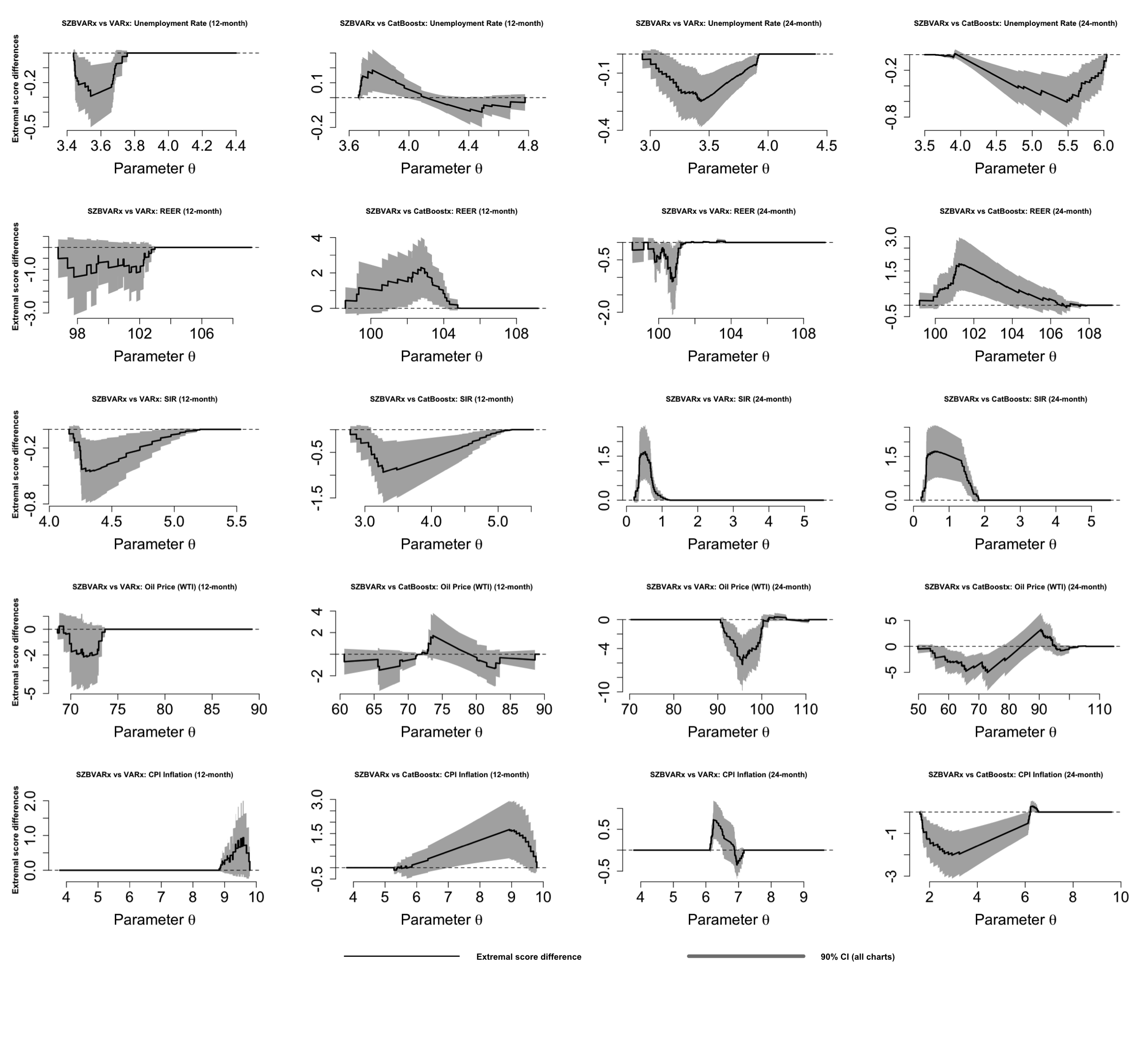}
   \caption{Murphy diagram difference plots comparing SZBVARx forecasting performance against VARx and CatBoostx baselines across 12-month-and 24-month-ahead forecast horizons for the UK. Panels in the first and second columns correspond to the 12-month horizon, while panels in the third and fourth columns display the results for the 24-month horizon. The analysis covers five key macroeconomic indicators: Unemployment Rate, REER, SIR, Oil Price (WTI), and CPI Inflation. Each subplot displays extremal score differences with 90\% HAC-based confidence bands (gray shaded areas) across threshold parameter values. Negative differences indicate superior SZBVARx performance, while the magnitude reflects the strength of the performance advantage at each decision threshold.}
   \label{fig:MD_uk}
\end{figure}
\FloatBarrier
\begin{figure}[htbp!]
 \centering
  \includegraphics[width=\textwidth]
  {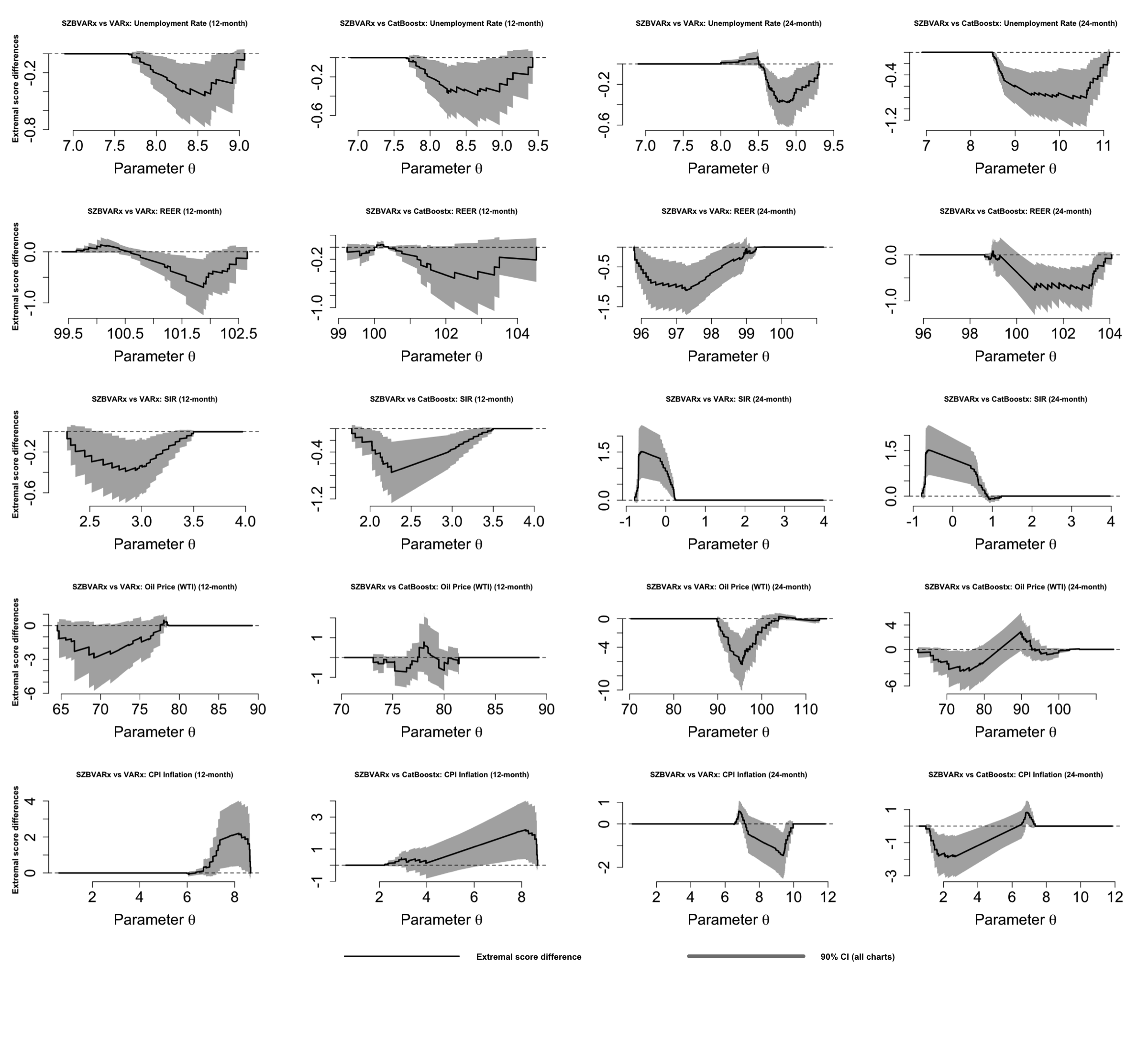}
   \caption{Murphy diagram difference plots comparing SZBVARx forecasting performance against VARx and CatBoostx baselines across 12-month-and 24-month-ahead forecast horizons for Italy. Panels in the first and second columns correspond to the 12-month horizon, while panels in the third and fourth columns display the results for the 24-month horizon. The analysis covers five key macroeconomic indicators: Unemployment Rate, REER, SIR, Oil Price (WTI), and CPI Inflation. Each subplot displays extremal score differences with 90\% HAC-based confidence bands (gray shaded areas) across threshold parameter values. Negative differences indicate superior SZBVARx performance, while the magnitude reflects the strength of the performance advantage at each decision threshold.}
   \label{fig:MD_italy}
\end{figure}
\FloatBarrier
\begin{table*}[htbp!]
\centering
\caption{Multivariate Giacomini-White test comparing SZBVARx with VARx and CatBoostx across G7 countries, five macroeconomic indicators (Unemployment Rate, RER, SIR, Oil Price (WTI), CPI Inflation), and 12-month-and 24-month-ahead forecast horizons. \textbf{Bold with *} denotes statistical significance at the 10\% level, indicating rejection of equal conditional predictive ability and highlighting SZBVARx’s robust gains.}
\label{tab:mult_GW_g7}
\begin{adjustbox}{width=0.95\textwidth}
\begin{tabular}{llcccc}
\toprule
\textbf{Country} & \textbf{Variable} 
& \begin{tabular}{@{}c@{}}\textbf{SZBVARx vs} \\ \textbf{VARx (12M)} \\
\textbf{(p-value)}\end{tabular} 
& \begin{tabular}{@{}c@{}}\textbf{SZBVARx vs} \\ \textbf{CatBoostx (12M)} \\
\textbf{(p-value)}\end{tabular}
& \begin{tabular}{@{}c@{}}\textbf{SZBVARx vs} \\ \textbf{VARx (24M)} \\
\textbf{(p-value)}\end{tabular}
& \begin{tabular}{@{}c@{}}\textbf{SZBVARx vs} \\ \textbf{CatBoostx (24M)} \\
\textbf{(p-value)}\end{tabular} \\
\midrule

Canada & Unemployment Rate   & 0.4471              & \textbf{0.0821*}    & 0.3416              & \textbf{0.0000*} \\
       & REER      & 0.2992              & 0.3922              & \textbf{0.0242*}    & \textbf{0.0120*} \\
       & SIR       & \textbf{0.0015*}    & \textbf{0.0000*}    & \textbf{0.0041*}    & 0.7414           \\
       & Oil Price (WTI)     & 0.1163              & \textbf{0.0004*}    & \textbf{0.0009*}    & \textbf{0.0072*} \\
       & CPI Inflation       & \textbf{0.0000*}    & 0.3139    & \textbf{0.0791*}    & \textbf{0.0396*} \\
\midrule

US & Unemployment Rate      & 0.2658              & \textbf{0.0099*}    & \textbf{0.0000*}    & \textbf{0.0000*} \\
    & REER         & \textbf{0.0158*}    & \textbf{0.0000*}    & 0.6372              & \textbf{0.0000*} \\
    & SIR          & \textbf{0.0024*}    & \textbf{0.0000*}    & \textbf{0.0046*}    & 0.6034           \\
    & Oil Price (WTI)        & \textbf{0.0724*}    & 0.1616              & \textbf{0.0005*}    & \textbf{0.0528*} \\
    & CPI Inflation          & 0.6139              & \textbf{0.0000*}    & \textbf{0.0121*}    & 0.5240           \\
\midrule

France & Unemployment Rate   & \textbf{0.0174*}    & \textbf{0.0071*}    & 0.6266              & \textbf{0.0000*} \\
       & REER      & 0.2960              & \textbf{0.0324*}    & 0.1162              & \textbf{0.0000*} \\
       & SIR       & \textbf{0.0183*}    & \textbf{0.0000*}    & \textbf{0.0059*}    & \textbf{0.0020*} \\
       & Oil Price (WTI)     & 0.4691              & \textbf{0.0739*}    & \textbf{0.0005*}    & \textbf{0.0356*} \\
       & CPI Inflation       & 0.1068              & 0.8804              & \textbf{0.0351*}    & \textbf{0.0011*} \\
\midrule

Germany & Unemployment Rate  & \textbf{0.0324*}    & \textbf{0.0000*}    & \textbf{0.0025*}    & \textbf{0.0063*} \\
        & REER     & \textbf{0.0036*}    & \textbf{0.0000*}    & 0.1992              & \textbf{0.0530*} \\
        & SIR      & 0.9062              & \textbf{0.0000*}    & \textbf{0.0608*}    & \textbf{0.0001*} \\
        & Oil Price (WTI)    & \textbf{0.0000*}    & 0.1119              & \textbf{0.0003*}    & \textbf{0.0645*} \\
        & CPI Inflation      & \textbf{0.0233*}    & 0.3276              & 0.8558              & \textbf{0.0101*} \\
\midrule

Japan & Unemployment Rate    & \textbf{0.0110*}    & 0.9039              & \textbf{0.0000*}    & \textbf{0.0000*} \\
      & REER       & \textbf{0.0048*}    & \textbf{0.0000*}    & \textbf{0.0007*}    & \textbf{0.0000*} \\
      & SIR        & 0.9373              & 0.4240              & \textbf{0.0349*}    & 0.6121           \\
      & Oil Price (WTI)      & \textbf{0.0010*}    & \textbf{0.0841*}    & \textbf{0.0003*}    & \textbf{0.0000*} \\
      & CPI Inflation        & 0.1240              & \textbf{0.0003*}    & \textbf{0.0000*}    & \textbf{0.0131*} \\
\midrule

UK & Unemployment Rate       & \textbf{0.0000*}    & 0.6738              & \textbf{0.0068*}    & \textbf{0.0000*} \\
   & REER          & \textbf{0.0005*}    & \textbf{0.0234*}    & \textbf{0.0074*}    & 0.1159           \\
   & SIR           & \textbf{0.0000*}    & \textbf{0.0000*}    & \textbf{0.0000*}    & \textbf{0.0001*} \\
   & Oil Price (WTI)         & 0.1846              & 0.3239              & \textbf{0.0001*}    & \textbf{0.0595*} \\
   & CPI Inflation           & \textbf{0.0008*}    & \textbf{0.0375*}    & 0.4879              & \textbf{0.0013*} \\
\midrule

Italy & Unemployment Rate    & \textbf{0.0000*}    & \textbf{0.0069*}    & \textbf{0.0557*}    & \textbf{0.0002*} \\
      & REER       & \textbf{0.0176*}    & \textbf{0.0913*}    & \textbf{0.0269*}    & \textbf{0.0000*} \\
      & SIR        & \textbf{0.0199*}    & \textbf{0.0000*}    & \textbf{0.0001*}    & \textbf{0.0028*} \\
      & Oil Price (WTI)      & \textbf{0.0401*}    & 0.4465              & \textbf{0.0000*}    & 0.5644           \\
      & CPI Inflation        & \textbf{0.0123*}    & 0.1351              & 0.3587              & 0.2495           \\
\bottomrule
\end{tabular}
\end{adjustbox}
\end{table*}
\FloatBarrier

\subsection{Empirical Results - Uncertainty Quantification through Predictive Credible Intervals}\label{app:Appendix_PPI_12M}
\begin{figure}[!htbp]
\centering
\includegraphics[width=0.95\textwidth,height=0.95\textheight,keepaspectratio]
{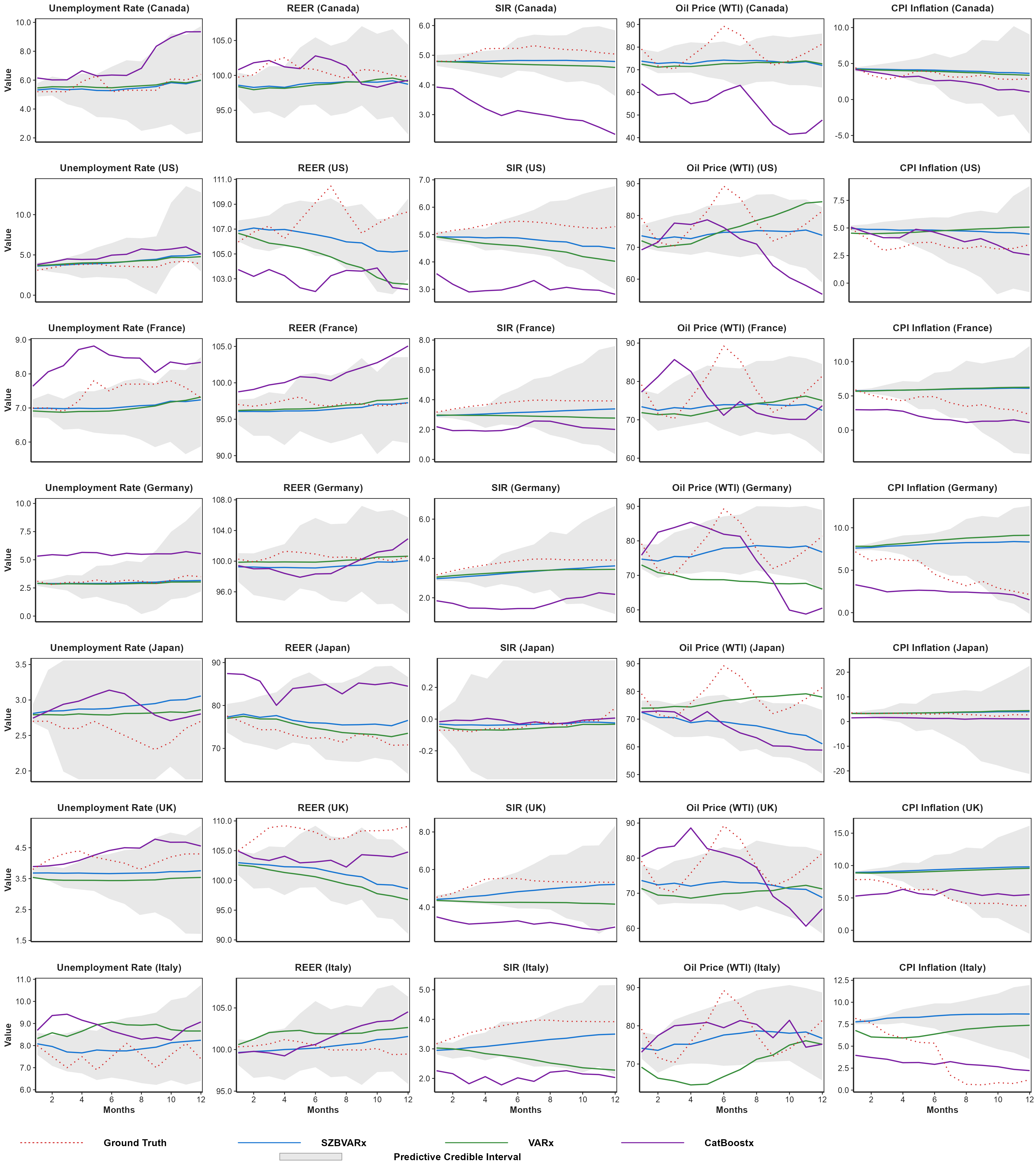}
\caption{Visualization of 12-month-ahead forecasts for five macroeconomic indicators across G7 countries. Each panel displays the ground truth (red dotted line), point forecasts from SZBVARx (blue), VARx (green), and CatBoostx (purple) models, along with predictive credible intervals from SZBVARx (grey shaded regions) quantifying forecast uncertainty. Variables shown are: Unemployment Rate, REER, SIR, Oil Price (WTI), and CPI Inflation Rate. Rows represent countries (Canada, US, France, Germany, Japan, UK, Italy from top to bottom), while columns represent the five macroeconomic variables.}
\label{fig:PPI_12M_G7}
\end{figure}
\FloatBarrier
Figure~\ref{fig:PPI_12M_G7} reports the 12-month-ahead forecast performance for all G7 countries. Overall, results are qualitatively similar to the 24-months horizon, with some notable quantitative differences. As it is standard, the credible intervals are narrower. In particular, the SZBVARx model exhibits, once again, a strong alignment with ground truth as compared to VARx and CatBoostx baselines and the predictive credible intervals cover most of the realized values. In general, the intervals are widening over time and this pattern is even more accentuated than for the 24 months ahead horizon. The information transmission through exchange rates has the overall same features as in the 24-month-ahead horizon. In particular, the unemployment rate displays the narrowest intervals in the United Kingdom and Italy and the widest in France and the United States. For the REER, the tightest intervals are estimated for Germany and the United States, while the widest are estimated for Japan and Canada. As for the 24-mont-ahead horizon, the forecasted SIR exhibits the narrowest uncertainty bands in the United States and Germany, while the wider intervals are obtained for France and Canada. This again points to important heterogeneity in monetary policy across countries, which is materializing in economic activity dynamics. Predictions of oil prices are instead fairly homogeneous across countries, with moderate uncertainty bands. Finally, the CPI inflation forecasts show large cross-country dispersion in predictive uncertainty, with the United Kingdom, the United States, and Germany being the countries with the narrowest intervals. France and Japan are instead the two countries with the largest predictive uncertainty, although the level of dispersion is smaller than for the 24-months horizon.  
\FloatBarrier

\end{document}